\documentclass[10pt]{article}
\usepackage{graphicx,float,latexsym,amsmath,amsfonts,amssymb,subeqnarray,colordvi,epstopdf}
\usepackage[latin1]{inputenc}
\usepackage{a4wide}

\def\R{\mathbb{R}}
\def\C{\mathbb{C}}

\newtheorem{theorem}{Theorem}

\renewcommand{\theequation}{\thesection.\arabic{equation}}

\newcommand{\lleft}{\rm{left}}
\newcommand{\rright}{\rm{right}}
\newcommand{\iin}{\rm{int}}
\newcommand{\rT}{{\rm{T}}}
\newcommand{\hU}{{\widehat U}}
\newcommand{\hl}{{\widehat \lambda}}
\newcommand{\tY}{{\widetilde Y}}

\title{ADI schemes for pricing American options\\under the Heston model}
\author{Tinne~Haentjens\footnote{Department of Mathematics and Computer Science,
University of Antwerp, Middelheimlaan 1, B-2020 Antwerp, Belgium.
\mbox{Email}: \texttt{\{tinne.haentjens,karel.inthout\}@uantwerpen.be}.}
~and Karel~J.~in 't Hout\footnotemark[\value{footnote}]
}
\date{\today}

\begin{document}

\maketitle

\begin{abstract}
\noindent
In this paper a simple, effective adaptation of Alternating
Direction Implicit (ADI) time discretization schemes is proposed
for the numerical pricing of American-style options under the
Heston model via a partial differential complementarity problem.
The stability and convergence of the new methods are extensively
investigated in actual, challenging applications.
In addition a relevant theoretical result is proved.
\end{abstract}
\vspace{0.2cm}\noindent
{\small\textbf{Key words:} Alternating Direction Implicit schemes,
American option pricing, Heston model,
linear complementarity problem, Ikonen--Toivanen splitting.}
\vspace{3mm}
\normalsize

%%%%%%%%%%%%%%%%%%%%%%%%%%%%%%%%%%%%%%%%%%%%%%%%%%%%%%%%%%%%%%%%%%%%%%%%%%%%%%%%%%%%
%%%%%%%%%%%%%%%%%%   SECTION 1   %%%%%%%%%%%%%%%%%%%%%%%%%%%%%%%%%%%%%%%%%%%%%%%%%%%
%%%%%%%%%%%%%%%%%%%%%%%%%%%%%%%%%%%%%%%%%%%%%%%%%%%%%%%%%%%%%%%%%%%%%%%%%%%%%%%%%%%%
\setcounter{equation}{0}
\section{Introduction}\label{intro}
This paper is concerned with the numerical valuation of
American-style options.
We consider the adaptation of the well-known class of
Alternating Direction Implicit (ADI) time discretization
schemes.
These splitting schemes have proven to be highly efficient,
stable and robust in the numerical pricing
of European-style options via multidimensional partial
differential equations (PDEs).
A study of their potential for American-style options is
still in its infancy, however.
In the present paper we propose and analyze an effective
adaptation of ADI schemes to the pricing of this important
type of options.
For the underlying asset price process the popular Heston
stochastic volatility model~\cite{H93} is considered.

Let $u(s,v,t)$ be the fair value of a vanilla American
put option under the Heston model if at $t$ time units
before the given maturity time the underlying asset price
equals $s\ge 0$ and its variance equals $v\ge 0$.
The Heston spatial differential operator $\mathcal{A}$
applied to the function $u$ is denoted by
\begin{equation}\label{Heston}
\mathcal{A} u  =
\tfrac{1}{2} s^2 v \frac{\partial^2 u}{\partial s^2}
+ \rho \sigma s v \frac{\partial^2 u}{\partial s \partial v}
+ \tfrac{1}{2}\sigma^2 v \frac{\partial^2 u}{\partial v^2}
+ r s \frac{\partial u}{\partial s} +
\kappa ( \eta - v) \frac{\partial u}{\partial v} - r u
\quad (s>0,~v>0).
\end{equation}
Here parameter $\kappa>0$ is the mean-reversion rate, $\eta>0$
is the long-term mean, $\sigma>0$ is the volatility-of-variance,
$\rho \in [-1,1]$ is the correlation between the two underlying
Brownian motions and $r$ is the risk-neutral interest rate.
We note that in the literature it is sometimes assumed that the
so-called Feller condition $2\kappa\eta > \sigma^2$ is fulfilled,
but in this paper we shall make no such assumption.
Let $K,\, T>0$ be the given strike price and maturity time
of the American put option and denote the payoff function by
\begin{equation}\label{payoff}
\phi(s) = \max(K-s,0) \quad (s\ge 0).
\end{equation}
It is well-known
that the option value function $u$ satisfies the following
so-called {\it partial differential complementarity problem
(PDCP)}:
\begin{equation}\label{H_pdcp}
\frac{\partial u}{\partial t}\ge \mathcal{A} u,
\quad u\ge \phi,
\quad (u-\phi)\left(\frac{\partial u}{\partial t} -
\mathcal{A} u\right)=0,
\end{equation}
valid pointwise for $(s,v,t)$ with $s>0$, \,$v>0$,\,
$0< t \le T$.
The Heston PDCP (\ref{H_pdcp}) is complemented with initial
and boundary conditions.
The initial condition is $u(s,v,0) = \phi(s)$ (for $s\ge 0$,
$v\ge 0$).
Boundary conditions are given in Section~\ref{space} below.

The three conditions in (\ref{H_pdcp}) naturally induce a
decomposition of the $(s,v,t)$-space:
the continuation region is the set of all points $(s,v,t)$
where the equality $\partial u/\partial t= \mathcal{A}u$
holds (and the option is held);
the exercise region is the set of all points $(s,v,t)$ where
the equality $u=\phi$ holds (and the option is exercised).
The joint boundary of these two regions is called the free
boundary or exercise boundary.
Both the option value function $u$ and the free boundary
are unknown in closed form.
Further, even though no rigorous proof appears to be
available at present, the function $u$ is expected to
suffer from a lack of smoothness on the free boundary,
as in the Black--Scholes case.

The numerical solution of the Heston PDCP for American-style
option prices has already been considered by various authors
in the literature.
We give here a brief overview of the main contributions.

Clarke \& Parrott~\cite{CP99} apply finite difference schemes
for the spatial discretization of (\ref{H_pdcp}) followed by
the $\theta$-method for the time discretization.
This gives rise to a fully discrete linear complementarity
problem (LCP) that needs to be solved in every time step.
As it turns out, the common projected SOR method (see
e.g.~\cite{TR00,WDH95}) is often too slow for the Heston
LCP and in \cite{CP99} the authors propose a multigrid
method.
It is based on the projected full approximation scheme
(PFAS) by Brandt \& Cryer~\cite{BC83}.

Oosterlee~\cite{O03} provides a detailed Fourier analysis
of the PFAS approach for the Heston LCP and concludes, in
particular, that an alternating line smoother is robust.
For the time discretization the second-order backward
differentiation formula (BDF2) was used in \cite{O03}.

Toivanen \& Oosterlee~\cite{TO12} present a projected
algebraic multigrid method for LCPs and show that this
is faster than geometric multigrid in the case of the
Heston LCP.

Zvan, Forsyth \& Vetzal \cite{ZFV98} view the American option
pricing problem as a nonlinear Heston PDE, where the early
exercise constraint is incorporated by a penalty method.~After
spatial discretization, by finite element/volume schemes,
and time discretization, by the $\theta$-method, they
obtain a nonlinear system of algebraic equations that
is numerically solved through an approximate Newton method
with a preconditioned CGSTAB iteration.
The penalty method is shown to yield the Heston LCP
as the penalty parameter tends to infinity.

Ikonen \& Toivanen~\cite{IT07} propose splitting methods for
the time discretization of the semidiscretized Heston PDCP.
The methods under consideration can be viewed as analogues of
well-known fractional step methods or locally one-dimensional
(LOD) methods for systems of ordinary differential equations.
In particular, the symmetric Strang-type splitting is adapted
to the semidiscretized Heston PDCP.
This leads to five LCPs per time step, each with a tridiagonal
matrix.
These simple LCPs are then exactly solved by the efficient
Brennan--Schwartz algorithm, introduced in~\cite{BS77}
for pricing American options under the (one-dimensional)
Black--Scholes model.
We remark that, since the methods in \cite{IT07} are based
on LOD methods, special attention must be given to the
treatment of the PDCP boundary conditions, otherwise order
reduction might occur.
Further, the Brennan--Schwartz algorithm is only applicable
under restrictive assumptions on the spatial discretization
as well as (the shape of) the free boundary for the option
price.

Somewhat related to \cite{IT07},
Villeneuve \& Zanette~\cite{VZ02} previously formulated
two adaptations of the original Peaceman--Rachford ADI
scheme~\cite{PR55} to the semidiscretized PDCP from the
pricing of American options on two assets under the
Black--Scholes model.
These authors first perform a coordinate transformation,
so as to arrive at an operator that is essentially the
standard two-dimensional Laplacian.
Correspondingly, a generalization of the approach from
\cite{VZ02} to the Heston model is not directly clear.

Ikonen \& Toivanen~\cite{IT09} propose a novel splitting
technique, which applies to the Heston LCP obtained after
space and time discretization.
Their idea, originally put forth in \cite{IT04} in the case
of the Black--Scholes model, is to introduce
an auxiliary variable such that a decoupling is achieved
between the underlying time discretization scheme and the
enforcement of the early exercise constraint.
The amount of computational work per time step of this
approach is governed by the solution of the pertinent
linear systems, for which the authors employ multigrid.
The computational cost of the update of the early exercise
constraint is negligible.
The time discretization schemes under consideration are
backward Euler, Crank--Nicolson and BDF2 as well as a
second-order, $L$-stable Runge--Kutta method.
It is shown \cite{IT08,IT09} that this splitting approach
performs well and is efficient.

Persson \& Von Sydow \cite{PS10} consider a tailored,
adaptive space-time discretization of the Heston PDCP
based on finite differences and the BDF2 method and
apply the splitting technique from \cite{IT09} where
for the solution of the linear systems a preconditioned
GMRES iteration is used.

The main aim of our present paper is to show that by
invoking the splitting idea from \cite{IT09} an effective
adaptation of ADI time discretization schemes to the
semidiscretized Heston PDCP for American-style options
is attained.
We refer to the newly obtained methods as ADI-IT methods.
The amount of computational work per time step of an
ADI-IT method is determined by the solution
method for the pertinent linear systems, as in
\cite{IT09}.
But, contrary to \cite{IT09}, these linear systems now
only involve matrices with a fixed, small bandwidth.
They can therefore be exactly and easily solved in an
efficient manner by using $LU$ factorization.
In our note \cite{HIHV10} this approach was already
briefly introduced.
In the present paper we shall give a comprehensive
study.
An outline of the rest of the paper is as follows.

Section~\ref{space} describes the discretization of the
Heston operator (\ref{Heston}) by finite difference
schemes on nonuniform Cartesian grids, leading to a
semidiscrete version of the Heston PDCP (\ref{H_pdcp}).
In Section~\ref{time_theta}, we first consider time
discretization by the common $\theta$-method.
For the resulting Heston LCP the splitting technique
by Ikonen \& Toivanen~\cite{IT09} is formulated and
discussed.
We prove a useful theorem on the closeness of the
numerical solutions with and without splitting.
Next, in Section~\ref{time_ADI} our adaptation of ADI
time discretization schemes to the semidiscrete Heston
PDCP is defined.
Four known ADI schemes are considered: the Douglas scheme,
the Craig--Sneyd scheme, the modified Craig--Sneyd scheme
and the Hundsdorfer--Verwer scheme.
Section~\ref{numexp} presents an extensive numerical
study of the acquired ADI-IT methods.
We investigate in detail their actual stability and
convergence behavior in a variety of representative,
challenging test cases - with short and long maturity
times, with zero and nonzero correlation, for cases
where the Feller condition holds and cases where it
is violated, and for vanilla American put options as
well as capped American put options.
Also, approximations obtained with the ADI-IT approach
are compared to known approximations from the literature
reviewed above and for all test cases the computed option
price surfaces and free boundaries are graphically displayed.
In Section~\ref{concl} conclusions are given.

%%%%%%%%%%%%%%%%%%%%%%%%%%%%%%%%%%%%%%%%%%%%%%%%%%%%%%%%%%%%%%%%%%%%%%%%%%%%%%%%%%%%
%%%%%%%%%%%%%%%%%%   SECTION 2   %%%%%%%%%%%%%%%%%%%%%%%%%%%%%%%%%%%%%%%%%%%%%%%%%%%
%%%%%%%%%%%%%%%%%%%%%%%%%%%%%%%%%%%%%%%%%%%%%%%%%%%%%%%%%%%%%%%%%%%%%%%%%%%%%%%%%%%%
\setcounter{equation}{0}
\section{Spatial discretization}\label{space}

The first step in numerically solving the Heston PDCP (\ref{H_pdcp}) is
the discretization of the Heston operator (\ref{Heston}).
For the spatial variable $s$ resp.~$v$ the domain is restricted to a
bounded set $[0,S_{\max}]$ resp.~$[0,V_{\max}]$ with fixed values
$S_{\max}$, $V_{\max}$ taken sufficiently large.
We deal with boundary conditions of Dirichlet and Neumann type, determined
by the specific option under consideration, or no condition, in the case of
a degenerate boundary.

For a vanilla American put option the following boundary conditions are
common in the literature.
\begin{itemize}
    \item In the $s$-direction:
        \begin{subeqnarray}\label{ibcs}
            \quad u(0,v,t) &=&K,\\
            \quad \frac{\partial u}{\partial s}(S_{\max},v,t) &=&0.
        \end{subeqnarray}
    \item In the $v$-direction:
        \begin{equation}\label{ibcv}
            \quad \frac{\partial u}{\partial v} (s,V_{\max},t)~~ = ~~\, 0.~
        \end{equation}
          Note the degeneracy feature of the Heston operator (\ref{Heston})
          in the $v$-direction that all second-order derivative terms vanish
          and the operator becomes convection-dominated for $v\downarrow0$.
          Relatedly, at $v=0$, it is assumed that the Heston PDCP (\ref{H_pdcp})
          is fulfilled.
\end{itemize}

For the discretization of (\ref{Heston}) a suitable Cartesian grid is chosen
on $[0,S_{\max}] \times [0,V_{\max}]$
with nonuniform meshes $0 = s_0 < s_1 < \ldots < s_{m_1} = S_{\max}$ and
$0 = v_0 < v_1 < \ldots < v_{m_2} = V_{\max}$ in the $s$- and $v$-directions.
The use of nonuniform meshes, instead of uniform ones, can substantially
improve the efficiency, cf.~e.g.~\cite{HH12}.
Moreover, when applying the nonuniform meshes defined below, taking larger
values for $S_{\max}$ and $V_{\max}$ will be computationally inexpensive.
\begin{itemize}
    \item In the $s$-direction: let integer $m_1 \ge 1$ and
    parameter $d_1>0$ and let equidistant
points
\mbox{$\xi_{\min}=\xi_0 < \xi_1 < \ldots < \xi_{m_1}=\xi_{\max}$}
be given with
\begin{align*}
\xi_{\min} &= \sinh^{-1}\left( \frac{- S_{\lleft}}{d_1} \right),\\
\xi_{\iin} &= \frac{S_{\rright}-S_{\lleft}}{d_1},\\
\xi_{\max} &= \xi_{\iin} + \sinh^{-1}\left( \frac{S_{\max} - S_{\rright}}{d_1} \right).
\end{align*}
Note that $\xi_{\min} < 0 < \xi_{\iin} < \xi_{\max}$.
We then define the mesh $0=s_0 < s_1 < \ldots < s_{m_1}=S_{\max}$ through the
transformation
\begin{equation*}
s_i = \varphi(\xi_i) \quad (0\le i \le m_1)
\end{equation*}
where
\begin{equation*}
\varphi(\xi) =
\begin{cases}
 S_{\lleft} + d_1\sinh(\xi) & ({\rm for}~\xi_{\min} \leq \xi < 0),\\
 S_{\lleft} + d_1\xi & ({\rm for}~0 \leq \xi \leq \xi_{\iin} ),\\
 S_{\rright} + d_1\sinh(\xi-\xi_{\iin}) & ({\rm for}~\xi_{\iin} < \xi \leq \xi_{\max}).
\end{cases}
\end{equation*}
The above mesh has relatively many points in a given fixed interval
$[S_{\lleft},S_{\rright}]\subset [0,S_{\max}]$.
This interval is taken in particular such that it lies in the region
of interest in applications and it contains the strike $K$.
Thus numerical difficulties due to the initial function (\ref{payoff}),
which possesses a discontinuous derivative at $K$, are alleviated.
The $s$-mesh is nonuniform outside $[S_{\lleft},S_{\rright}]$ and
uniform inside.
\item In the $v$-direction:
let integer $m_2 \ge 1$ and parameter $d_2>0$ and let equidistant points
$\psi_0 < \psi_1 < \ldots < \psi_{m_2}$ be given by
\[
\psi_j = j\cdot \Delta \psi \quad (0\le j \le m_2) ~~~{\rm with}~~~
\Delta \psi = \frac{1}{m_2} \sinh^{-1}\left(\frac{V_{\max}}{d_2}\right).
\]
We then define the mesh $0=v_0 < v_1 < \ldots < v_{m_2}=V_{\max}$ by
\[
v_j = d_2~{\rm sinh}(\psi_j) \quad (0\le j \le m_2).
\]
This nonuniform mesh has relatively many points in the neighborhood
of the important degenerate boundary $v=0$.
\end{itemize}
\noindent
Let $\Delta \xi = \xi_1-\xi_0$. It is readily verified that the $s$- and
$v$-meshes defined above are smooth, in the sense that there exist real
constants $C_0, C_1, C_2$, $C_0^\prime, C_1^\prime, C_2^\prime$ $>0$ such
that the mesh widths $\Delta s_i = s_i-s_{i-1}$ and $\Delta v_j = v_j-v_{j-1}$
satisfy
\begin{align*}\label{smooth}
  &C_0\, \Delta \xi \le \Delta s_i \le C_1\, \Delta \xi ~~ &&{\rm and} ~~
  &|\Delta s_{i+1} - \Delta s_i| \le C_2 \left( \Delta \xi \right)^2 ~~
  &&({\rm uniformly~in}~\, i,\, m_1),\\
  &C_0^\prime\, \Delta \psi \le \Delta v_j \le C_1^\prime\, \Delta \psi ~~ &&{\rm and} ~~
  &|\Delta v_{j+1} - \Delta v_j| \le C_2^\prime \left( \Delta \psi \right)^2 ~~
  &&({\rm uniformly~in}~\, j,\, m_2). \\
\end{align*}
\vskip-0.3cm\noindent
The actual choice of parameters in our numerical experiments,
in Section~\ref{numexp} below, are $d_1=K/20$, $d_2=V_{\max}/500$ and
with $r=\frac{1}{4}$,
\[
S_{\lleft}=\max\left(\tfrac{1}{2},e^{-rT}\right)K~,~~S_{\rright}=K,~~
S_{\max} = 14K,~~V_{\max} = 5.
\]
The above values for $S_{\max}$, $V_{\max}$ might be considered as large.
They were heuristically determined, so as to guarantee that the error
induced by the restriction of the spatial domain to a bounded set is
negligible in all our experiments.
As already mentioned, with the nonuniform meshes under consideration,
increasing the upper bounds $S_{\max}$, $V_{\max}$ is harmless for the
overall efficiency.
The interval $[S_{\lleft},S_{\rright}]$ has been chosen such that,
in addition to containing the strike $K$, the $s$-points of the exercise
boundary are expected to be contained in it for all practical values $v$,
$t$.
A further investigation into possibly better parameter values than above
may be interesting, but this is beyond the scope of the present paper.
\\

We discuss next the discretization of the spatial derivatives on the
selected nonuniform grid.
In view of the boundary conditions, given above, the pertinent spatial
grid is
\begin{align*}
\mathcal{G} = \{(s_i, v_j): 1\leq i \leq m_1\,,\, 0\leq j\leq m_2\}.
\end{align*}
Write $u_{i,j}=u(s_i,v_j,t)$.
We employ the following well-known finite difference (FD) schemes
for discretizing the convection and diffusion terms in the Heston
operator (\ref{Heston}).
\begin{itemize}
    \item In the $s$-direction:\\
        the \textit{forward scheme for convection}
\begin{align*}
    \frac{\partial u}{\partial s}(s_i,v_j,t) &\approx \frac{u_{i+1,j}-u_{i,j}}{\Delta s_{i+1}},
\end{align*}
\noindent
the \textit{central scheme for diffusion}
\begin{align*}
    \frac{\partial^2 u}{\partial s^2}(s_i,v_j,t) &\approx
    \frac{2}{\Delta s_i(\Delta s_i+\Delta s_{i+1})}u_{i-1,j}
    - \frac{2}{\Delta s_i\Delta s_{i+1}} u_{i,j}
    +  \frac{2}{\Delta s_{i+1}(\Delta s_i+\Delta s_{i+1})}u_{i+1,j}.
\end{align*}
    \item In the $v$-direction:\\
     the \textit{forward scheme for convection}
    \begin{align*}
    \frac{\partial u}{\partial v}(s_i,v_j,t) &\approx
    \frac{u_{i,j+1}-u_{i,j}}{\Delta v_{j+1}} \qquad \text{whenever}\quad v_j \leq \eta,
    \end{align*}
    the \textit{backward scheme for convection}
\begin{align*}
    \frac{\partial u}{\partial v}(s_i,v_j,t) &\approx
    \frac{u_{i,j}-u_{i,j-1}}{\Delta v_{j}}   \qquad \text{whenever}\quad v_j>\eta,
\end{align*}
\noindent
the \textit{central scheme for diffusion}
\begin{align*}
    \frac{\partial^2 u}{\partial v^2}(s_i,v_j,t) &\approx
    \frac{2}{\Delta v_j(\Delta v_j+\Delta v_{j+1})}u_{i,j-1}
    - \frac{2}{\Delta v_j\Delta v_{j+1}} u_{i,j}
    +  \frac{2}{\Delta v_{j+1}(\Delta v_j+\Delta v_{j+1})}u_{i,j+1}.
\end{align*}
\end{itemize}

\noindent
For arbitrary meshes the forward and backward FD schemes for convection
and central FD scheme for diffusion all possess a first-order truncation
error (whenever $u$ is sufficiently often differentiable).
For smooth meshes as in our paper, the truncation error of the central
scheme is of second-order.
\\

Special attention is needed for the FD discretization at the important
degenerate boundary as well as at the boundaries with a Neumann condition.
\begin{itemize}
\item In the $s$-direction: the Neumann condition (\ref{ibcs}b) at
$s=S_{\max}$ directly renders the first derivative $\partial u /\partial s$.
Using the central scheme, together with linear extrapolation to obtain
the required approximation at a virtual mesh point beyond $S_{\max}$,
the second derivative $\partial^2 u /\partial s^2$ is approximated.
\item In the $v$-direction: the Neumann condition (\ref{ibcv}) at $v=V_{\max}$
is treated analogously to~(\ref{ibcs}b) above.
At the degenerate boundary $v=0$ we approximate $\partial u/\partial v$
by the forward FD scheme.\footnote{Some authors deal with the $v=0$ boundary
separately, by applying an explicit time stepping scheme.
But this yields an unpractical discretization, where an excessively large
number of time steps is necessary for stability.}
Next, the term in (\ref{Heston}) involving $\partial^2 u/\partial v^2$
vanishes if $v=0$ and is thus trivially dealt with.
\end{itemize}

If the underlying asset price and variance processes are correlated,
so if $\rho \neq 0$, then the Heston operator (\ref{Heston}) contains
a mixed derivative term.
For the mixed derivative $\partial^2 u/\partial s\partial v$ we
consider a standard FD discretization based on a centered 9-point
stencil formed by successive application of the following central
FD schemes in the $s$- and $v$-directions:
\begin{align*}
    \frac{\partial u}{\partial s}(s_i,v_j,t) &\approx
    \frac{-\Delta s_{i+1}}{\Delta s_i (\Delta s_i + \Delta s_{i+1})}u_{i-1,j}
    + \frac{\Delta s_{i+1}-\Delta s_i}{\Delta s_i \Delta s_{i+1}}u_{i,j}
    + \frac{\Delta s_i}{\Delta s_{i+1} (\Delta s_i + \Delta s_{i+1})} u_{i+1,j},\\
    \frac{\partial u}{\partial v}(s_i,v_j,t) &\approx
    \frac{-\Delta v_{j+1}}{\Delta v_j (\Delta v_j + \Delta v_{j+1})}u_{i,j-1}
    + \frac{\Delta v_{j+1}-\Delta v_j}{\Delta v_j \Delta v_{j+1}}u_{i,j}
    + \frac{\Delta v_j}{\Delta v_{j+1} (\Delta v_j + \Delta v_{j+1})} u_{i,j+1}.
\end{align*}
It is readily verified that at the degenerate boundary as well as
the two Neumann boundaries the mixed derivative term in (\ref{Heston})
vanishes and, hence, is trivially dealt with.
\\

We remark that with the simple first-order FD schemes for convection
above, it is easily proved that the obtained semidiscrete Heston matrix
$A$ is such that $-A$ is always an M-matrix\footnote{For the definition
of an M-matrix see e.g.~\cite{BP79}.} if the correlation $\rho=0$.
In the literature on the pricing of financial options, this type of
condition has been used for deriving favorable properties of numerical
methods.
In general, $-A$ is not an M-matrix whenever $\rho\not=0$ and standard
FD discretizations of the mixed derivative, such as above, are applied.
More advanced discretizations of the mixed derivative have been
constructed in this case, see e.g.~\cite{IT07,IT08, IT09}.
In the present paper we shall adhere to the above standard choice,
however.
\\

For an accurate approximation of the option value function $u$ it
is beneficial to smoothen the payoff function (\ref{payoff}) at
the strike $K$, where it is discontinuous in the first derivative.
Accordingly, we replace the value of the payoff function at the mesh
point $s_i$ nearest to $K$ by its cell average, see e.g.~\cite{TR00}:
\[
\frac{1}{h}
\int_{s_{i-1/2}}^{s_{i+1/2}} \phi(s)\, ds \quad
{\rm with}~~
s_{i-1/2} = \tfrac{1}{2}(s_{i-1}+s_i),~
s_{i+1/2} = \tfrac{1}{2}(s_i+s_{i+1}),~
h = s_{i+1/2}-s_{i-1/2}.
\]

By spatial discretization, the option values $u(s,v,t)$ are
approximated at the spatial grid points $(s,v) \in \mathcal{G}$.
These approximations form the entries of a vector $U(t)$,
which is given as the solution of a \textit{semidiscrete PDCP},
\begin{equation}\label{lcp_ode}
U'(t) \ge AU(t) + g,
\quad U(t) \ge U_0,
\quad (U(t) - U_0)^\rT \left(U'(t) - AU(t)- g\right)=0\quad (0 < t \le T).
\end{equation}
Here inequalities are to be interpreted componentwise.
The size of the system (\ref{lcp_ode}) equals $M=m_1(m_2+1)$,
$A$ is a given real $M\times M$ matrix and $U_0$ and $g$ are
given real $M\times 1$ vectors determined by the initial and
boundary conditions, respectively.
Further, $^\rT$ stands for transpose.
The next main step in numerically solving (\ref{H_pdcp}) is the
time discretization of (\ref{lcp_ode}).

%%%%%%%%%%%%%%%%%%%%%%%%%%%%%%%%%%%%%%%%%%%%%%%%%%%%%%%%%%%%%%%%%%%%%%%%%%%%%%%%%%%%
%%%%%%%%%%%%%%%%%%   SECTION 3   %%%%%%%%%%%%%%%%%%%%%%%%%%%%%%%%%%%%%%%%%%%%%%%%%%%
%%%%%%%%%%%%%%%%%%%%%%%%%%%%%%%%%%%%%%%%%%%%%%%%%%%%%%%%%%%%%%%%%%%%%%%%%%%%%%%%%%%%
\setcounter{equation}{0}
\section{Time discretization: $\theta$-method}\label{time_theta}
After the spatial discretization of the Heston PDCP (\ref{H_pdcp})
in the previous section, we now consider the time discretization
of the obtained semidiscrete PDCP (\ref{lcp_ode}).
An often-used scheme for its time discretization is the $\theta$-method,
with parameter $\theta\in [\frac{1}{2}, 1]$,
see e.g.~\cite{CP99,HP98,IT07,IT08,IT09}.
The choices $\theta=\frac{1}{2}$ and $\theta=1$ represent, respectively,
the well-known Crank--Nicolson (CN) and backward Euler (BE) methods.
These methods have classical orders of consistency
equal to two and one, respectively, in the numerical solution of
systems of ordinary differential equations (ODEs) and possess
favorable linear stability properties, cf.~e.g.~\cite{HW02}.
Let $I$ denote the $M\times M$ identity matrix, let $\Delta t = T/N$
with integer $N\ge 1$ be a given time step and let temporal grid
points $t_n = n\cdot\Delta t$ for integers $0\le n \le N$.
Then application of the $\theta$-method to the semidiscrete Heston PDCP
(\ref{lcp_ode}) defines approximations $U_n \approx U(t_n)$ successively
for $n=1,2,\ldots,N$
by
\begin{subeqnarray}\label{lcp_cn}
&&(I-\theta\Delta t A)U_n \ge (I+(1-\theta)\Delta t A)U_{n-1} + \Delta t\, g,\\
\nonumber \\
&&U_n \ge U_0,~~(U_n - U_0)^\rT\left((I-\theta\Delta t A)U_n - (I+(1-\theta)\Delta t A)U_{n-1}
- \Delta t\, g\right)=0.
\end{subeqnarray}
The fully discrete PDCP (\ref{lcp_cn}) constitutes, for each given
$n$, a so-called \textit{linear complementarity problem (LCP)}.
By introducing an auxiliary vector $\lambda_n$, it can clearly
be rewritten as
\begin{subeqnarray}\label{lcp_cn2}
&&(I-\theta\Delta t A)U_n = (I+(1-\theta)\Delta t A)U_{n-1} +
\Delta t\, g + \Delta t\, \lambda_n,\\
\nonumber \\
&&\lambda_n\ge 0,~~U_n \ge U_0,~~(U_n - U_0)^\rT \lambda_n =0.
\end{subeqnarray}
If the $i$-th component $\lambda_{n,i}$ of $\lambda_n$ is equal to
zero, then the corresponding spatial grid point $(s,v) \in \mathcal{G}$
is assumed to lie in the continuation region at time $t_n$, and otherwise
in the exercise region.
We consider in this paper the numerical solution of the LCPs (\ref{lcp_cn2})
for $1\le n\le N$ by employing a splitting technique proposed by Ikonen
\& Toivanen \cite{IT04,IT09}:
\begin{subeqnarray}\label{lcp_cnIT}
\quad (I-\theta\Delta t A)\Bar{U}_n = (I+(1-\theta)\Delta t A) \hU_{n-1}
+ \Delta t\, g + \Delta t\,\bar{\lambda}_n,\label{step1}\\\nonumber\\
\left\{\begin{array}{l}
\hU_n-\Bar{U}_n-\Delta t\,(\hl_n-\bar{\lambda}_n)=0,\\\\
\hl_n\geq 0,\quad \hU_n\geq U_0,\quad (\hU_n-U_0)^\rT\, \hl_n =0,
\end{array}\right.\label{step2}
\end{subeqnarray}
where $\hU_0 = U_0$.
The vector $\bar{\lambda}_n$ is given at the start of each time step.
Here the basic choice from \cite{IT04,IT09} is taken:
\begin{equation}\label{choice}
\bar{\lambda}_n = \hl_{n-1} ~~\text { with } \hl_0 \text{ the zero vector}.
\end{equation}
We refer to the above technique as \textit{Ikonen--Toivanen (IT) splitting}
and call (\ref{lcp_cnIT}) the {\it $\theta$-IT method}.
Special cases are the {\it CN-IT method} and {\it BE-IT method} given by
$\theta=\frac{1}{2}$ and $\theta=1$, respectively.
The IT splitting approach has been inspired by similar techniques in
computational fluid dynamics~\cite{G03}.
The vectors $\hU_n$ and $\hl_n$ defined by (\ref{lcp_cnIT}) constitute
approximations to $U_n$ and $\lambda_n$ defined by (\ref{lcp_cn2}).
These vectors are computed in two, successive stages.
In the first stage, an intermediate approximation $\Bar{U}_n$ is
determined by solving the system of linear equations (\ref{lcp_cnIT}a).
Notice that this system can be viewed as obtained from application of
the classical $\theta$-method to a system of ODEs.
In the second stage, $\Bar{U}_n$ and $\bar{\lambda}_n$ are updated to
$\hU_n$ and $\hl_n$ through (\ref{lcp_cnIT}b).
It is readily verified that these updates are given by the simple,
explicit formulas
\begin{equation}\label{updates}
\hU_n = \max\left\{\Bar{U}_n-\Delta t\, \bar{\lambda}_{n}\,,\,U_0\right\},\quad
\hl_{n} = \max\left\{0\,,\,\bar{\lambda}_{n}+(U_0-\Bar{U}_n)/\Delta t\right\}.
\end{equation}
Here the maximum of two vectors is taken componentwise.

The following useful theorem concerns the difference between
the LCP (\ref{lcp_cn2})
and its approximate version (\ref{lcp_cnIT}) obtained by IT splitting
if $\theta =1$.
For any given diagonal matrix $D\in \R^{M\times M}$ with positive
diagonal entries, let the scaled inner product be given by
\[
\langle x , y \rangle_D = y^{\rm T} D x
~~{\rm whenever}~x, y \in \R^M
\]
and let $\| \cdot \|_D$ denote both the induced vector and matrix
norms.
We have
\begin{theorem}\label{thm}
Consider the processes {\rm (\ref{lcp_cn2})} and {\rm (\ref{lcp_cnIT})}
with $\theta = 1$.
Assume there exists a positive diagonal matrix $D$ such that
\begin{equation}\label{assA}
DA + A^{\rm T}D ~{\it is~negative~semidefinite}
\end{equation}
and assume there is a real constant $\nu$ independent of $\Delta t >0$
such that
\begin{equation}\label{lambda}
\|\lambda_1\|_D + \sum^{N}_{n=2} \|\lambda_n - \lambda_{n-1}\|_D \leq \nu.
\end{equation}
Then
\begin{equation}
\max_{1\le n\le N} \|U_n - \hU_n\|_D \leq \nu \, \Delta t
\end{equation}
whenever $\Delta t = T/N$, integer $N\ge 1$.
\end{theorem}
\vskip0.2cm\noindent
\textbf{Proof}~
(i)
From assumption (\ref{assA}) and Berman \& Plemmons
\cite[Chs.~6,~10]{BP79} it first follows that $Q=I-\Delta t A$ is a
P-matrix\footnote{A square matrix is called a P-matrix if
all its principal minors are positive, see e.g. \cite{BP79,HP98}.
} and that (\ref{lcp_cn}) always possesses a unique solution $U_n$.
By (\ref{lcp_cn2}a),
\[
Q U_n = U_{n-1} + \Delta t\, g + \Delta t\, \lambda_n.
\]
By (\ref{lcp_cnIT}b),
\[
\Bar{U}_n = \hU_n - \Delta t\,(\hl_n-\bar{\lambda}_n)
\]
and inserting this into (\ref{lcp_cnIT}a) yields
\[
Q \hU_n = \hU_{n-1} + \Delta t\, g + \Delta t\,\bar{\lambda}_n + \Delta t\, Q (\hl_n-\bar{\lambda}_n).
\]
Define $V_n = U_n -  \hU_n$.
Then, with $R = Q^{-1}$,
\[
Q V_n = V_{n-1} + \Delta t\,(\lambda_n - \bar{\lambda}_n) - \Delta t\, Q (\hl_n-\bar{\lambda}_n),
\]
\[
V_n = R V_{n-1} + \Delta t\,R (\lambda_n - \bar{\lambda}_n) - \Delta t\, (\hl_n-\bar{\lambda}_n).
\]
Define $W_n = \Delta t\,(\lambda_n - \hl_n)$.
Upon writing $\hl_n-\bar{\lambda}_n = \hl_n-\lambda_n+\lambda_n-\bar{\lambda}_n$ there follows
\[
V_n - W_n = R V_{n-1} + \Delta t\, S (\lambda_n - \bar{\lambda}_n),
\]
with $S= R-I$.
Inserting the choice (\ref{choice}) for $\bar{\lambda}_n$ leads to
\begin{equation}\label{VW1}
V_n - W_n = R V_{n-1} + S W_{n-1} + \Delta t\, S (\lambda_n - \lambda_{n-1}),
\end{equation}
where we put $\lambda_0 = 0$.
\vskip0.2cm\noindent
(ii) Conditions (\ref{lcp_cn2}b) and (\ref{lcp_cnIT}b) imply for the components
of the vectors $V_n$, $W_n$ that
\begin{equation}\label{VW2}
V_{n,i} W_{n,i} \le 0 ~~~{\rm for~all}~ i.
\end{equation}
\noindent
To see this, consider four cases:
\vskip0.2cm\noindent
If $\lambda_{n,i} =0$ and $\hl_{n,i} = 0$, then $W_{n,i}=0$.
\vskip0.2cm\noindent
If $\lambda_{n,i}>0$ and $\hl_{n,i} = 0$, then $W_{n,i}>0$ and
$V_{n,i} = U_{0,i} - \hU_{n,i} \le 0$.
\vskip0.2cm\noindent
If $\lambda_{n,i}=0$ and $\hl_{n,i} > 0$, then $W_{n,i}<0$ and
$V_{n,i} = U_{n,i} - U_{0,i} \ge 0$.
\vskip0.2cm\noindent
If $\lambda_{n,i}>0$ and $\hl_{n,i} > 0$, then
$V_{n,i} = U_{0,i} - U_{0,i} = 0$.
\vskip0.4cm\noindent
(iii)
Define on $\R^{2M}$ the norm
\[
\Big\|
\left(
     \begin{array}{c}
       x \\
       y \\
     \end{array}
   \right)
\Big\|_D =
\sqrt{\| x \|_D^2 + \| y \|_D^2}
~~~{\rm whenever}~x, y \in \R^M
\]
with induced matrix norm denoted the same.
By (\ref{VW2}), we have
\begin{equation*}
\Big\|
\left(
     \begin{array}{c}
       V_n \\
       W_n \\
     \end{array}
   \right)
\Big\|_D^2 =
\| V_n \|_D^2 + \| W_n \|_D^2 \le
\| V_n \|_D^2 + \| W_n \|_D^2 - 2 \langle V_n , W_n \rangle_D =
\| V_n - W_n \|_D^2.
\end{equation*}
Using (\ref{VW1}) this yields
\begin{equation}\label{est}
\Big\|
\left(
     \begin{array}{c}
       V_n \\
       W_n \\
     \end{array}
\right)
\Big\|_D \le
\Big\|
\left(
  \begin{array}{cc}
    R & S \\
    O & O \\
  \end{array}
\right)
\left(
     \begin{array}{c}
       V_{n-1} \\
       W_{n-1} \\
     \end{array}
\right)
\Big\|_D
+
\Delta t\, \| S (\lambda_n - \lambda_{n-1})\|_D.
\end{equation}
Consider the $2\times 2$ matrix-valued rational function $\Phi$
given by
\[
\Phi(z) =
\left(
  \begin{array}{cc}
    \frac{1}{1-z} & \frac{z}{1-z} \\
    0 & 0 \\
  \end{array}
\right)
\]
for $z\in\C$.
Then
\[
\Phi(\Delta t A) =
\left(
  \begin{array}{cc}
    R & S \\
    O & O \\
  \end{array}
\right).
\]
For the spectral norm of $\Phi(z)$ it is easily shown that
\[
\| \Phi(z) \|_2^2 =
\lambda_{\max} [\Phi(z)^\ast \Phi(z)] =
\frac{1+|z|^2}{|1-z|^2}\,,
\]
which yields
\begin{equation}\label{Phi}
\| \Phi(z) \|_2 \le 1 ~~{\rm if~and~only~if}~~ \Re z\le 0.
\end{equation}
Next, the condition (\ref{assA}) on matrix $A$ is equivalent to
\begin{equation}\label{assA2}
{\rm Re}\,\langle Ax , x \rangle_D \le 0 ~~{\rm whenever}~x \in \C^M,
\end{equation}
where $\langle x , y \rangle_D = y^\ast D x$ for any given vectors
$x, y \in \C^M$.
In view of (\ref{Phi}) and (\ref{assA2}), we can invoke a matrix-valued
version of a well-known theorem due to von Neumann,
see e.g.~\cite[Sect.~V.7]{HW02}.
This directly leads to
\[
\| \Phi(\Delta t A)\|_D \le 1.
\]
By (\ref{est}), it thus follows that
\[
\Big\|
\left(
     \begin{array}{c}
       V_n \\
       W_n \\
     \end{array}
\right)
\Big\|_D
\le
\Big\|
\left(
     \begin{array}{c}
       V_{n-1} \\
       W_{n-1} \\
     \end{array}
\right)
\Big\|_D
+
\Delta t\, \| \lambda_n - \lambda_{n-1} \|_D
\le
\ldots
\le
\Delta t\, \sum_{j=1}^n \| \lambda_j - \lambda_{j-1} \|_D
\le
\nu \, \Delta t,
\]
which completes the proof.
\begin{flushright}
$\Box$
\end{flushright}
\vskip0.3cm

Theorem \ref{thm} provides the useful result that the sequence
$\{ \hU_n \}$ generated by (\ref{lcp_cnIT}) is ${\cal O}(\Delta t)$
close to the sequence $\{ U_n \}$ defined by (\ref{lcp_cn2}) if
$\theta=1$.
Observe that there is no restriction on the time step $\Delta t$.

The matrix condition (\ref{assA}), or the equivalent condition
(\ref{assA2}), are well-known.
In the numerical ODE literature they are often referred to by
saying that a scaled logarithmic 2-norm of $A$ is less than
or equal to zero.
Recently In 't Hout \& Volders \cite{iHV12} investigated
this condition in the case of the Heston PDE.
Even though the FD discretization studied in there differs
somewhat from the one considered here, it is interesting to
note that a positive result concerning (\ref{assA2}) was
proved \cite{iHV12} for arbitrary correlation factors
$\rho\in [-1,1]$ with a natural scaling matrix $D$.
The extension of this (nontrivial) result to the present
semidiscretization will be left as a topic for future
research.

The condition (\ref{lambda}), on the $\lambda_n$ defined
by (\ref{lcp_cn2}), is analogous to a condition by Ikonen
\& Toivanen~\cite{IT09}, except that these authors dealt
with the maximum norm.
Theoretical and numerical evidence indicates that a
moderate constant $\nu$ exists that is valid uniformly
in the spatial grid and the time step such that
(\ref{lambda}) is fulfilled.

Theorem \ref{thm} is closely related to \cite[Thm.~1]{IT09}.
The latter theorem provides an upper bound on the maximum
norm of $U_n - \hU_n$ if $\theta=\frac{1}{2}$.
Unfortunately, however, the derivation of this result is not
clear as the last statement in the proof of \cite[Lemma~2]{IT09}
does not hold in general.
Also, a restriction on the time step has been assumed that is
often too severe for practical applications.

An analogue of Theorem \ref{thm} for the important case of
the maximum norm is not obvious.
Note that our proof above relies in an essential way on the
use of inner product norms.
Nevertheless, in the numerical experiments below we shall
always deal with the maximum norm.

Theorem \ref{thm} can be extended, along the
same line of proof above, to any given parameter value
$\theta \in [0,1)$ if the condition (\ref{assA}) is replaced
by
\[
\|\Delta t A  + \gamma I \|_D \le \gamma ~~~{\rm with}~~~
\gamma = (1-\theta)^{-2}.
\]
This is often called a circle condition on $\Delta t\, A$.
It is substantially stronger than (\ref{assA}).
In particular it implies that $\|\Delta t A \|_D \le 2\gamma$,
which yields an upper bound on the time step that is often
too severe in practice.
In the numerical experiments below however, we shall consider
the $\theta$-IT method (\ref{lcp_cnIT}) both with $\theta=1$
and $\theta=\frac{1}{2}$.

%%%%%%%%%%%%%%%%%%%%%%%%%%%%%%%%%%%%%%%%%%%%%%%%%%%%%%%%%%%%%%%%%%%%%%%%%%%%%%%%%%%%
%%%%%%%%%%%%%%%%%%   SECTION 4   %%%%%%%%%%%%%%%%%%%%%%%%%%%%%%%%%%%%%%%%%%%%%%%%%%%
%%%%%%%%%%%%%%%%%%%%%%%%%%%%%%%%%%%%%%%%%%%%%%%%%%%%%%%%%%%%%%%%%%%%%%%%%%%%%%%%%%%%
\setcounter{equation}{0}
\section{Time discretization: ADI schemes}\label{time_ADI}
When numerically solving multidimensional problems, the spatial
discretization rapidly leads to very large systems of semidiscrete
PDCPs.
Applying the $\theta$-IT method then renders very large linear
systems with a large bandwidth that need to be solved.
This can be computationally very demanding.
One good possibility is to employ tailored multigrid methods,
as is done e.g.~in~\cite{IT09}.
In the present paper we propose to combine ADI time discretization
schemes with the IT splitting approach for solving (\ref{lcp_ode}).
In (\ref{lcp_cnIT}) the $\theta$-method is thus replaced with
an ADI scheme.

We shall consider four different schemes of the ADI type: the
Douglas scheme (Do), the Craig--Sneyd scheme (CS), the modified
Craig--Sneyd scheme (MCS) and the Hundsdorfer--Verwer scheme (HV).
These four schemes have recently been elaborately investigated
in the numerical pricing of European-style vanilla and barrier
options under the Heston model~\cite{IHF10} as well as under the
three-dimensional Heston--Hull--White and
Heston--Cox--Ingersoll--Ross models, see~\cite{HH12} and
\cite{H13}, respectively.
It was found that in particular the MCS and HV schemes, with
a proper choice of their parameter, are highly efficient,
stable and robust.
A first brief numerical study of ADI schemes adapted to the
pricing of American-style options under the Heston model was
carried out in~\cite{HIHV10}.
In this paper we shall substantially extend the promising
initial results obtained in loc.~cit.

When considering schemes of the ADI type, the semidiscrete
matrix $A$ is split into several convenient parts.
In the case of the Heston model we have
\begin{equation*}
    A = A_0+A_1+A_2.
\end{equation*}
Here the matrix $A_0$ is the part of $A$ that stems from the
FD discretization of the mixed derivative term and $A_1$ and
$A_2$ are given by the parts of $A$ that correspond to the
FD discretization of all spatial derivatives in the $s$-
and $v$-directions, respectively, and further contain
an equal part of the $ru$ term from (\ref{Heston}).
Note that the matrices $A_1$ and $A_2$ are essentially
tridiagonal and that the matrix $A_0$ is nonzero whenever
the correlation $\rho$ is nonzero.

Let $\theta>0$ be a given real parameter.
Let $\Delta t=T/N$ with integer $N\geq 1$ and set $t_n=n\,\Delta t$.
The following four methods are given by combination of the ADI
schemes mentioned above with the IT splitting stage (\ref{lcp_cnIT}b),
or equivalently, (\ref{updates}).
Each method defines successive approximations $\hU_n$ to the
solution vectors $U(t_n)$ of (\ref{lcp_ode}) for $n=1,2,\ldots,N$.
We refer to them as \textit{ADI-IT methods}.
\\
\\
{\it Do-IT}\,:
\vspace{0.2cm}
\begin{equation}\label{Do}
\begin{cases}
Y_0 = \hU_{n-1} + \Delta t(A \hU_{n-1} + g) + \Delta t\, \bar{\lambda}_n,\\
Y_j = Y_{j-1}+\theta \Delta t A_j \left(Y_j-\hU_{n-1}\right)~~~~~~~~~~~~~~~~~~~~~~~(j=1,2),\\
\Bar{U}_n = Y_2,\\\\
\hU_n = \max\left\{\Bar{U}_n-\Delta t\, \bar{\lambda}_{n}\,,\,U_0\right\},\\
\hl_{n} = \max\left\{0\,,\,\bar{\lambda}_{n}+(U_0-\Bar{U}_n)/\Delta t\right\}.
\end{cases}
\end{equation}
\\
\\
{\it CS-IT}\,:
\vspace{0.2cm}
\begin{equation}\label{CS}
\begin{cases}
Y_0 = \hU_{n-1} + \Delta t(A \hU_{n-1} + g) + \Delta t\, \bar{\lambda}_n,\\
Y_j = Y_{j-1}+\theta \Delta t A_j \left(Y_j-\hU_{n-1}\right)~~~~~~~~~~~~~~~~~~~~~~~(j=1,2),\\
\tY_0 = Y_0+\tfrac{1}{2} \Delta t A_0 \left(Y_2-\hU_{n-1}\right),\\
\tY_j = \tY_{j-1}+\theta \Delta t A_j \left(\tY_j-\hU_{n-1}\right)~~~~~~~~~~~~~~~~~~~~~~~(j=1,2),\\
\Bar{U}_n = \tY_2,\\\\
\hU_n = \max\left\{\Bar{U}_n-\Delta t\, \bar{\lambda}_{n}\,,\,U_0\right\},\\
\hl_{n} = \max\left\{0\,,\,\bar{\lambda}_{n}+(U_0-\Bar{U}_n)/\Delta t\right\}.
\end{cases}
\end{equation}
\\
\\
\noindent
{\it MCS-IT}\,:
\vspace{0.2cm}
\begin{equation}\label{MCS}
\begin{cases}
Y_0 = \hU_{n-1} + \Delta t(A \hU_{n-1} + g) + \Delta t\, \bar{\lambda}_n,\\
Y_j = Y_{j-1}+\theta \Delta t A_j \left(Y_j-\hU_{n-1}\right)~~~~~~~~~~~~~~~~~~~~~~~(j=1,2),\\
\tY_0 = Y_0+\left( \theta \Delta t\, A_0 +(\tfrac{1}{2}-\theta) \Delta t A \right)\left(Y_2-\hU_{n-1}\right),\\
\tY_j = \tY_{j-1}+\theta \Delta t A_j \left(\tY_j-\hU_{n-1}\right)~~~~~~~~~~~~~~~~~~~~~~~(j=1,2),\\
\Bar{U}_n = \tY_2,\\\\
\hU_n = \max\left\{\Bar{U}_n-\Delta t\, \bar{\lambda}_{n}\,,\,U_0\right\},\\
\hl_{n} = \max\left\{0\,,\,\bar{\lambda}_{n}+(U_0-\Bar{U}_n)/\Delta t\right\}.
\end{cases}
\end{equation}
\\
\\
\noindent
{\it HV-IT}\,:
\vspace{0.2cm}
\begin{equation}\label{HV}
\begin{cases}
Y_0 = \hU_{n-1} + \Delta t(A \hU_{n-1} + g) + \Delta t\, \bar{\lambda}_n,\\
Y_j = Y_{j-1}+\theta \Delta t A_j \left(Y_j-\hU_{n-1}\right)~~~~~~~~~~~~~~~~~~~~~~~(j=1,2),\\
\tY_0 = Y_0+ \tfrac{1}{2} \Delta t A \left(Y_2-\hU_{n-1}\right),\\
\tY_j = \tY_{j-1}+\theta \Delta t A_j \left(\tY_j-Y_2\right)~~~~~~~~~~~~~~~~~~~~~~~~~~~(j=1,2),\\
\Bar{U}_n = \tY_2,\\\\
\hU_n = \max\left\{\Bar{U}_n-\Delta t\, \bar{\lambda}_{n}\,,\,U_0\right\},\\
\hl_{n} = \max\left\{0\,,\,\bar{\lambda}_{n}+(U_0-\Bar{U}_n)/\Delta t\right\}.
\end{cases}
\end{equation}
\\
\\
\noindent
As in Section~\ref{time_theta}, the vector $\bar{\lambda}_n$ is
given at the start of each time step by the choice (\ref{choice}).

The Do-IT method can be viewed as a natural analogue of the
$\theta$-IT method:
upon formally setting $A_0=A_2=0$ and $A_1=A$, one recovers
(\ref{lcp_cnIT}) from (\ref{Do}).
The CS-IT, MCS-IT, HV-IT methods form different extensions to the
Do-IT method.
Indeed, their first two lines are identical to those of Do-IT.
They require about twice the amount of computational work per
time step.
Observe that if $A_0=0$, then the CS-IT method reduces to the
Do-IT method.

For each ADI-IT method, the {\it underlying ADI scheme} for the
ODE system $U'(t) = AU(t) + g$ is given by the first part,
defining $\Bar{U}_n$, where one just omits the
$\Delta t\, \bar{\lambda}_n$ term from the first line and
replaces $\hU_{n-1}$ and $\Bar{U}_n$ by $U_{n-1}$ and $U_n$,
respectively.
The above adaptation of the ADI technique from European- to
American-style options is thus very simple.

In the underlying ADI schemes, the matrix $A_0$ is always handled
in an explicit fashion, while the matrices $A_1$ and $A_2$ are
treated in an implicit fashion.
Application of each of the four ADI-IT methods (\ref{Do}), (\ref{CS}),
(\ref{MCS}), (\ref{HV}) requires solving linear systems with the
two matrices $(I-\theta \Delta t A_j)$ for $j=1,2$.
Since these matrices are both tridiagonal, the solution can be
done very efficiently by computing once, beforehand, their $LU$
factorizations and then use these in all time steps.
Thus the computational cost per time step of each ADI-IT method
is directly proportional to the number of spatial grid points $M$,
i.e., the same as in the case of European-style options.
Note that the computational cost of the second part of each
method, the update (\ref{updates}), is negligible.

The order of consistency in the nonstiff sense of the underlying
ADI schemes is always one for the Do scheme whenever $A_0$ is
nonzero; it is two for the CS scheme provided that
$\theta = \frac{1}{2}$ and two for the MCS and HV schemes for any
given $\theta$.
In the literature substantial attention has recently been given
to the study of unconditional von Neumann stability for ADI
schemes when applied to multidimensional convection-diffusion
equations with mixed spatial derivative terms,
cf.~\cite{CS88,IHM11,IHW07,IHW09,MKM70,MKWW96}.
Here positive results were proved, guaranteeing unconditional
stability on various convection-diffusion problem classes under
sharp lower bounds on the parameter $\theta$ of each ADI scheme.
Based on the obtained stability results for two-dimensional
problems with mixed derivative term, we select the following
values of $\theta$:
\begin{itemize}
    \item the Do-IT method: $\theta = \frac{1}{2}$
    \item the CS-IT method: $\theta = \frac{1}{2}$
    \item the MCS-IT method: $\theta = \frac{1}{3}$
    \item the HV-IT method: $\theta = \frac{1}{2} + \frac{1}{6}\sqrt{3}$.
\end{itemize}

At present a rigorous theoretical stability and convergence
analysis of ADI-IT methods for semidiscrete PDCPs is beyond
reach.
In the following section we carry out an extensive numerical
investigation.
We shall study the four methods selected above, in the application
to a variety of representative, challenging Heston test cases.

%%%%%%%%%%%%%%%%%%%%%%%%%%%%%%%%%%%%%%%%%%%%%%%%%%%%%%%%%%%%%%%%%%%%%%%%%%%%%%%%%%%%
%%%%%%%%%%%%%%%%%%   SECTION 5   %%%%%%%%%%%%%%%%%%%%%%%%%%%%%%%%%%%%%%%%%%%%%%%%%%%
%%%%%%%%%%%%%%%%%%%%%%%%%%%%%%%%%%%%%%%%%%%%%%%%%%%%%%%%%%%%%%%%%%%%%%%%%%%%%%%%%%%%
\setcounter{equation}{0}
\section{Numerical experiments}\label{numexp}
For the $\theta$-IT and ADI-IT methods define the \textit{global
temporal discretization error} by
\begin{equation}\label{gte}
\widehat{e}\,(\Delta t;m_1,m_2) =
\max\{\,|U_{l}(T)-\hU_{N,l}|:\,(s_i, v_j) \in ROI \}.
\end{equation}
Here $U(T)$ represents the exact solution to the semidiscrete
Heston PDCP (\ref{lcp_ode}) at time $T$ and
\begin{equation*}\label{ROI}
ROI = (\tfrac{1}{2}K,\tfrac{3}{2}K) \times (0,1)
\end{equation*}
is a natural region of interest.
The index $l$ in the above is such that the $l$-th component of
a vector corresponds to the spatial grid point $(s_i, v_j)$.
Clearly, (\ref{gte}) represents a maximum norm on the {\it ROI}.

In this section we numerically investigate the actual behavior
of the global temporal errors as a function of $\Delta t$ for
the four ADI-IT methods selected at the end of Section~\ref{time_ADI}.
Along with this, the BE-IT and CN-IT methods from
Section~\ref{time_theta} are briefly considered.
For the numerical experiments we choose six cases of parameter
sets, listed in Table~\ref{cases}.
The cases A, B, C stem from~\cite{BQF05,SST04,WAW02},
respectively.
Here the Feller condition is always satisfied.
The cases D, E, F all stem from~\cite{A08}.
Here the Feller condition is always strongly violated and,
in addition, the maturity times are relatively long.
The six test cases of Table~\ref{cases} have, all or in part,
been recently considered in the numerical PDE pricing of
European options under the Heston model in~\cite{IHF10}, under
the Heston--Hull--White model in~\cite{HH12} and under the
Heston--Cox--Ingersoll--Ross model in~\cite{H13}.\\

\begin{table}[H]
\begin{center}
\begin{tabular}{|c|l|l|l|l|l|l|}
        \hline
        & Case A & Case B & Case C & Case D & Case E & Case F \\
        \hline \hline
        $\kappa$    & 3    & 0.6067 & 2.5  & 0.5   & 0.3  & 1\\
        $\eta$      & 0.12 & 0.0707 & 0.06 & 0.04  & 0.04 & 0.09\\
        $\sigma$  & 0.04 & 0.2928 & 0.5  & 1     & 0.9  & 1\\
        $\rho$  & 0.6~(0) & -0.7571~(0) & -0.1~(0) & -0.9~(0) & -0.5~(0) & -0.3~(0) \\
        $r$ &0.01&0.03&0.0507&0.05&0.04&0.03\\
        $T$ & 1   & 3   & 0.25 & 10  & 15  & 5\\
        $K$ & 100 & 100 & 100  & 100 & 100 & 100\\
        \hline
\end{tabular}
\end{center}
\caption{Parameter sets for the Heston model and American put options}
\label{cases}
\end{table}
\vskip0.2cm

Numerical experience has shown that, for efficiency, one can
use much less spatial grid points in the $v$-direction than
in the $s$-direction.
Accordingly, we always set
\[
m_1 = 2m_2 = 2m.
\]
Since a closed-form analytic solution to the semidiscrete
Heston PDCP (\ref{lcp_ode}) is not at hand, we computed in
each case a reference solution for $U(T)$ by applying either
the CN-IT method or the MCS-IT method using $N=20~000$ time
steps.

In the following we first consider vanilla American put
options.
The global temporal errors (\ref{gte}) are studied in
detail for the six methods under consideration in all
six cases of Table~\ref{cases}.
We then compare American put option price approximations
given in the literature to approximations computed using
the ADI-IT technique.
Next, for all cases A--F, the obtained option price surfaces
and free boundaries are displayed.
We conclude the section with experiments for a more exotic
American-style option, the capped American put.

\begin{figure}[H]
\begin{center}
\begin{tabular}{c c}
         \includegraphics[width=0.5\textwidth]{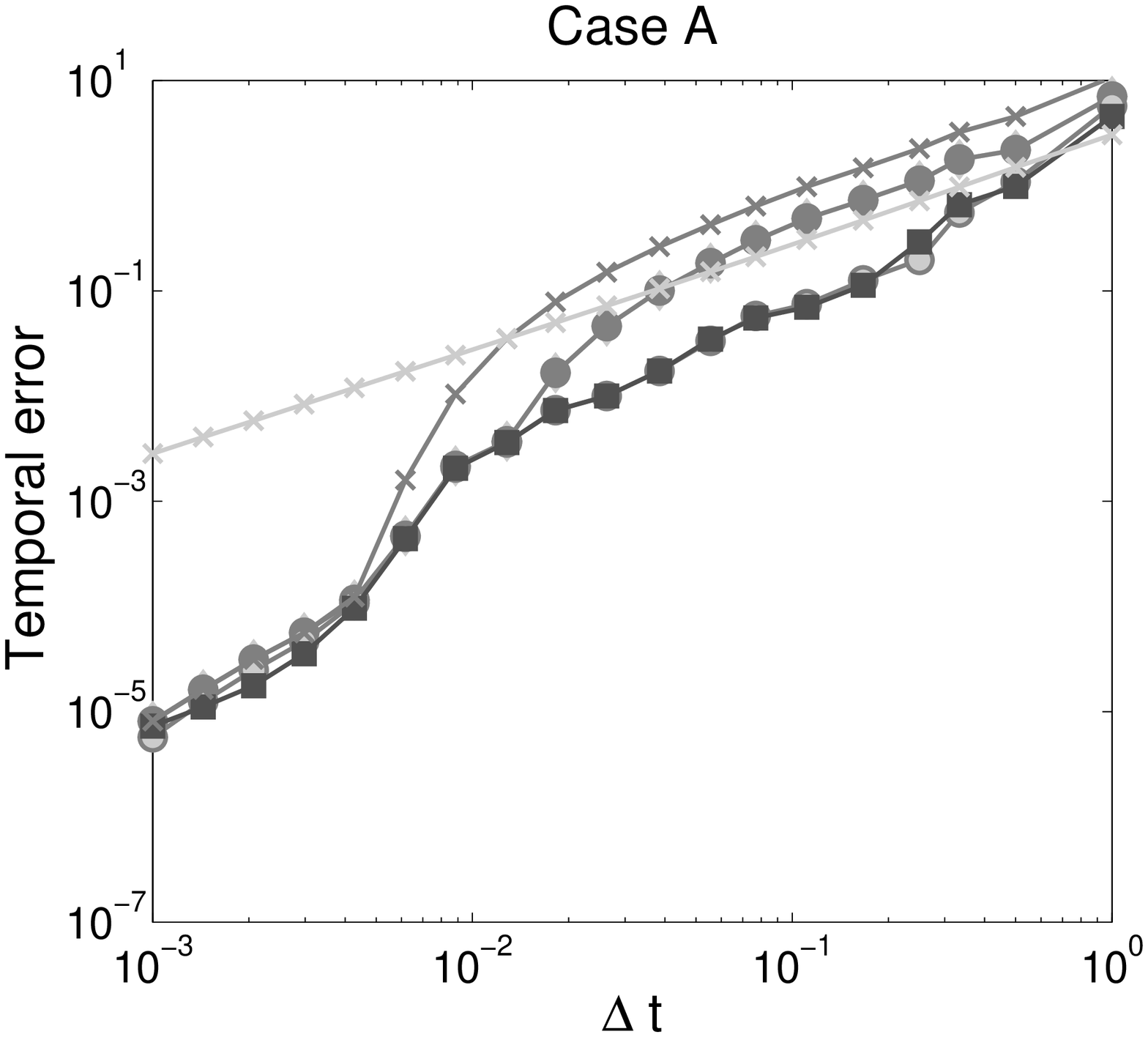}&
         \includegraphics[width=0.5\textwidth]{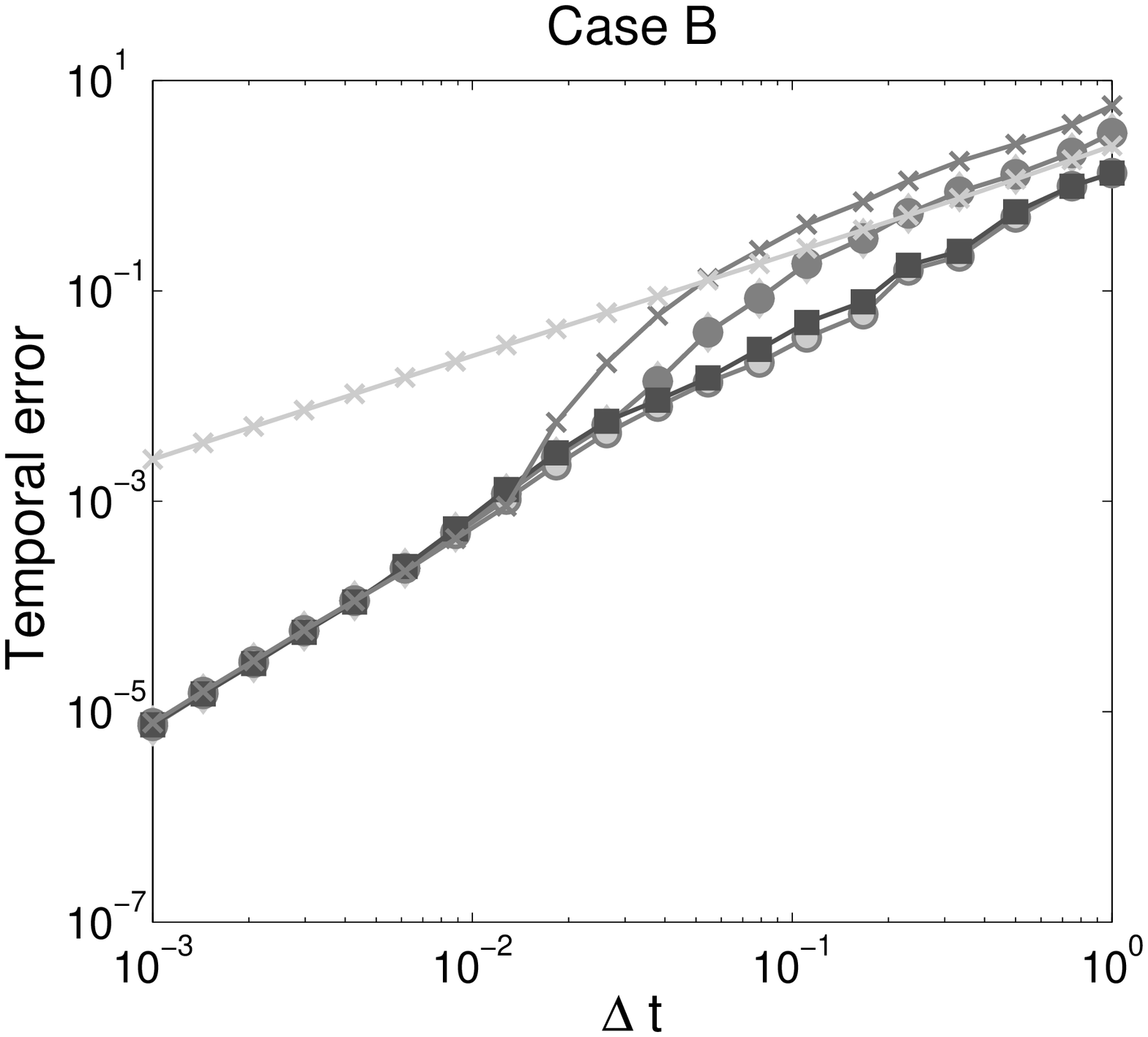}\\
         \includegraphics[width=0.5\textwidth]{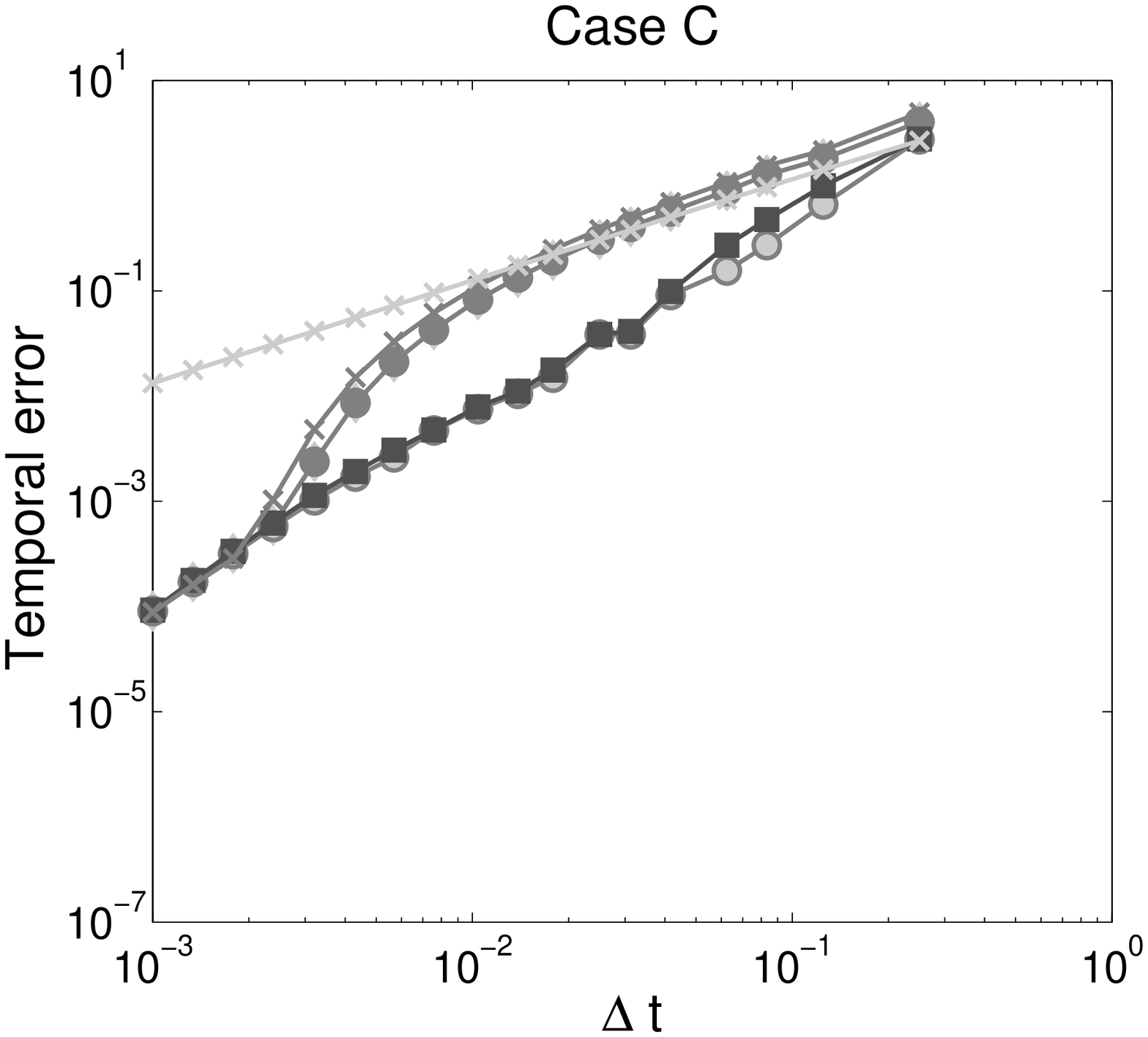}&
         \includegraphics[width=0.5\textwidth]{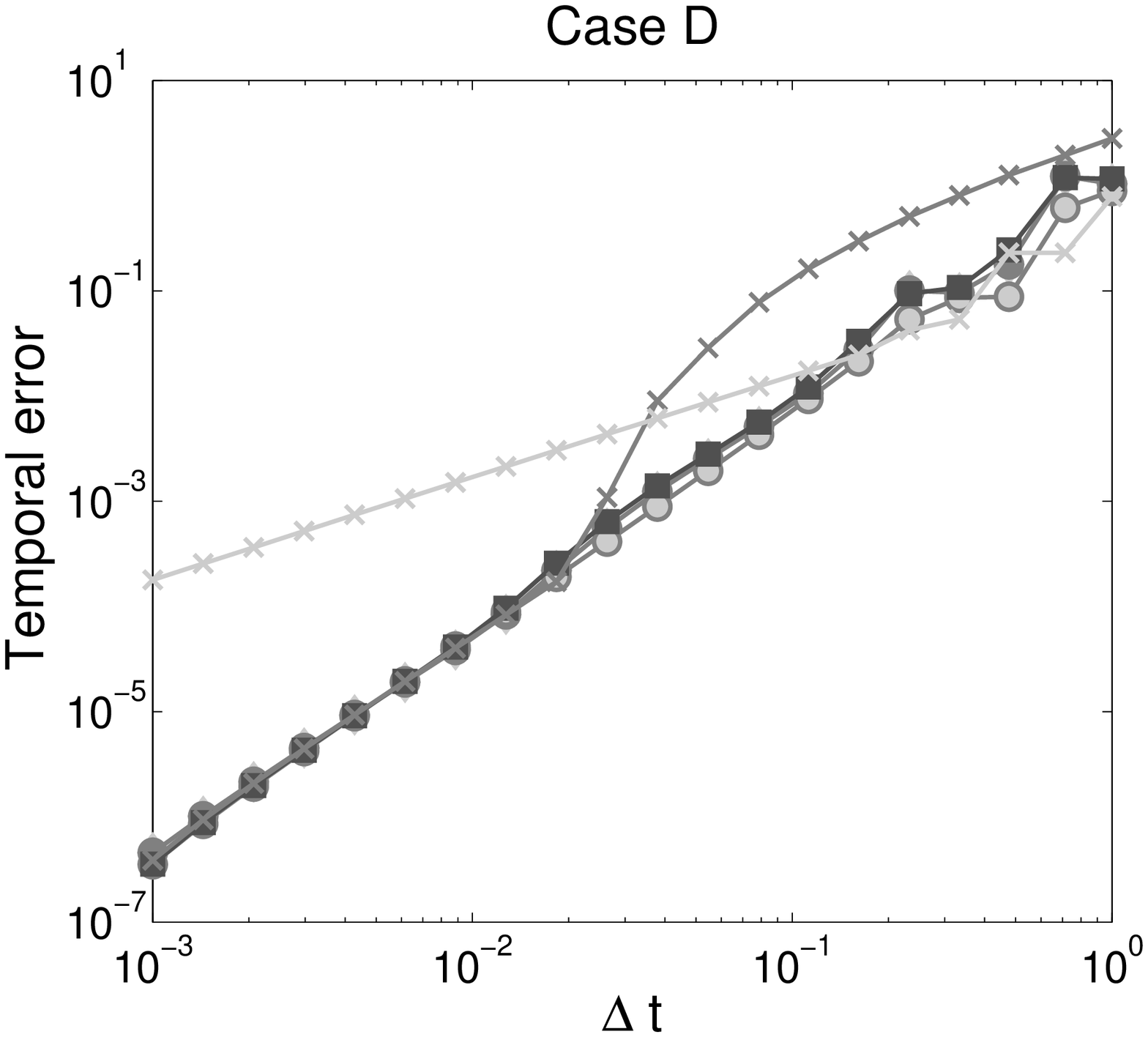}\\
         \includegraphics[width=0.5\textwidth]{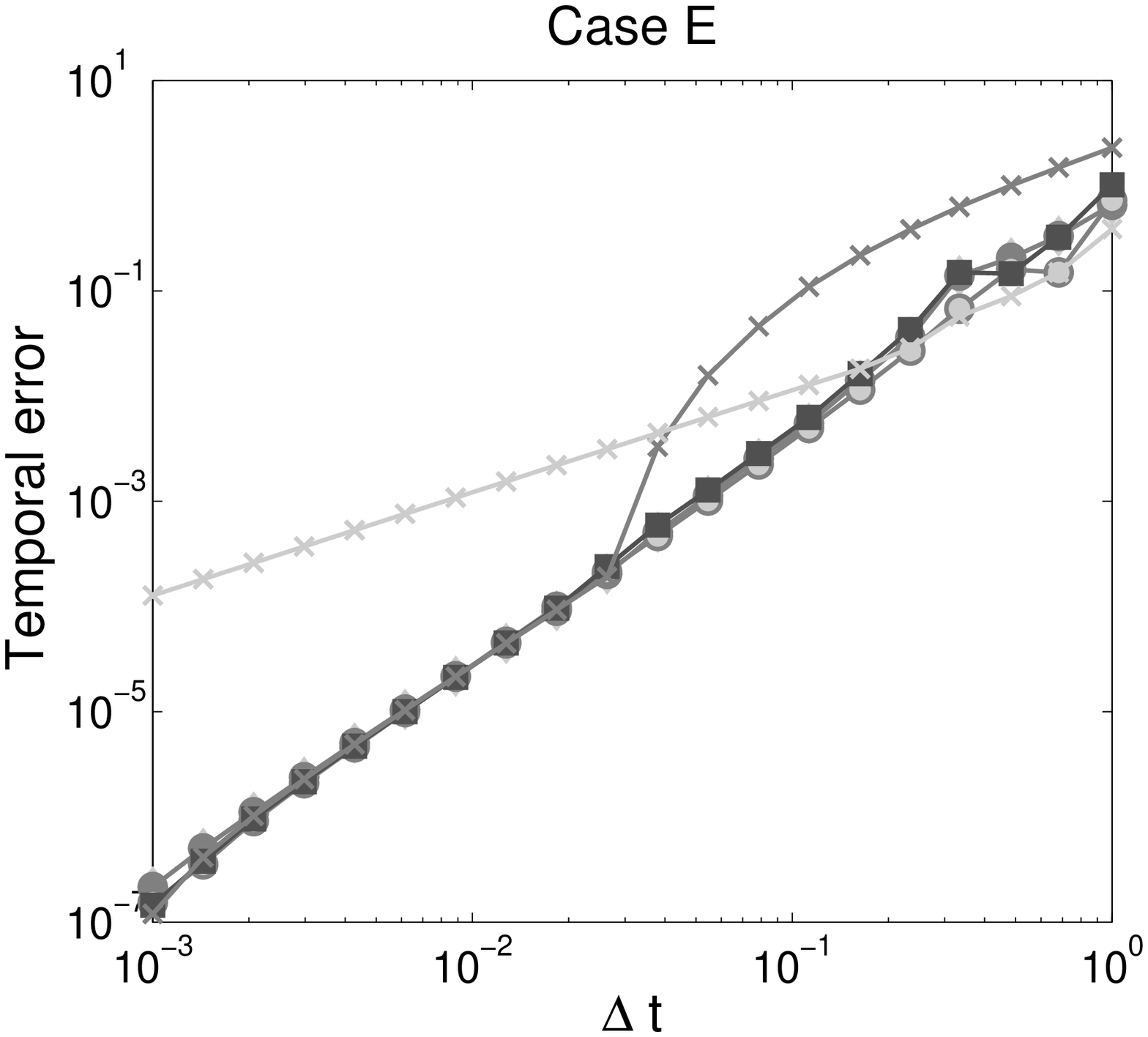}&
         \includegraphics[width=0.5\textwidth]{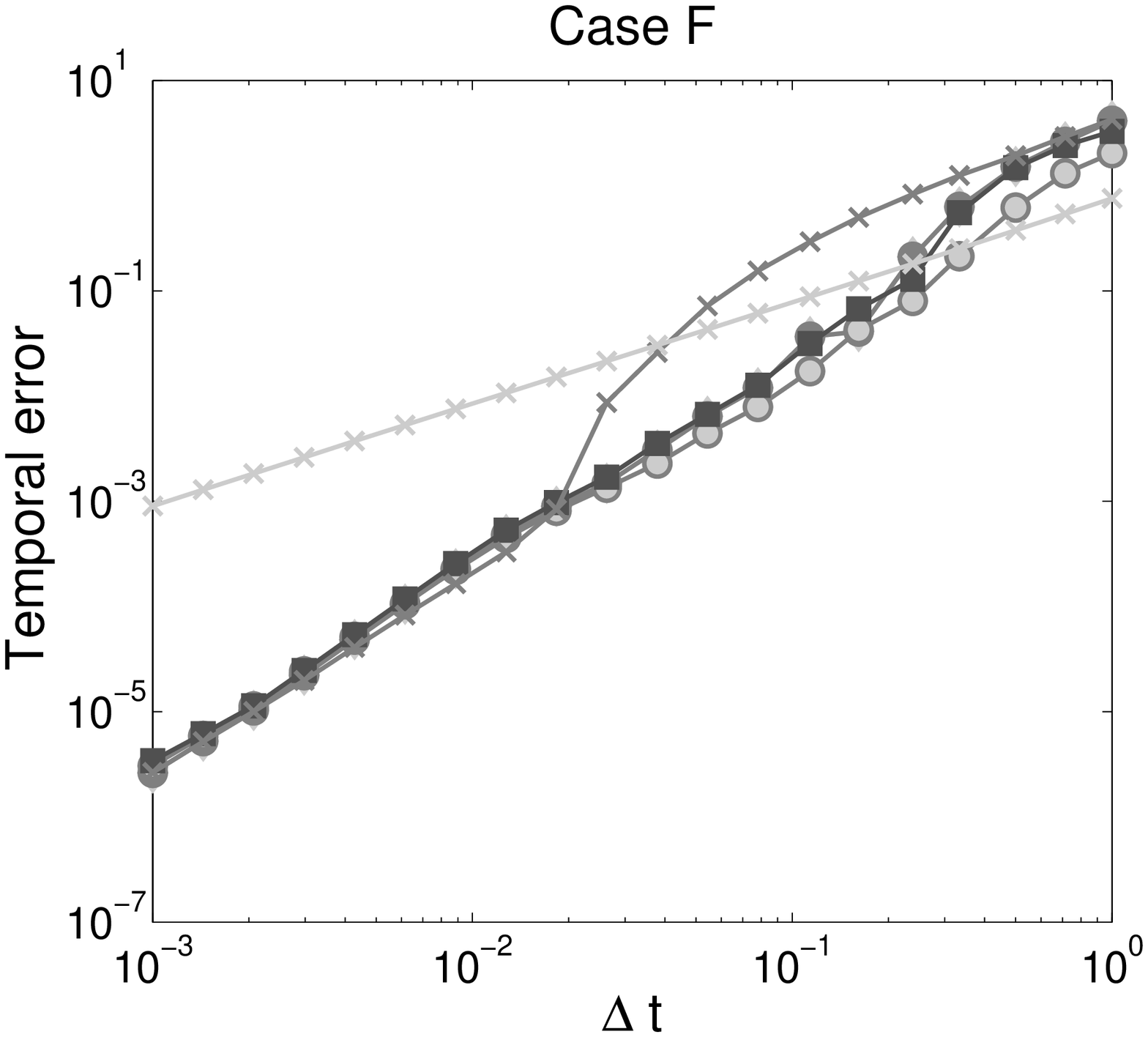}
\end{tabular}
\end{center}
\caption{Temporal errors $\widehat{e}\,(\Delta t;100,50)$ vs.~$\Delta t$
for vanilla American put options in the six cases
of Table~\ref{cases} with $\rho=0$ and reference solution by CN-IT with
$N=20~000$.
Two $\theta$-IT methods: BE-IT (light x), CN-IT (dark x).
Four ADI-IT methods: Do-IT with $\theta=\frac{1}{2}$ (light diamond), CS-IT
with $\theta=\frac{1}{2}$ (dark circle), MCS-IT with $\theta=\frac{1}{3}$
(light circle) and HV-IT with $\theta=\frac{1}{2}+\frac{1}{6}\sqrt{3}$ (dark
square). No damping applied.}
\label{TemporalErrorCN}
\end{figure}

For vanilla American put options,
Figure~\ref{TemporalErrorCN} displays for the Do-IT, CS-IT,
MCS-IT, HV-IT methods as well as the BE-IT and CN-IT methods
the global temporal errors $\widehat{e}\,(\Delta t; 2m,m)$ versus
$\Delta t$ in all cases A--F with $\rho=0$ for a sequence of 20
step sizes $10^{-3}\leq \Delta t\leq 10^0$ with $m=50$.
Note that the results for Do-IT and CS-IT are the same in
this experiment, since $A_0=0$.

As a first main observation, for each method the temporal
errors all remain below a moderate value in each case, and
further, they decrease monotonically as $\Delta t$ decreases.
This indicates an unconditionally stable behavior of each
method, which is a favorable and nontrivial result.

Concerning the actual convergence behavior, it is readily seen
from Figure~\ref{TemporalErrorCN} that the temporal errors as
a function of $\Delta t$ are bounded from above in each case
by $C(\Delta t)^p$ with some moderate constant $C$ where
$p\approx 1$ for the BE-IT method and $p\approx 2$ for the
MCS-IT and HV-IT methods.
The observed orders of convergence $p$ for these three methods
thus agree with the classical (nonstiff) orders of consistency
of their underlying time-discretization schemes for ODEs.
This constitutes a second positive and nontrivial result.

In all cases A--F for the CN-IT method and in the cases A--C
for the Do-IT and CS-IT methods, one observes in
Figure~\ref{TemporalErrorCN} relatively large temporal errors
for moderate values of $\Delta t$, compared to what one may
expect based on their error behavior for small $\Delta t$.
This undesirable phenomenon is related to the nonsmoothness
of the initial (payoff) function.
It is already well-known in the literature and as a remedy it
is common to apply backward Euler damping, also called Rannacher
time stepping.
In line with this damping procedure, we consider taking
first two BE-IT substeps with $\Delta t/2$ at $t=0$ and then
proceed onwards from $t=\Delta t$ with the method under
consideration.

% Figure 1

% Figure 2
\begin{figure}[H]
\begin{center}
\begin{tabular}{c c}
         \includegraphics[width=0.5\textwidth]{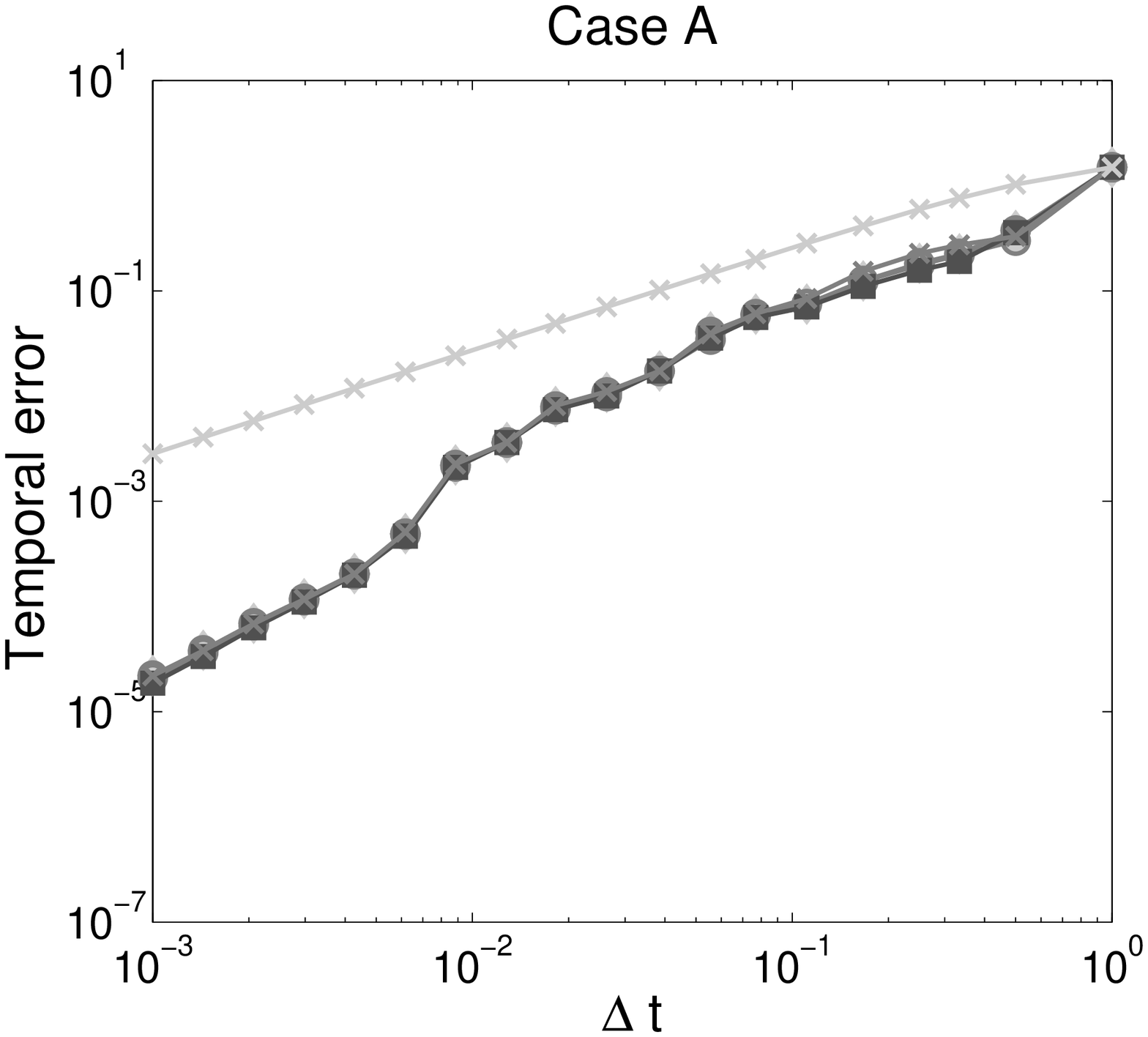}&
         \includegraphics[width=0.5\textwidth]{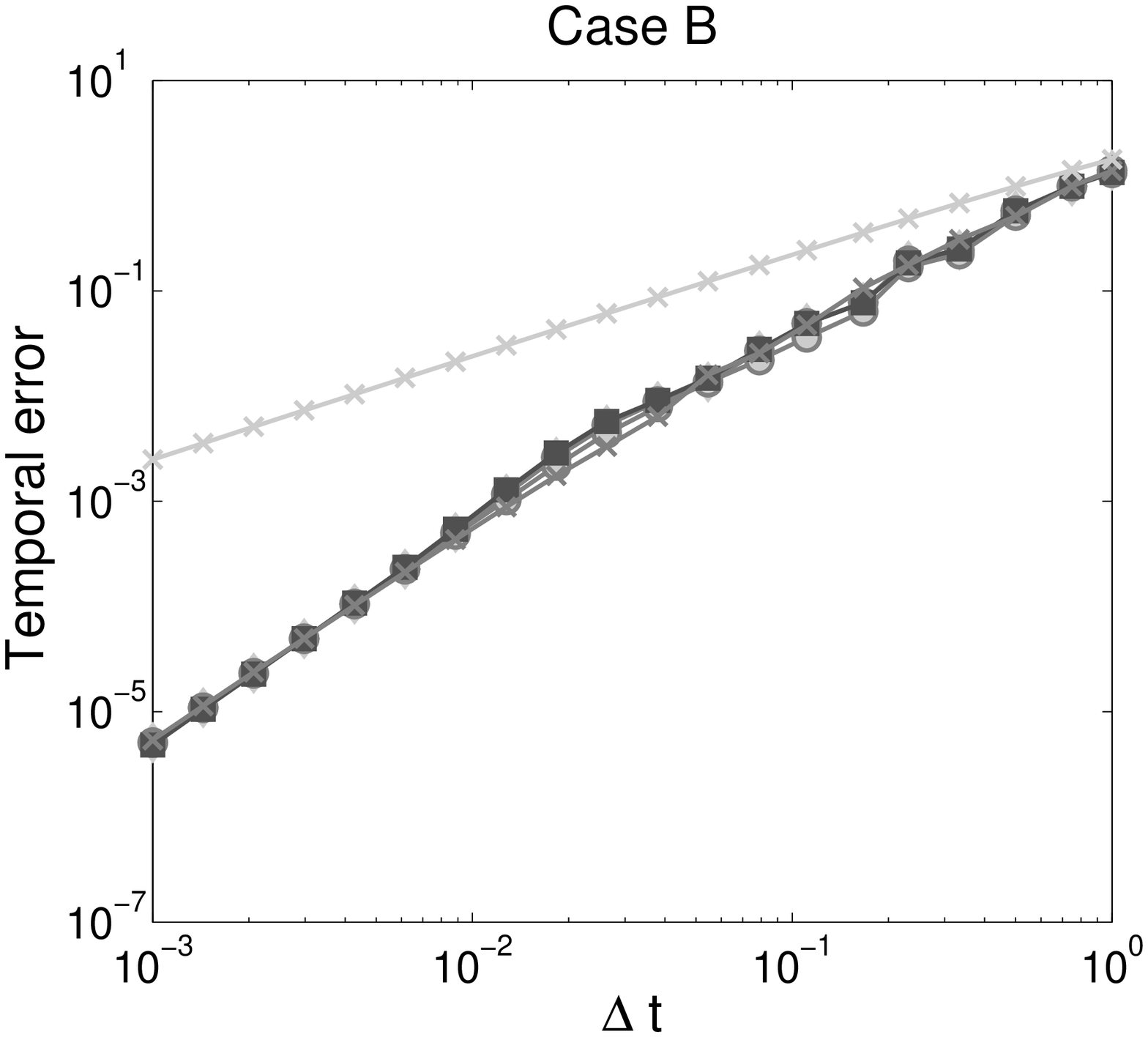}\\
         \includegraphics[width=0.5\textwidth]{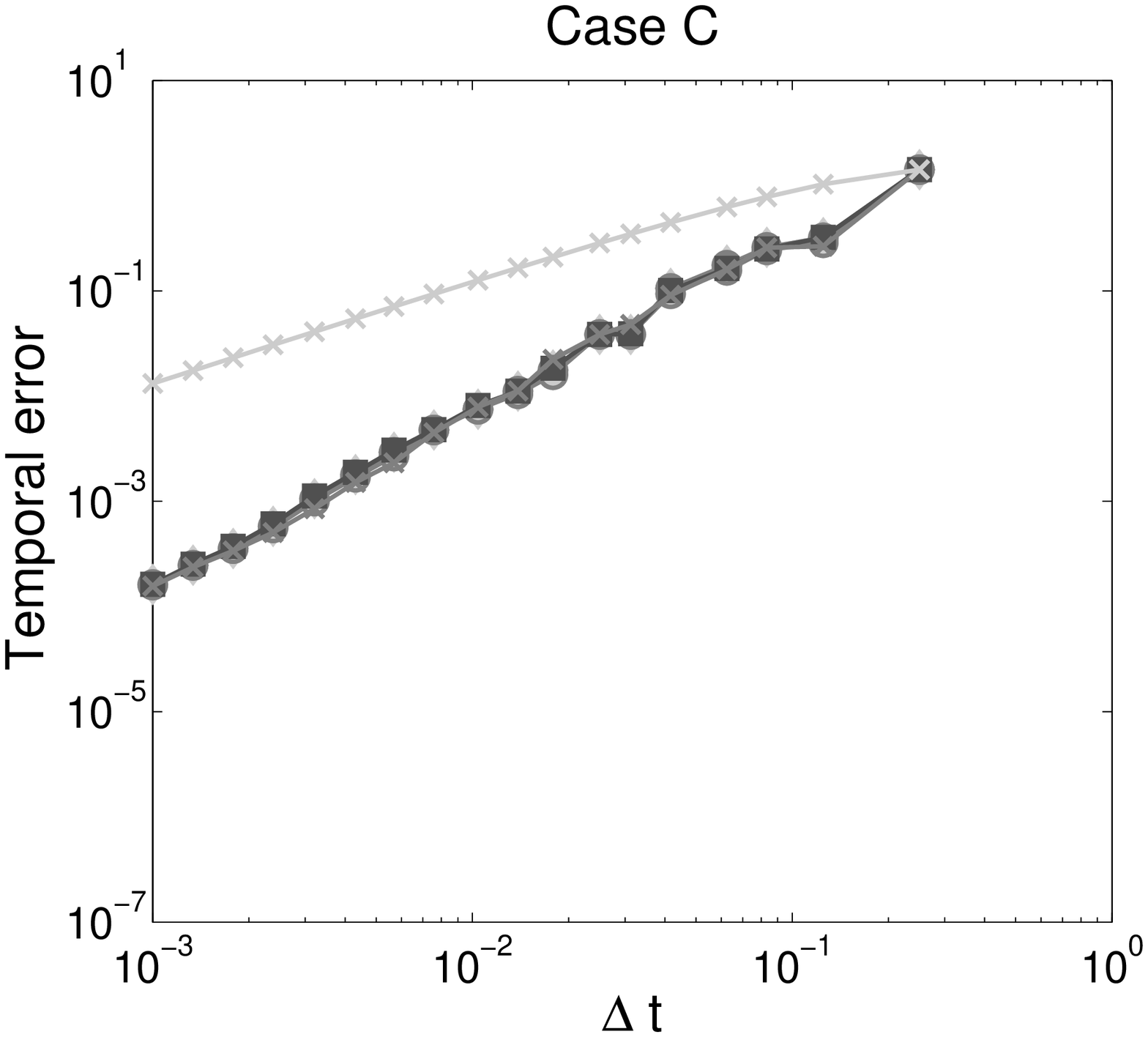}&
         \includegraphics[width=0.5\textwidth]{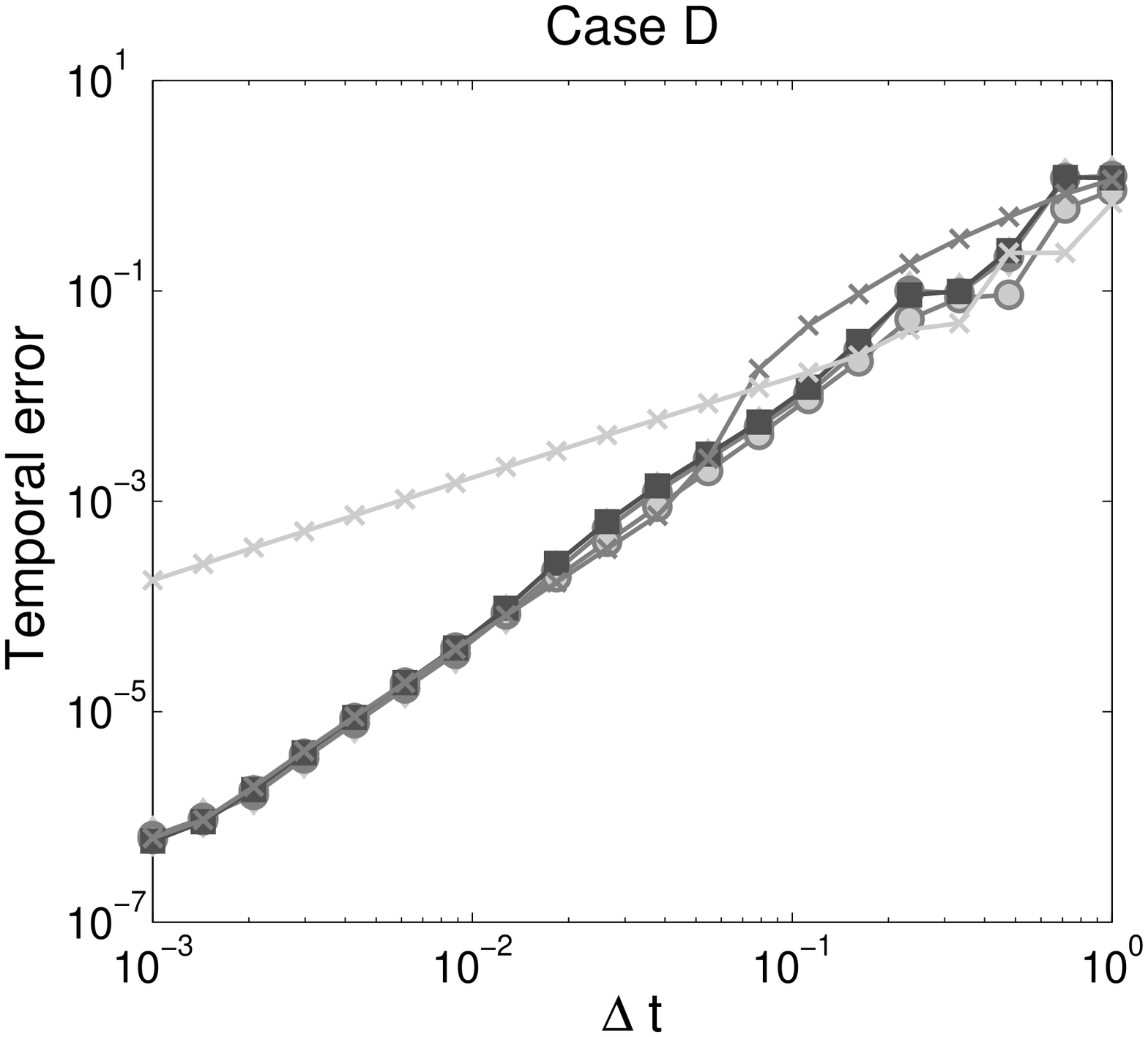}\\
         \includegraphics[width=0.5\textwidth]{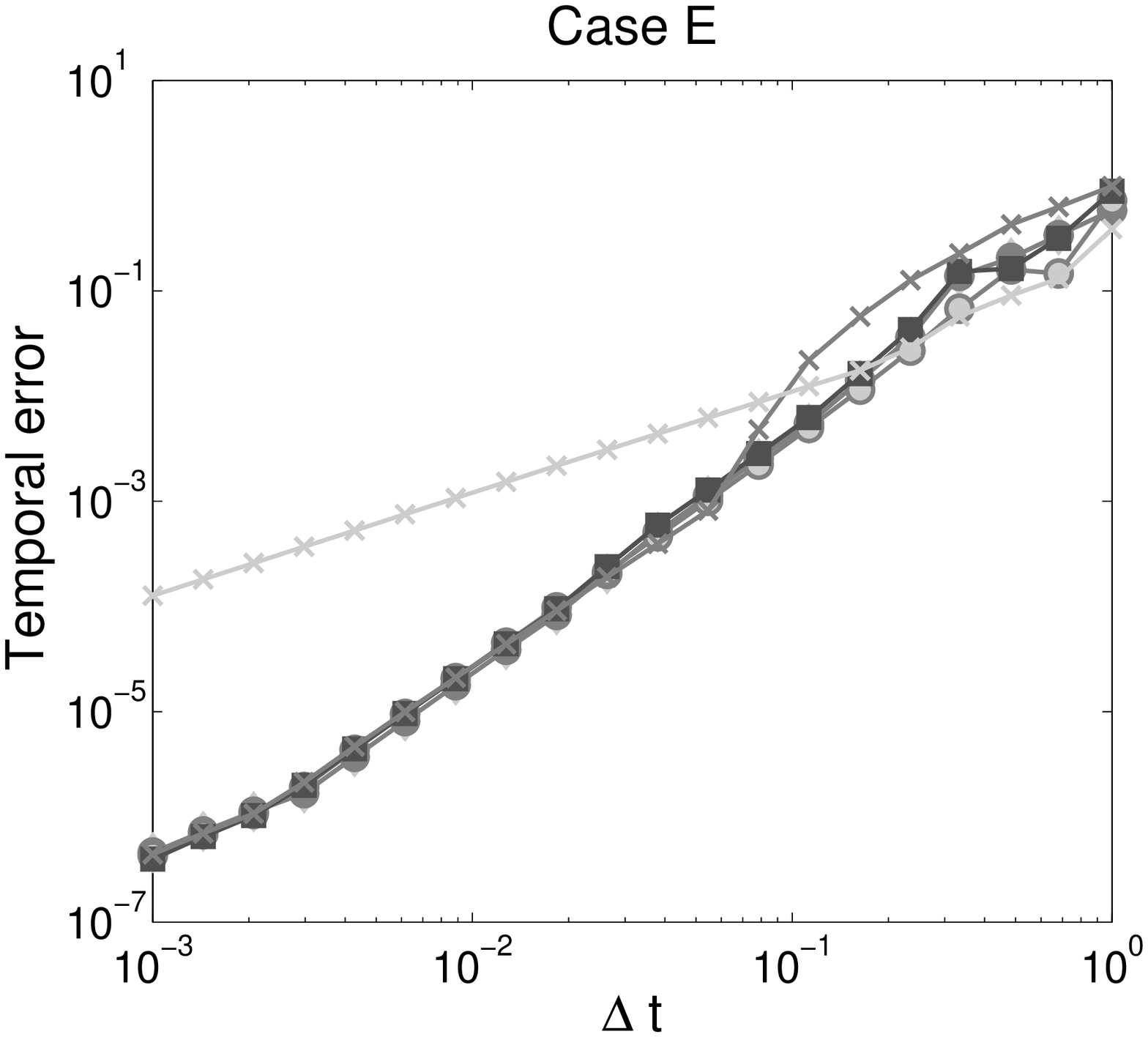}&
         \includegraphics[width=0.5\textwidth]{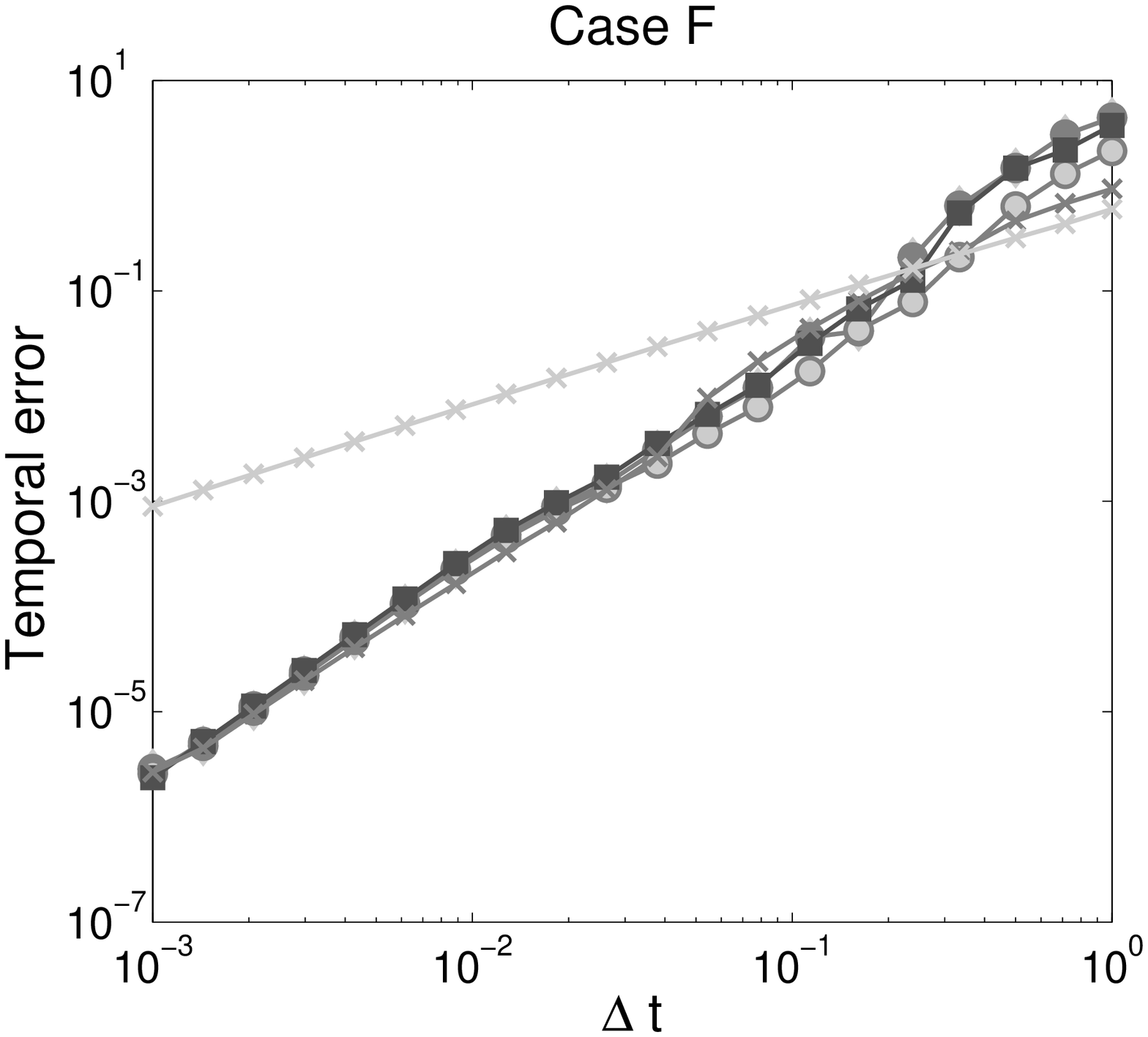}
\end{tabular}
\end{center}
\caption{Temporal errors $\widehat{e}\,(\Delta t;100,50)$ vs.~$\Delta t$
for vanilla American put options in the six cases
of Table~\ref{cases} with $\rho=0$ and reference solution by CN-IT with
$N=20~000$.
Two $\theta$-IT methods: BE-IT (light x), CN-IT (dark x).
Four ADI-IT methods: Do-IT with $\theta=\frac{1}{2}$ (light diamond), CS-IT
with $\theta=\frac{1}{2}$ (dark circle), MCS-IT with $\theta=\frac{1}{3}$
(light circle) and HV-IT with $\theta=\frac{1}{2}+\frac{1}{6}\sqrt{3}$ (dark
square). All methods with damping - two steps $\Delta t/2$ with BE-IT.}
\label{TemporalErrorCNdamp}
\end{figure}

% Figure 3
\begin{figure}[H]
\begin{center}
\begin{tabular}{c c}
         \includegraphics[width=0.5\textwidth]{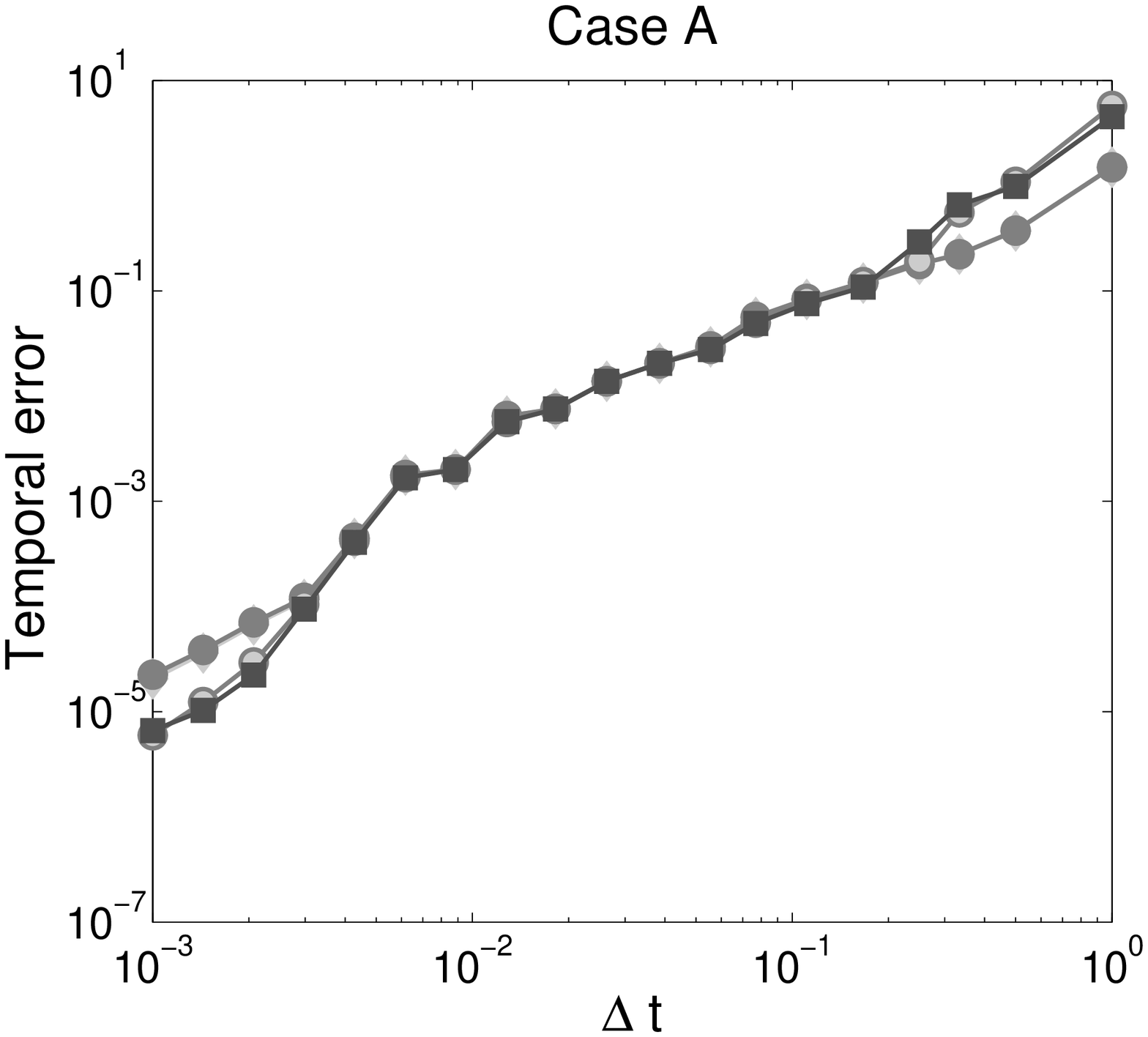}&
         \includegraphics[width=0.5\textwidth]{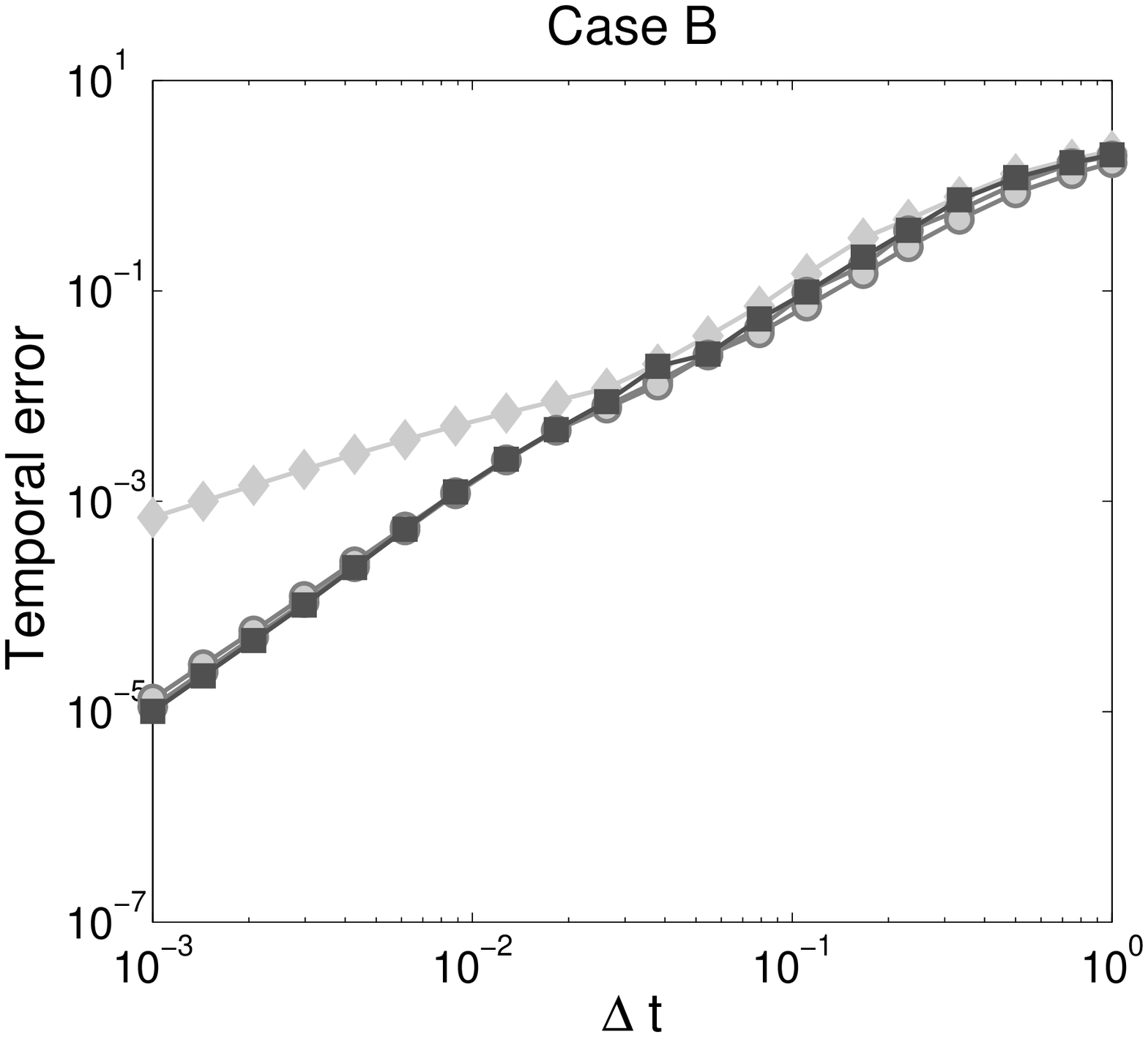}\\
         \includegraphics[width=0.5\textwidth]{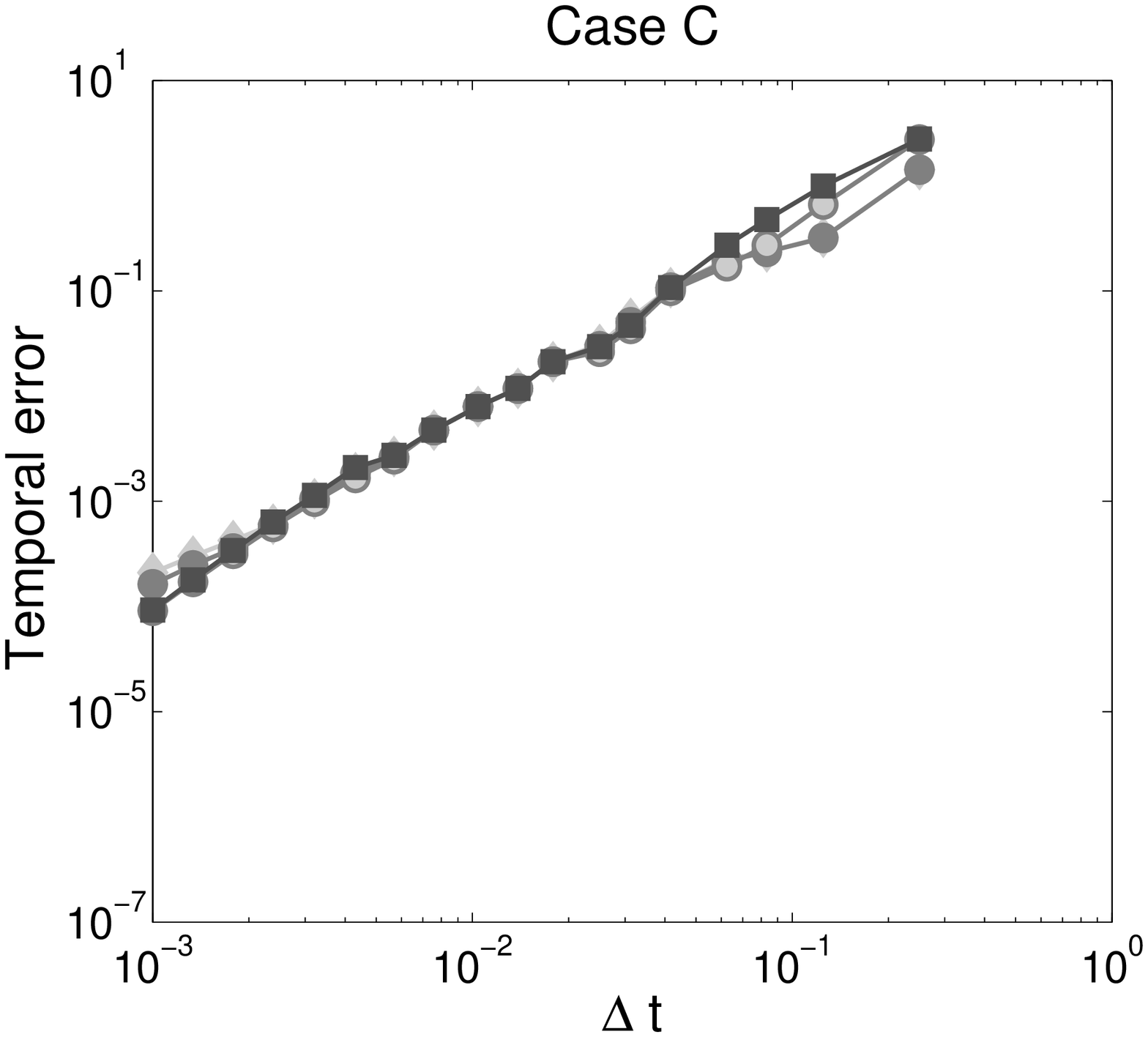}&
         \includegraphics[width=0.5\textwidth]{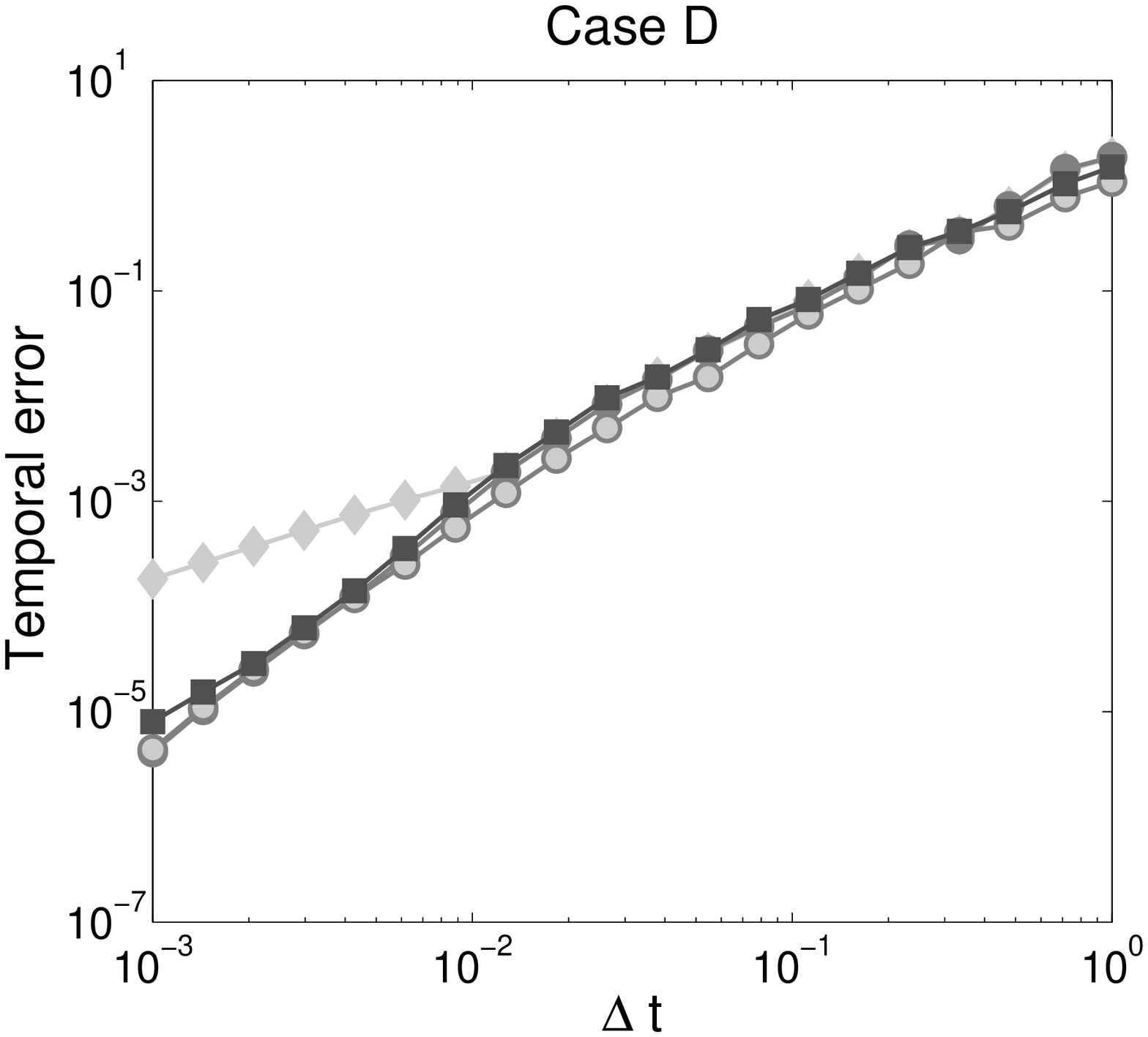}\\
         \includegraphics[width=0.5\textwidth]{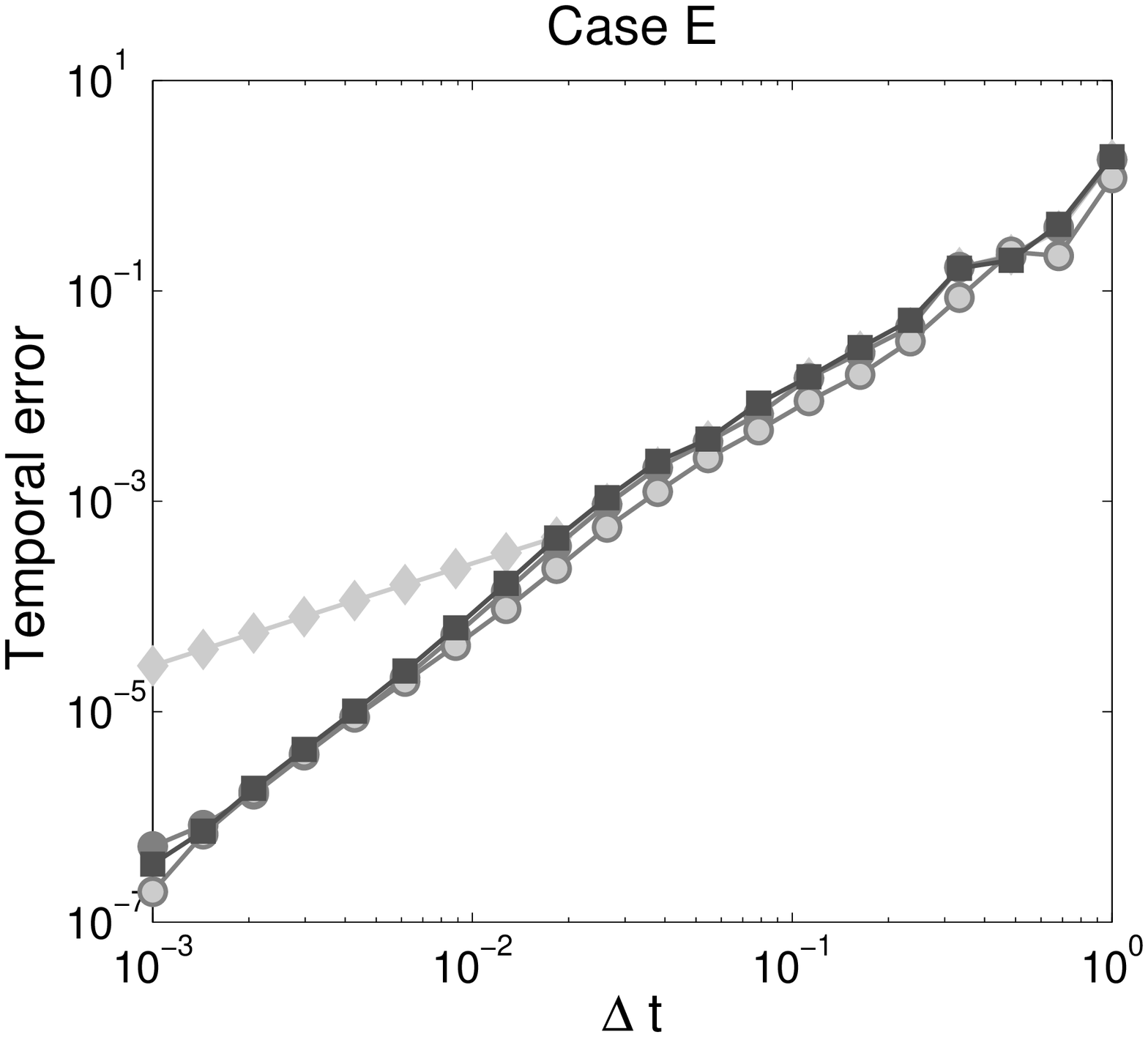}&
         \includegraphics[width=0.5\textwidth]{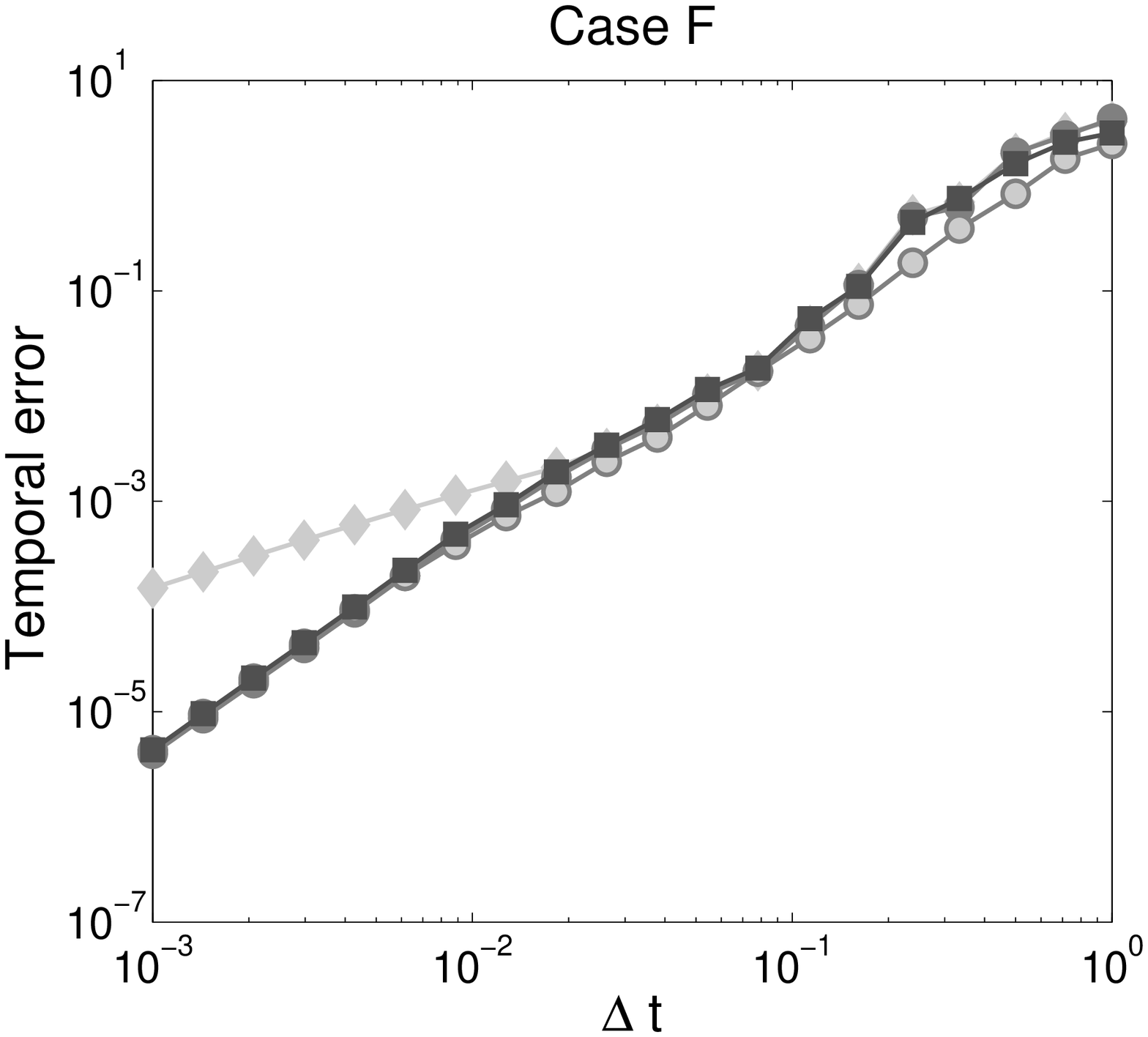}
\end{tabular}
\end{center}
\caption{Temporal errors $\widehat{e}\,(\Delta t;100,50)$ vs.~$\Delta t$
for vanilla American put options in the six cases in
Table~\ref{cases} with $\rho \neq 0$ and reference solution by MCS-IT with
$N=20~000$.
Four ADI-IT methods: Do-IT with $\theta=\frac{1}{2}$ (light diamond), CS-IT
with $\theta=\frac{1}{2}$ (dark circle), MCS-IT with $\theta=\frac{1}{3}$
(light circle) and HV-IT with $\theta=\frac{1}{2}+\frac{1}{6}\sqrt{3}$ (dark
square). Do-IT and CS-IT with damping - two steps $\Delta t/2$ with BE-IT.}
\label{TemporalErrorADICorrd}
\end{figure}

% Figure 4
\begin{figure}[H]
\begin{center}
\begin{tabular}{c c}
         \includegraphics[width=0.5\textwidth]{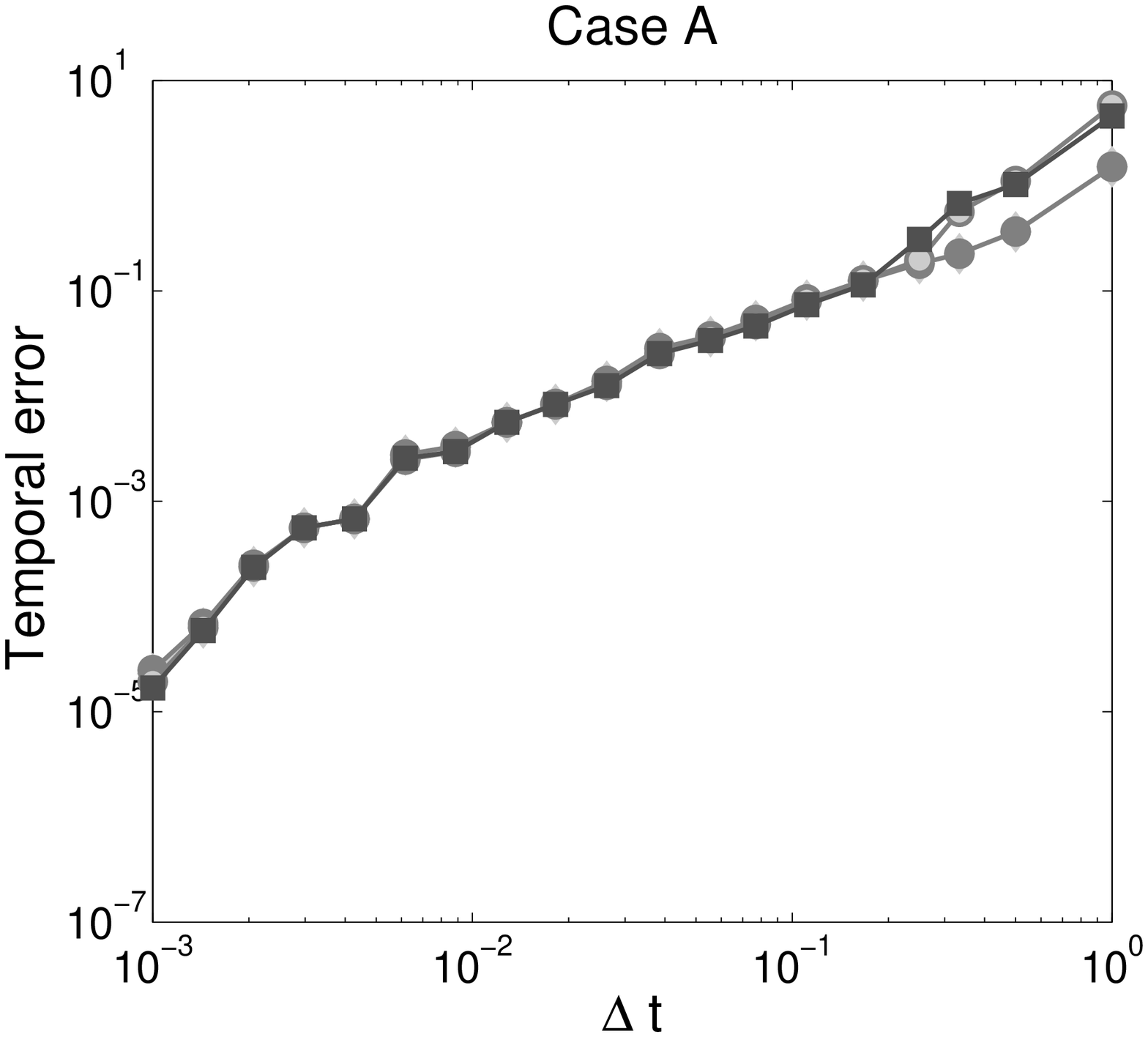}&
         \includegraphics[width=0.5\textwidth]{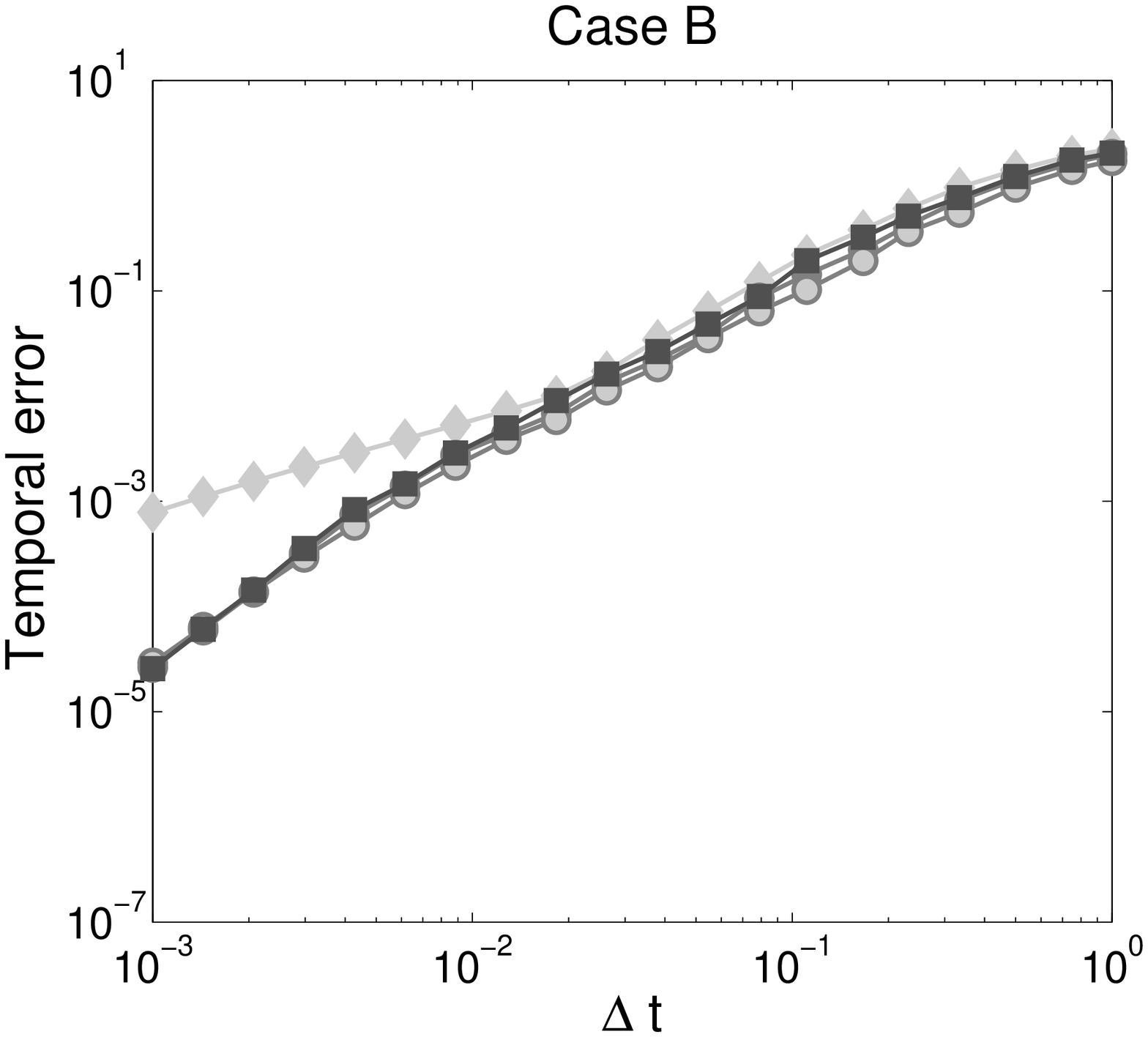}\\
         \includegraphics[width=0.5\textwidth]{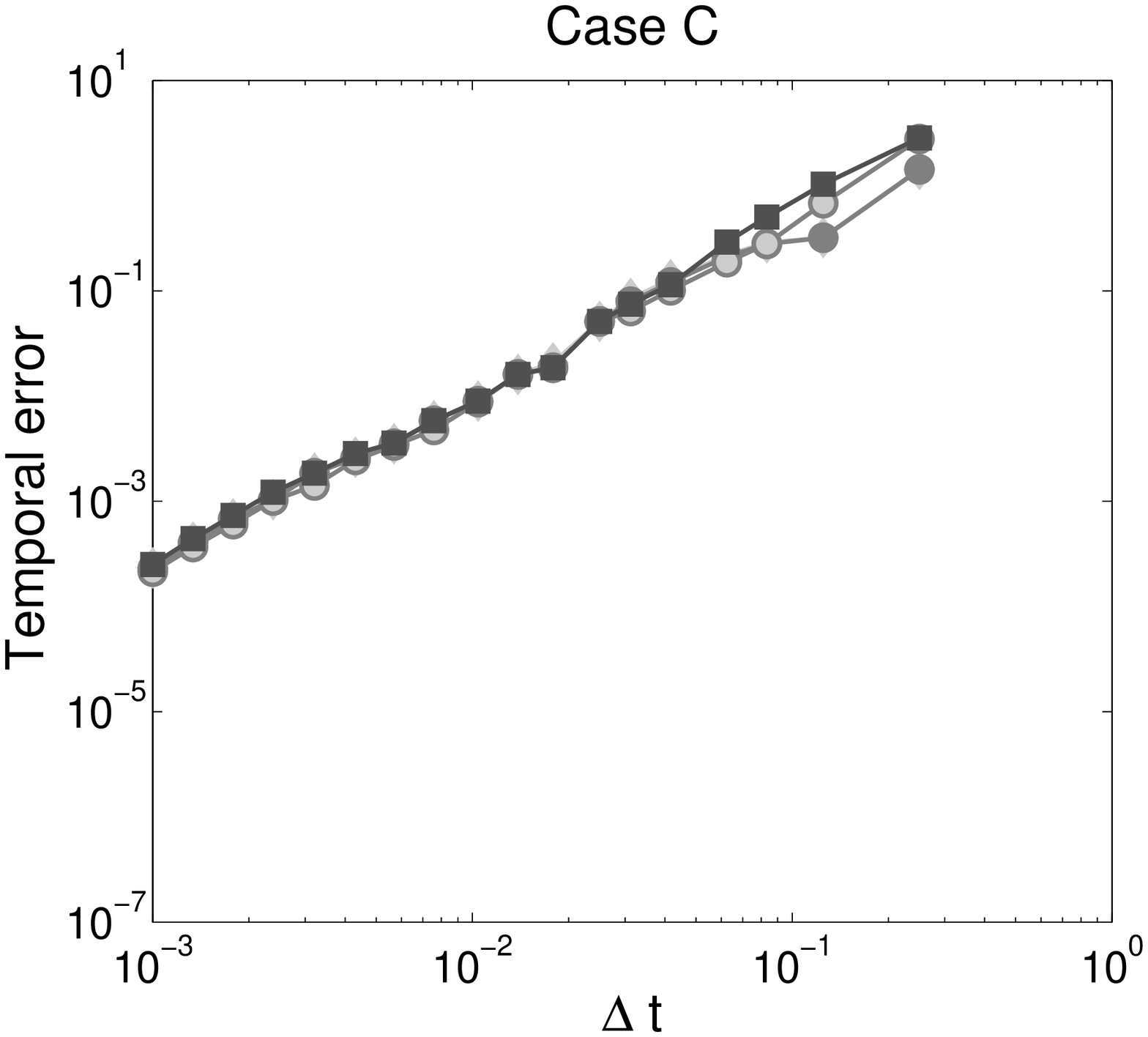}&
         \includegraphics[width=0.5\textwidth]{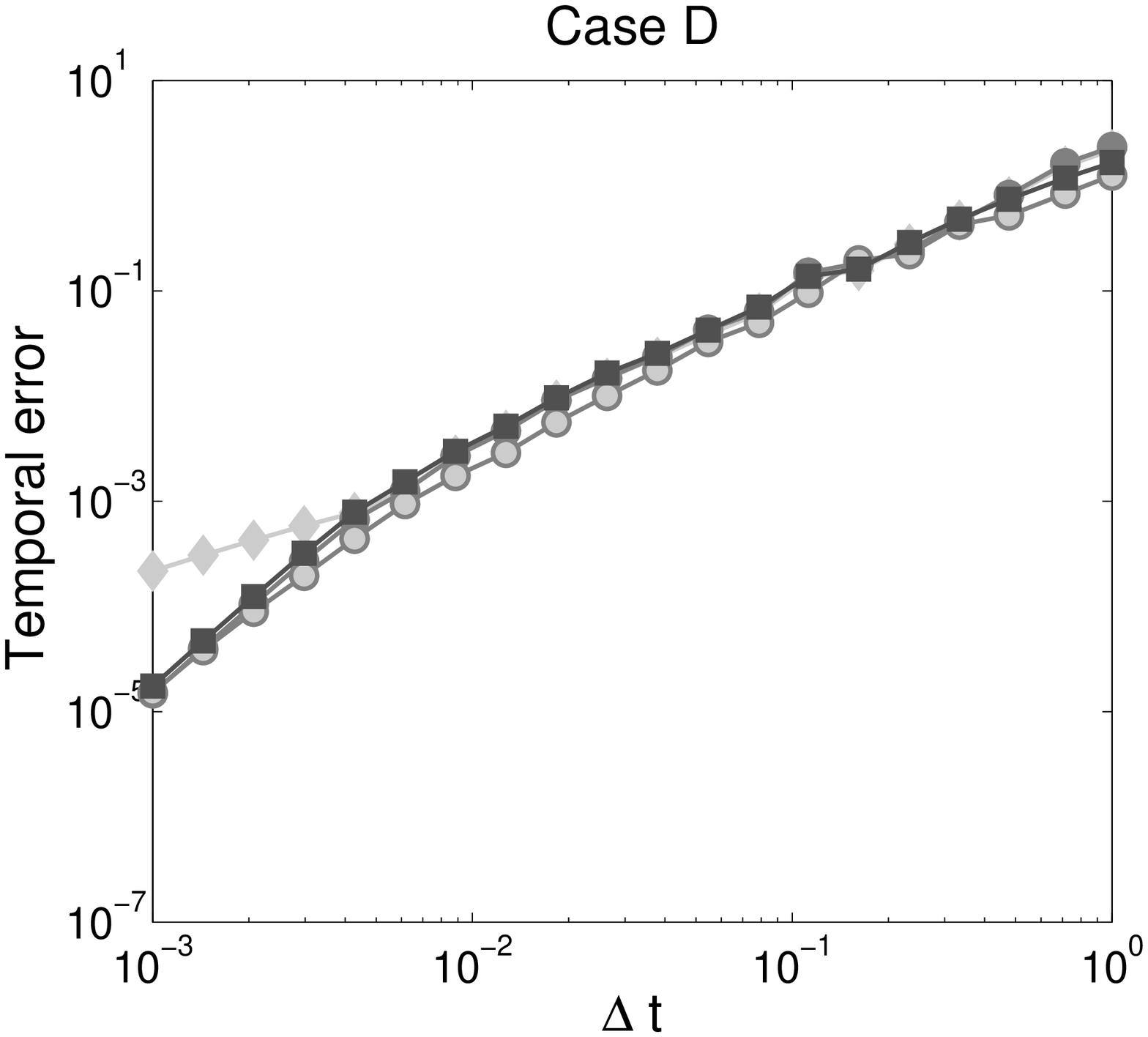}\\
         \includegraphics[width=0.5\textwidth]{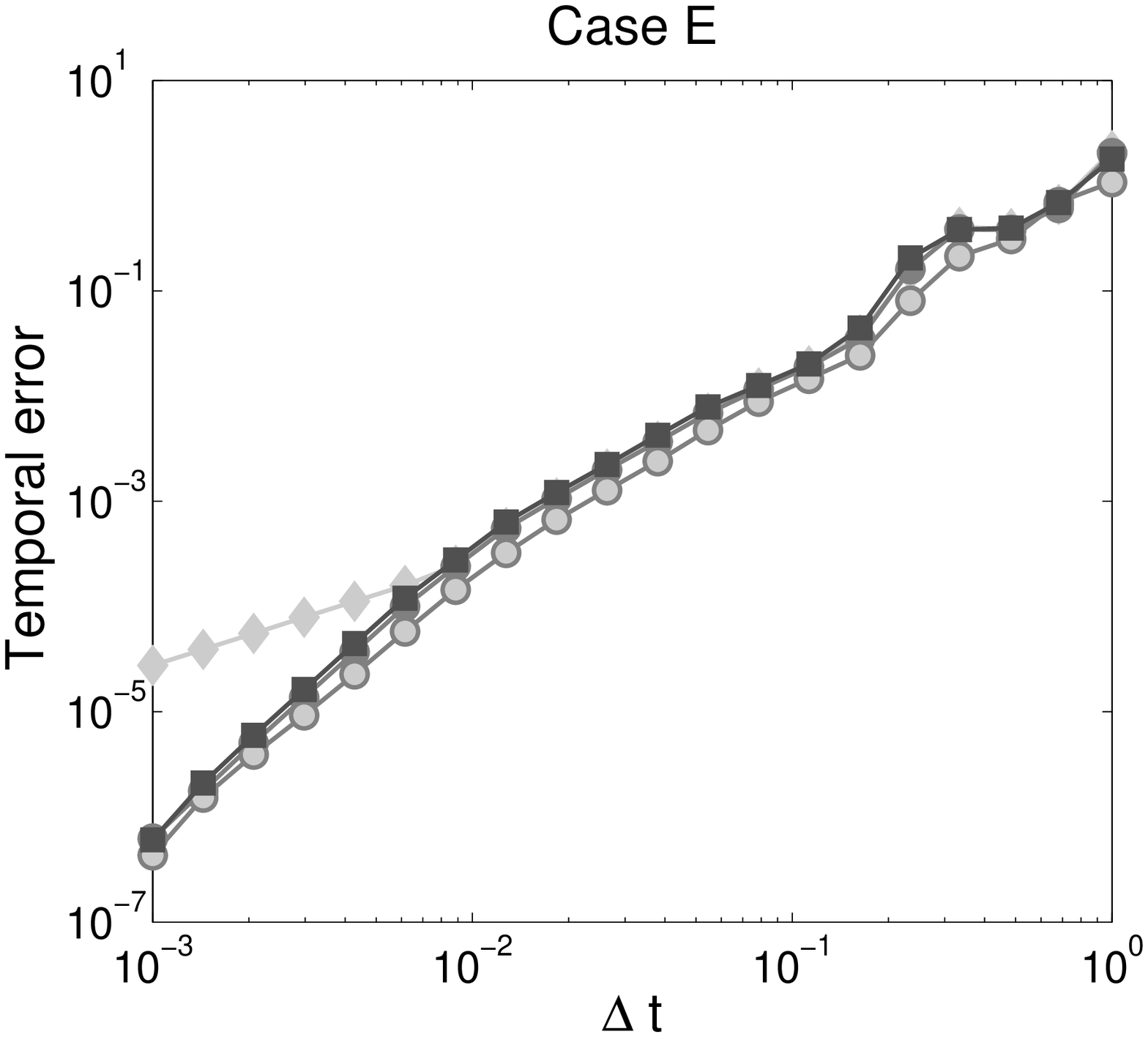}&
         \includegraphics[width=0.5\textwidth]{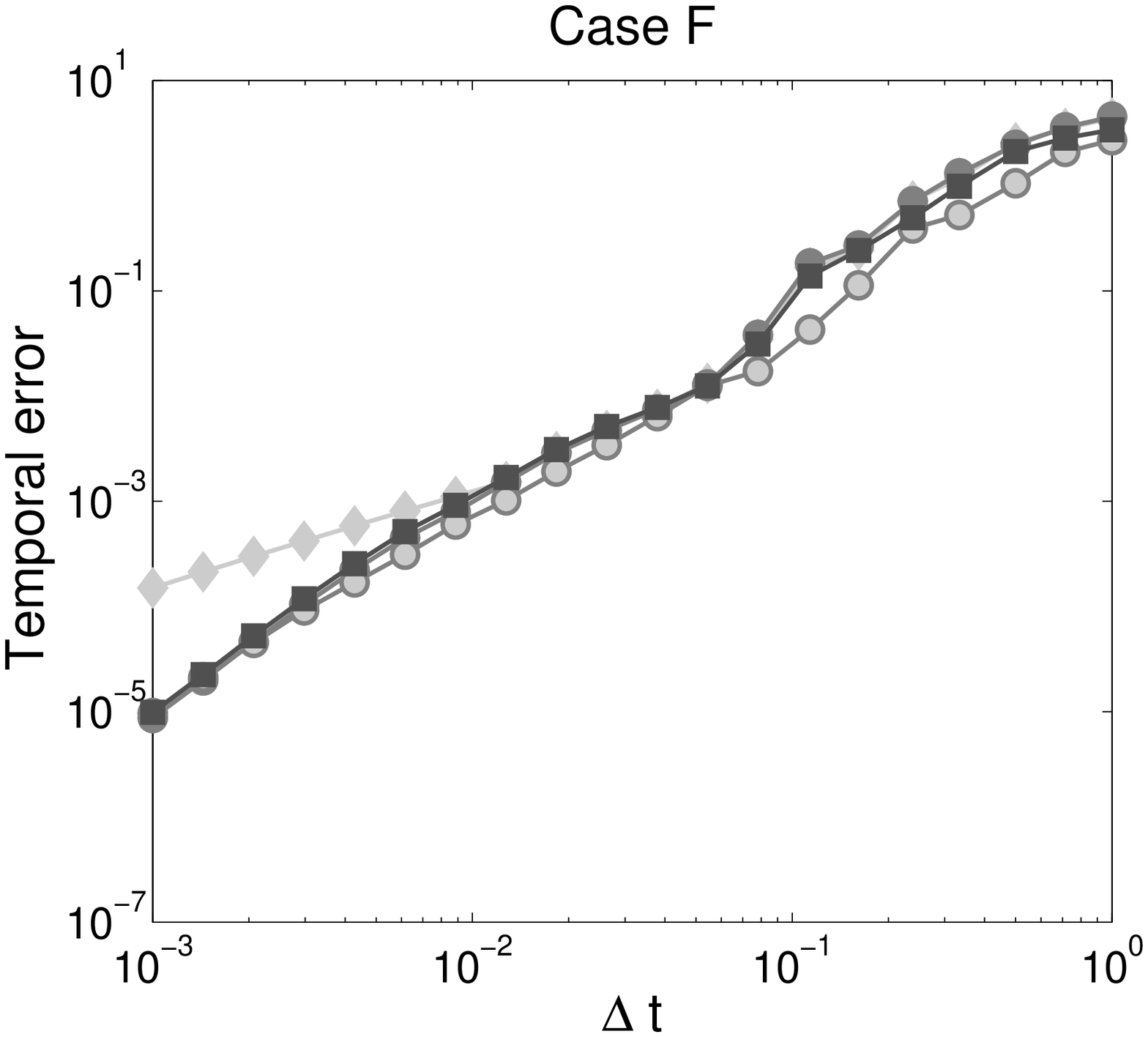}
\end{tabular}
\end{center}
\caption{Temporal errors $\widehat{e}\,(\Delta t;200,100)$ vs.~$\Delta t$
for vanilla American put options in the six cases in
Table~\ref{cases} with $\rho \neq 0$ and reference solution by MCS-IT with
$N=20~000$.
Four ADI-IT methods: Do-IT with $\theta=\frac{1}{2}$ (light diamond), CS-IT
with $\theta=\frac{1}{2}$ (dark circle), MCS-IT with $\theta=\frac{1}{3}$
(light circle) and HV-IT with $\theta=\frac{1}{2}+\frac{1}{6}\sqrt{3}$ (dark
square). Do-IT and CS-IT with damping - two steps $\Delta t/2$ with BE-IT.}
\label{TemporalErrorADICorrdm100}
\end{figure}

\noindent
The result, for all six methods in all six cases, is displayed
in Figure~\ref{TemporalErrorCNdamp}.
Clearly the undesirable phenomenon for the CN-IT, Do-IT and
CS-IT methods is mitigated, and for each $\Delta t$ their
temporal errors are now almost the same as for the MCS-IT
and HV-IT methods.

A careful inspection of case C in Figure~\ref{TemporalErrorCNdamp}
shows a convergence order that is slightly smaller than 2, namely
about 1.7, for all methods (except BE-IT which has only order 1).
The difference between convergence orders 2 and 1.7 is not readily
visible in a figure and the significance of the observation is not
completely clear to us.
A possible explanation may lie in the short maturity time and
corresponding nonsmoothness of the option pricing function in
this case.
We note that the lower order in case C did not improve by using
the variable time step strategy from~\cite{IT09}.

In the subsequent experiments we focus on the ADI-IT methods.
Here, in view of the above, the Do-IT and CS-IT methods are
applied with the BE-IT damping procedure.
Thus we consider:
\begin{itemize}
\item the Do-IT method with $\theta = \frac{1}{2}$ and damping
\item the CS-IT method with $\theta = \frac{1}{2}$ and damping
\item the MCS-IT method with $\theta = \frac{1}{3}$
\item the HV-IT method with $\theta = \frac{1}{2} + \frac{1}{6}\sqrt{3}$.
\end{itemize}
Figure~\ref{TemporalErrorADICorrd} displays for these four methods
the global temporal errors $\widehat{e}\,(\Delta t;100,50)$ versus
$\Delta t$ in all cases A--F, now with correlation $\rho \neq 0$.
The figure demonstrates that also with nonzero correlation all
ADI-IT methods show an unconditionally stable behavior.
For the CS-IT, MCS-IT and HV-IT methods, the temporal errors
are bounded from above by $C(\Delta t)^p$ with $p\approx 2$,
except in case C where again $p\approx 1.7$.
The Do-IT method often has temporal errors that are almost
the same as for the latter three methods.
A further investigation reveals that the regions of time steps
where this occurs are precisely those where Do-IT possesses
large temporal errors if no damping would have been applied.
Once $\Delta t$ gets sufficiently small, then a first-order
convergence behavior for this method sets in, as expected.
We note that in cases A, C this is not observed in the figure;
here it happens for $\Delta t < 10^{-3}$.

To gain insight into the dependence of the temporal
convergence behavior on the number of spatial grid
points $M$, we have doubled the number of mesh points
the $s$- and $v$-directions.
Figure~\ref{TemporalErrorADICorrdm100} displays the
obtain errors $\widehat{e}\,(\Delta t;200,100)$
versus $\Delta t$.
Comparing with Figure~\ref{TemporalErrorADICorrd}
we see that the temporal errors are at most mildly
affected by the strong increase in $M$.
This suggests that the convergence behavior of the
four ADI-IT methods is valid in the so-called
stiff sense, which forms a key property of
effective time discretization methods.

The implementation of all methods has been done in
Matlab where all matrices have been defined as sparse.
As an indication for the CPU times, the CS-IT, MCS-IT and HV-IT
methods each took about 0.003, 0.01, 0.02 seconds per time
step if $m=50, 100, 150$, respectively, on one Intel Core
Duo T7250 2.00 GHz processor with 4 GB memory.
For the Do-IT method these times are about halved.

We next validate the American option pricing method from
this paper by comparing its results with approximations
already given in the literature.
The MCS-IT method is chosen as a representative from the ADI-IT
class.
To test the numerical valuation of American put options in
the Heston model, many authors have considered the parameter
set
\begin{equation}\label{val1}
\kappa = 5,~ \eta=0.16,~ \sigma=0.9,~ \rho=0.1,~
r=0.1,~ T=0.25,~ K=10,
\end{equation}
with spot asset prices and variances given by
$s = S_0 \in \{8,9,10,11,12\}$ and
$v = V_0 \in \{0.0625,0.25\}$.
We approximated the pertinent (unknown) exact option prices
by FD discretization with $m=50, 100, 150$ and applying the
MCS-IT method with $N=25, 50, 75$ time steps, respectively.
Next spline interpolation
was used to compute approximations at the
off-grid points $(s,v)=(S_0,V_0)$.
Our results are given in the upper part of Tables~\ref{VN1}
(for $V_0=0.0625$) and \ref{VN2} (for $V_0=0.25$).
In the lower part, the two tables show approximations
obtained by
Zvan, Forsyth \& Vetzal \cite[Table~2]{ZFV98},
Ikonen \& Toivanen \cite[Table~1]{IT09},
Persson \& Von Sydow \cite[Tables~2,\,3]{PS10},
Oosterlee \cite[Table~5.2]{O03},
Clarke \& Parrott \cite{CP96} and
Vellekoop \& Nieuwenhuis \cite[Table~4]{VN09}.
The latter paper employs tree-based methods.
The reference prices taken from \cite{IT09} were computed
using a second-order, $L$-stable Runge--Kutta (RK) scheme
in the IT splitting approach.
Tables~\ref{VN1} and \ref{VN2} show that our option prices
are nicely in line with those obtained, by different
discretization techniques, in the literature.

\begin{table}[H]
\begin{center}
\begin{tabular}{|l|l|l|l|l @{.} l |l @{.} l |l @{.} l |l @{.} l |l @{.} l |}
        \hline
        & $m_1$ & $m_2$ & $N$ & \multicolumn{2}{|l}{8} & \multicolumn{2}{l}{9} & \multicolumn{2}{l}{10} &\multicolumn{2}{l}{11}&\multicolumn{2}{l|}{12}\\
        \hline \hline
        MCS-IT  & 100  & 50   & 25  & 2&0001 & 1&1088 & 0&5209 & 0&2152 & 0&0836\\
                & 200  & 100  & 50  & 2&0000 & 1&1083 & 0&5206 & 0&2146 & 0&0830\\
                & 300  & 150  & 75  & 2&0000 & 1&1081 & 0&5204 & 0&2143 & 0&0827\\
        \hline
        \cite{ZFV98} ZFV  & 177  & 103  &     & 2&0000 & 1&1076 & 0&5202 & 0&2138 & 0&0821\\
        \cite{IT09} RK-IT & 320  & 128  & 64  & 2&0000 & 1&1076 & 0&5199 & 0&2135 & 0&0820\\
        \cite{PS10} PS    & 81   & 21   & 329 & 1&9998 & 1&1085 & 0&5195 & 0&2150 & 0&0822\\
        \cite{O03} O      & 256  & 256  &     & 2&00   & 1&107  & 0&517  & 0&212  & 0&0815\\
        \cite{CP96} ~\,CP &      &      &     & 2&0000 & 1&1080 & 0&5316 & 0&2261 & 0&0907\\
        \cite{VN09} VN    & 1000 & 48   & 71  & 1&9968 & 1&1076 & 0&5202 & 0&2134 & 0&0815\\
        \hline
\end{tabular}
\end{center}
\caption{Vanilla American put prices for
parameter set (\ref{val1}), $S_0 \in \{8,9,10,11,12\}$, \mbox{$V_0 = 0.0625$}.
Upper part: approximations using MCS-IT method with $\theta=\frac{1}{3}$.
Lower part: approximations from the literature.}
\label{VN1}
\end{table}

\begin{table}[H]
\begin{center}
\begin{tabular}{|l|l|l|l|l @{.} l |l @{.} l |l @{.} l |l @{.} l |l @{.} l |}
        \hline
        & $m_1$ & $m_2$ & $N$ & \multicolumn{2}{|l}{8} & \multicolumn{2}{l}{9} & \multicolumn{2}{l}{10} &\multicolumn{2}{l}{11}&\multicolumn{2}{l|}{12}\\
        \hline \hline
        MCS-IT  & 100  & 50   & 25  & 2&0793 & 1&3342 & 0&7963 & 0&4488 & 0&2438\\
                & 200  & 100  & 50  & 2&0789 & 1&3340 & 0&7963 & 0&4487 & 0&2435\\
                & 300  & 150  & 75  & 2&0788 & 1&3339 & 0&7962 & 0&4486 & 0&2433\\
        \hline
        \cite{ZFV98} ZFV  & 177  & 103  &     & 2&0784 & 1&3337 & 0&7961 & 0&4483 & 0&2428\\
        \cite{IT09} RK-IT & 320  & 128  & 64  & 2&0785 & 1&3336 & 0&7959 & 0&4482 & 0&2427\\
        \cite{PS10} PS    & 81   & 21   & 329 & 2&0784 & 1&3333 & 0&7955 & 0&4479 & 0&2426\\
        \cite{O03} O      & 256  & 256  &     & 2&079  & 1&334  & 0&796  & 0&449  & 0&243 \\
        \cite{CP96} ~\,CP &      &      &     & 2&0733 & 1&3290 & 0&7992 & 0&4536 & 0&2502\\
        \hline
\end{tabular}
\end{center}
\caption{Vanilla American put prices for
parameter set (\ref{val1}), $S_0 \in \{8,9,10,11,12\}$, $V_0 = 0.25$.
Upper part: approximations using MCS-IT method with $\theta=\frac{1}{3}$.
Lower part: approximations from the literature.}
\label{VN2}
\end{table}

For the parameter set (\ref{val1}) the Feller condition is satisfied.
A test set where Feller is violated, and reference prices for American
put options are given, is not included in the references above.
We choose here a parameter set that has recently been considered by
Fang \& Oosterlee \cite{FO11} in the numerical pricing of Bermudan
options under the Heston model, where the Feller condition does not
hold:
\begin{equation}\label{val2}
\kappa = 1.15,~ \eta=0.0348,~ \sigma=0.39,~ \rho=-0.64,~
r=0.04,~ T=0.25,~ K=100,
\end{equation}
with spot asset prices and variance given by
$s = S_0 \in \{90,100,110\}$ and $v = V_0 =0.0348$.
The upper part of Table~\ref{FO} displays the pertinent Bermudan
put option prices approximated by the COS method from
\cite[Table~5]{FO11}.
Here $N$ represents the number of exercise dates.
If $N$ increases, then the Bermudan prices tend to those of
their American counterpart.
The lower part of Table~\ref{FO} shows our approximations to
the American put option prices using the FD discretization
with fixed $m=150$ and applying the MCS-IT method
with $N=20, 40, 60$ time steps.
Clearly the prices obtained by both methods agree well,
in particular those for the largest $N$.

Note that the values in the lower part of Table~\ref{FO}
could also be viewed as approximations to Bermudan put
option prices with $N$ exercise dates.
The time step $\Delta t$ is then equal to the full period
between two successive exercise dates.
More accurate approximations to the Bermudan prices are
(readily) obtained by reducing this time step.

\begin{table}[H]
\begin{center}
\begin{tabular}{|l|l|l|l|l @{.} l |l @{.} l |l @{.} l |}
        \hline
        & $m_1$ & $m_2$ & $N$ &\multicolumn{2}{l}{~\,90} & \multicolumn{2}{l}{100} &\multicolumn{2}{l|}{110}\\
        \hline
        \hline
        \cite{FO11} FO  & & &20 & \phantom{0}9&9784 & 3&2047 & 0&9274\\
                        & & &40 & \phantom{0}9&9916 & 3&2073 & 0&9281\\
                        & & &60 & \phantom{0}9&9958 & 3&2079 & 0&9280\\
        \hline
        \hline
        MCS-IT  & 300 & 150 &20 & \phantom{0}9&9984 & 3&2121 & 0&9301\\
                & 300 & 150 &40 & 10&0015           & 3&2125 & 0&9304\\
                & 300 & 150 &60 & 10&0039           & 3&2126 & 0&9305\\
        \hline
\end{tabular}
\end{center}
\caption{
Parameter set (\ref{val2}), $S_0 \in \{90,100,110\}$, $V_0 = 0.0348$.
Upper part: approximations to vanilla Bermudan put prices
from \cite{FO11}.
Lower part: approximations to vanilla American put prices
using MCS-IT method with $\theta=\frac{1}{3}$.}
\label{FO}
\end{table}

\begin{figure}[H]
\begin{center}
\hspace*{-1.2cm}
\begin{tabular}{c c}
         \includegraphics[width=0.6\textwidth]{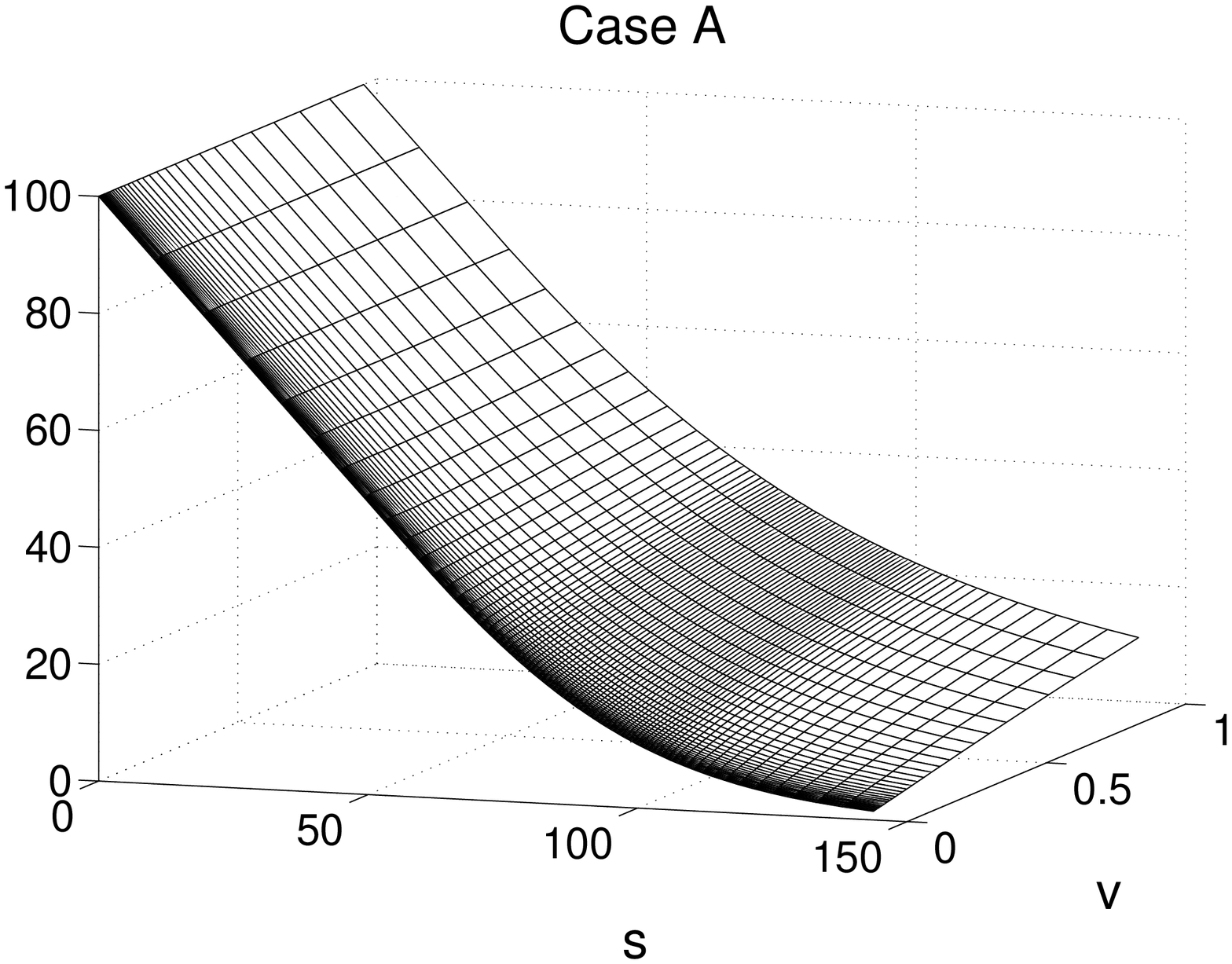}& \hspace*{-1cm}
         \includegraphics[width=0.6\textwidth]{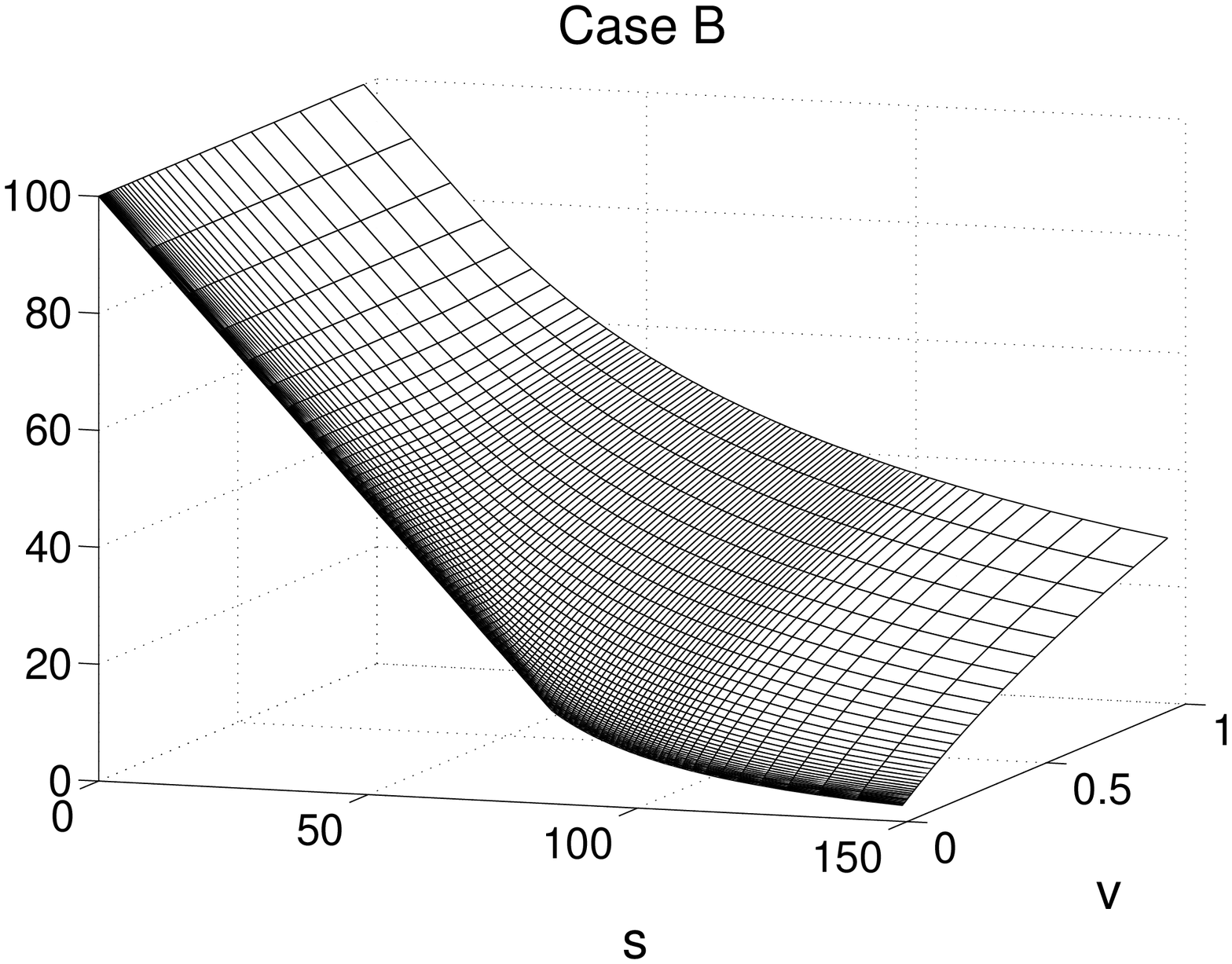}\\
         \includegraphics[width=0.6\textwidth]{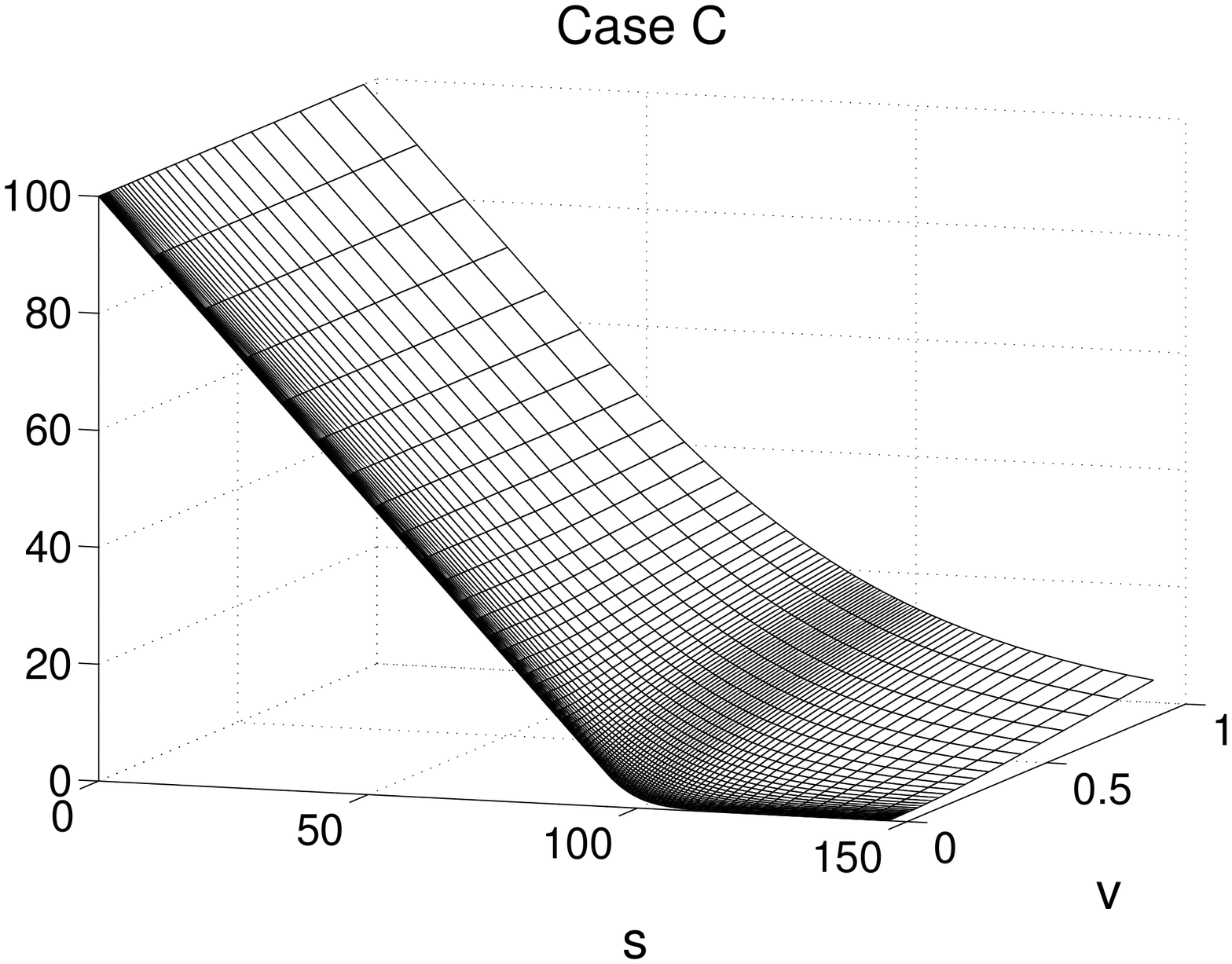}&\hspace*{-1cm}
         \includegraphics[width=0.6\textwidth]{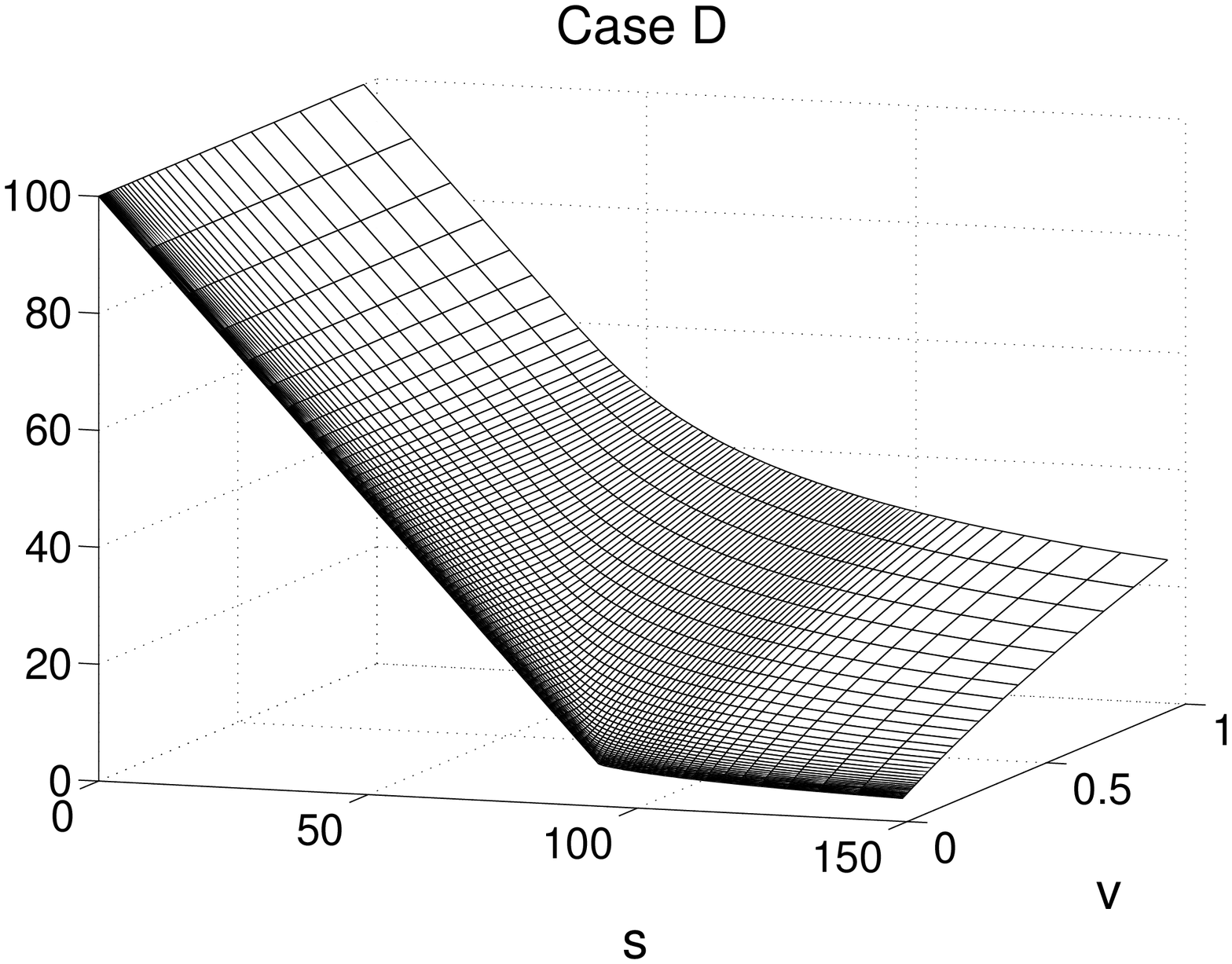}\\
         \includegraphics[width=0.6\textwidth]{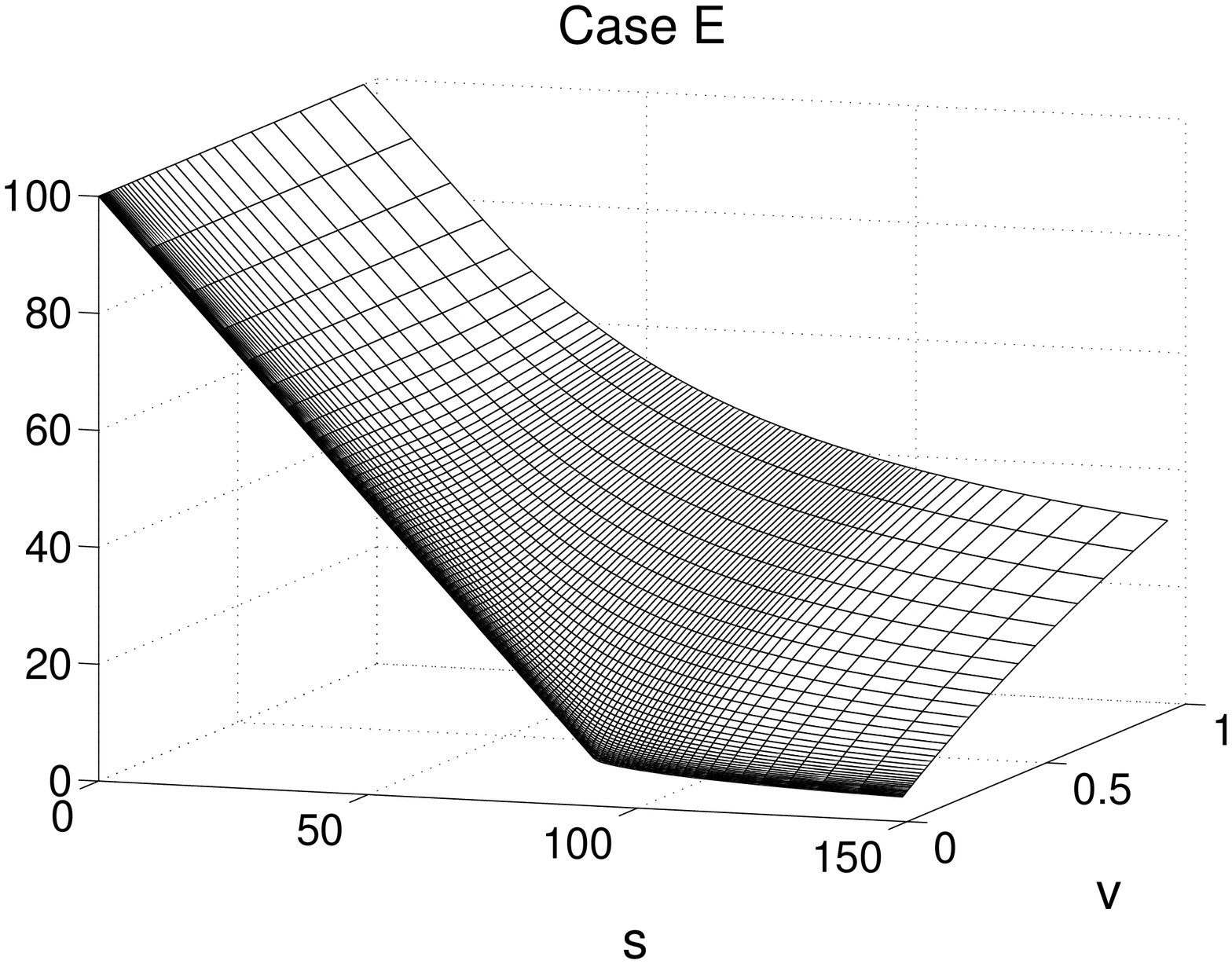}&\hspace*{-1cm}
         \includegraphics[width=0.6\textwidth]{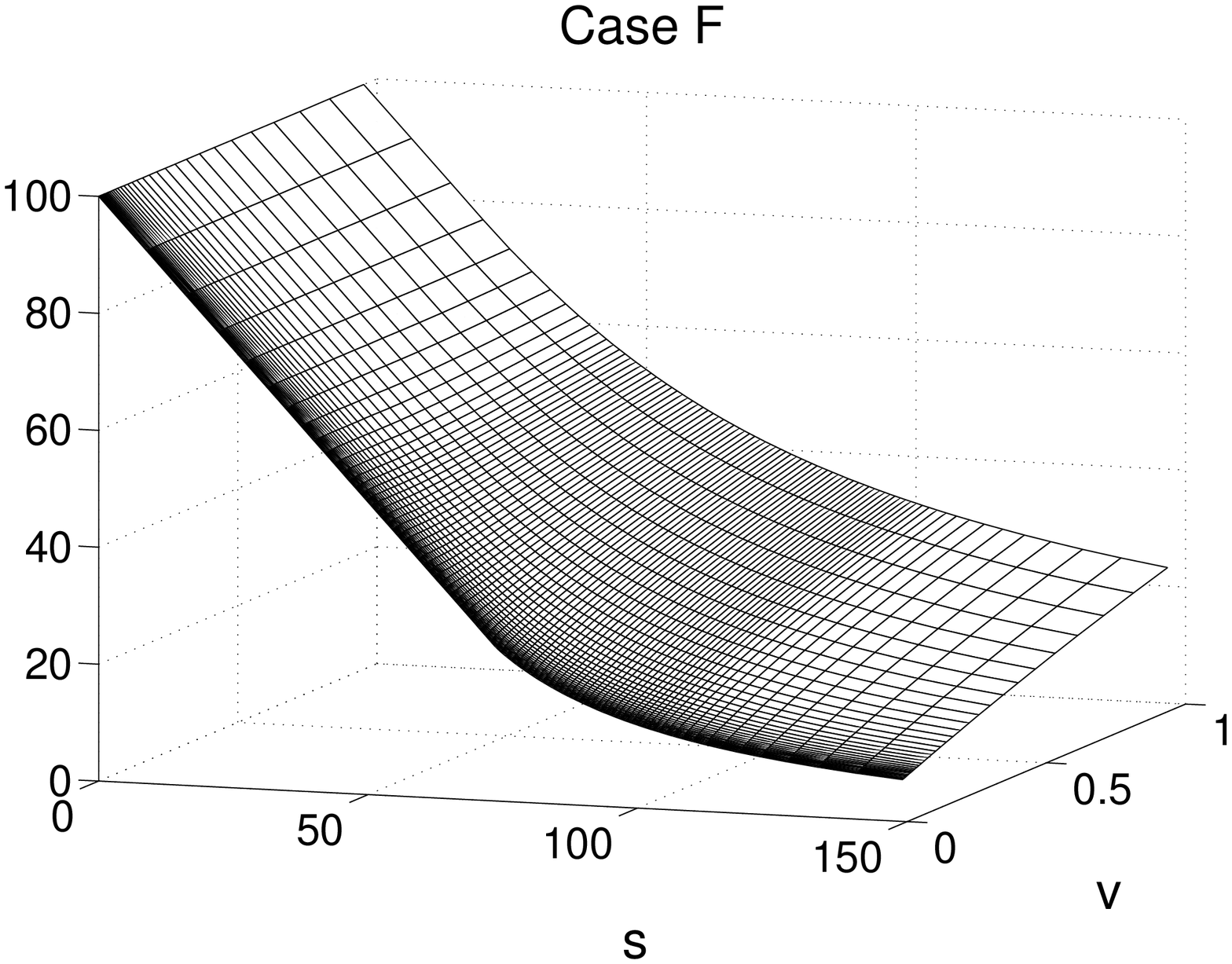}
\end{tabular}
\end{center}
\caption{
Approximated vanilla American put price surfaces in all cases
of Table~\ref{cases} with $\rho \not= 0$.
}
\label{AmPutSolution}
\end{figure}

For future reference we give in Table~\ref{Refvalues}
approximations for vanilla American put option prices in
all six cases of Table~\ref{cases} for spot asset prices
$S_0 \in \{90,100,110\}$ and spot variance $V_0 = 0.05$.
These have been computed using the FD discretization with
$m=250$ and the MCS-IT method with $\theta=\frac{1}{3}$
and $N=125$.
The full option price surfaces are displayed
in Figure~\ref{AmPutSolution}.

\begin{table}[H]
\begin{center}
\begin{tabular}{|c|l @{.} l |l @{.} l |l @{.} l |}
        \hline
        Case & \multicolumn{2}{l}{90} & \multicolumn{2}{l}{100} &\multicolumn{2}{l|}{110}\\
        \hline \hline
        A & 16&9245 & 11&9442           & \phantom{0}8&2270\\
        B & 16&0470 & 12&4326           & \phantom{0}9&8746\\
        C & 10&4054 & \phantom{0}3&9235 & \phantom{0}1&1784\\
        D & 10&9554 & \phantom{0}8&6273 & \phantom{0}7&4999\\
        E & 12&8442 & \phantom{0}9&8116 & \phantom{0}8&4312\\
        F & 18&9325 & 15&6696           & 13&2838\\
        \hline
\end{tabular}
\end{center}
\caption{
Vanilla American put option price approximations in all
cases of Table~\ref{cases} with $\rho \not= 0$ and
$S_0 \in \{90,100,110\}$, $V_0 = 0.05$.}
\label{Refvalues}
\end{table}

\begin{figure}[H]
\begin{center}
\begin{tabular}{c c}
         \includegraphics[width=0.5\textwidth]{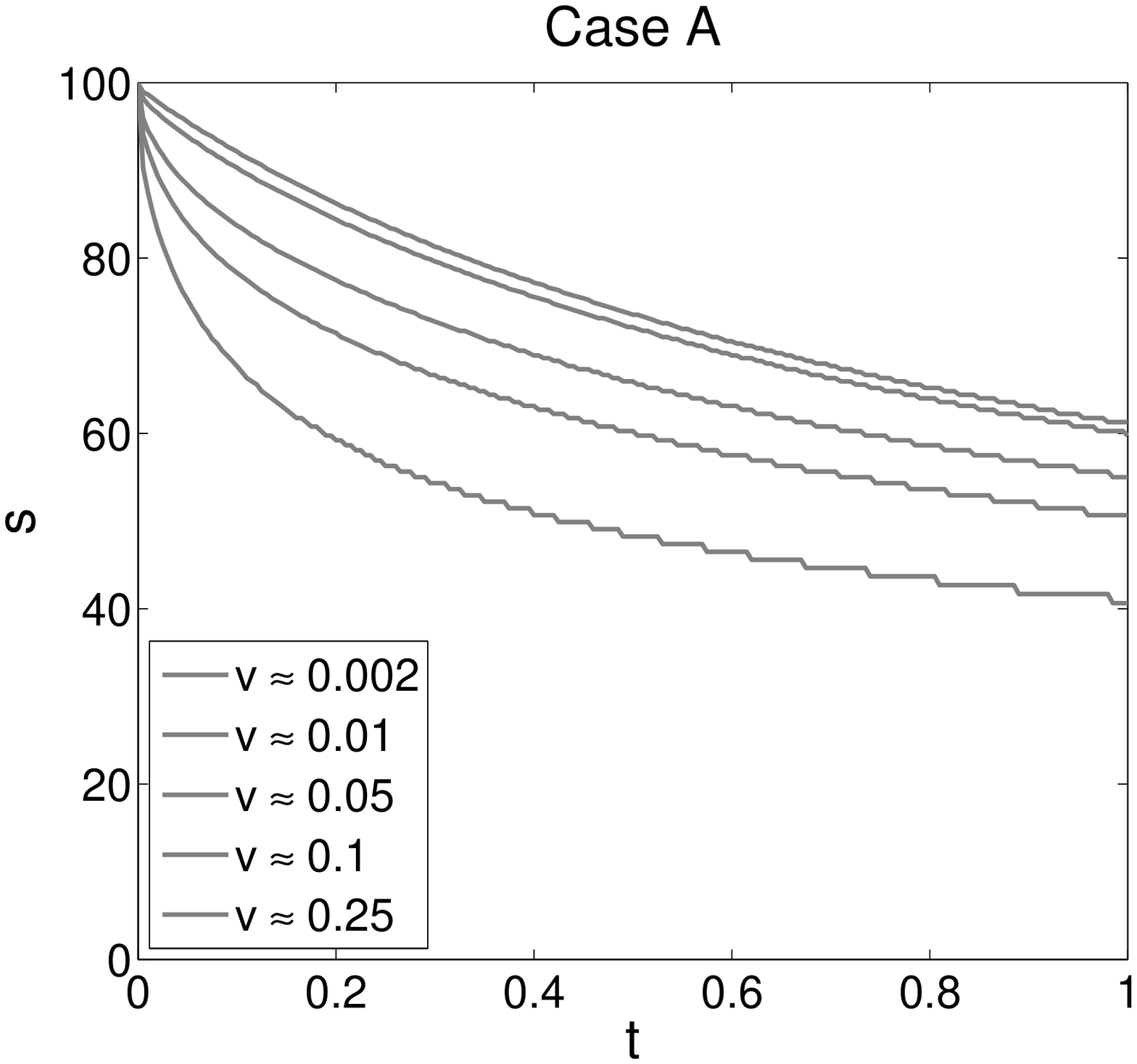}&
         \includegraphics[width=0.5\textwidth]{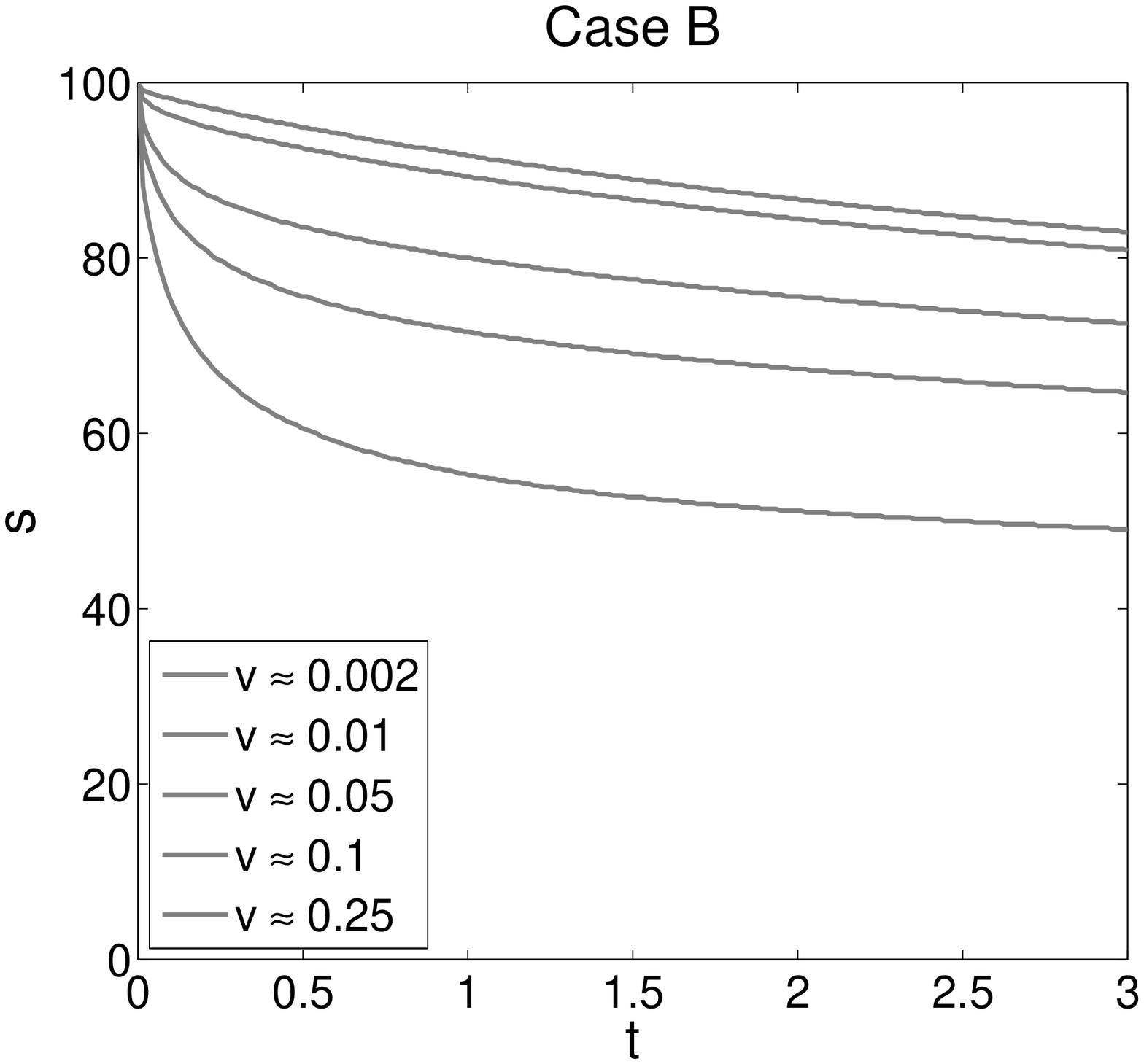}\\
         \includegraphics[width=0.5\textwidth]{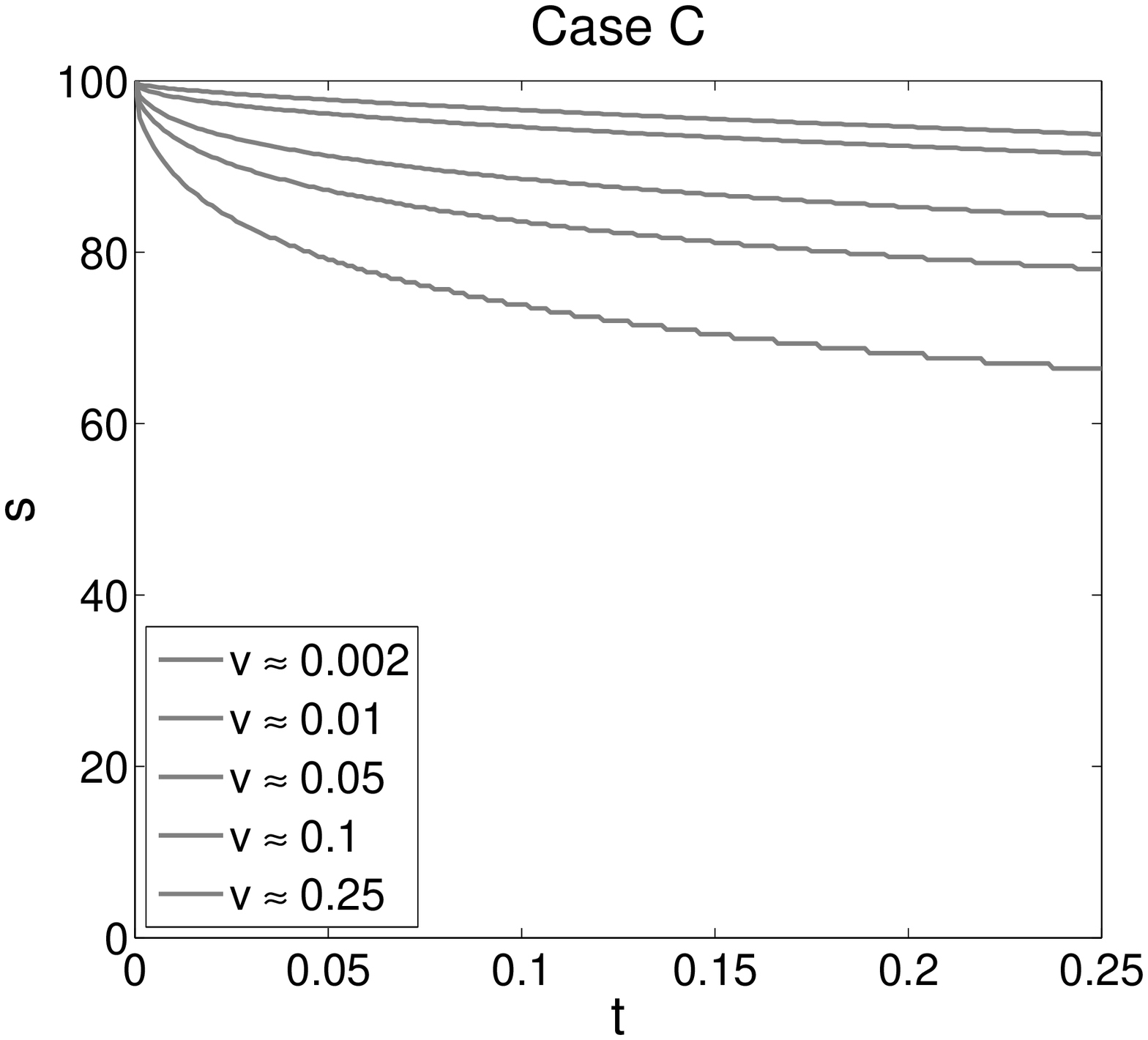}&
         \includegraphics[width=0.5\textwidth]{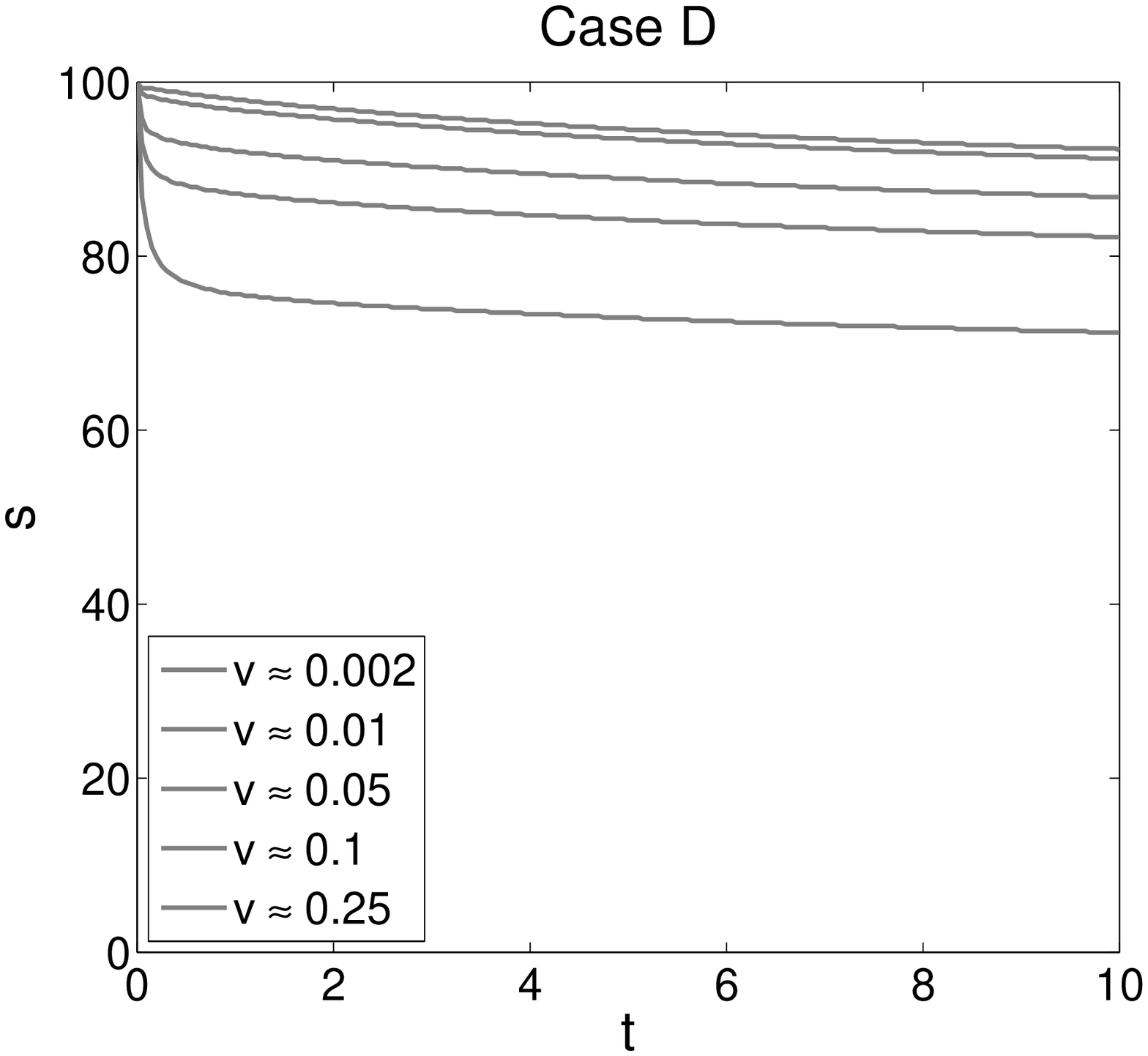}\\
         \includegraphics[width=0.5\textwidth]{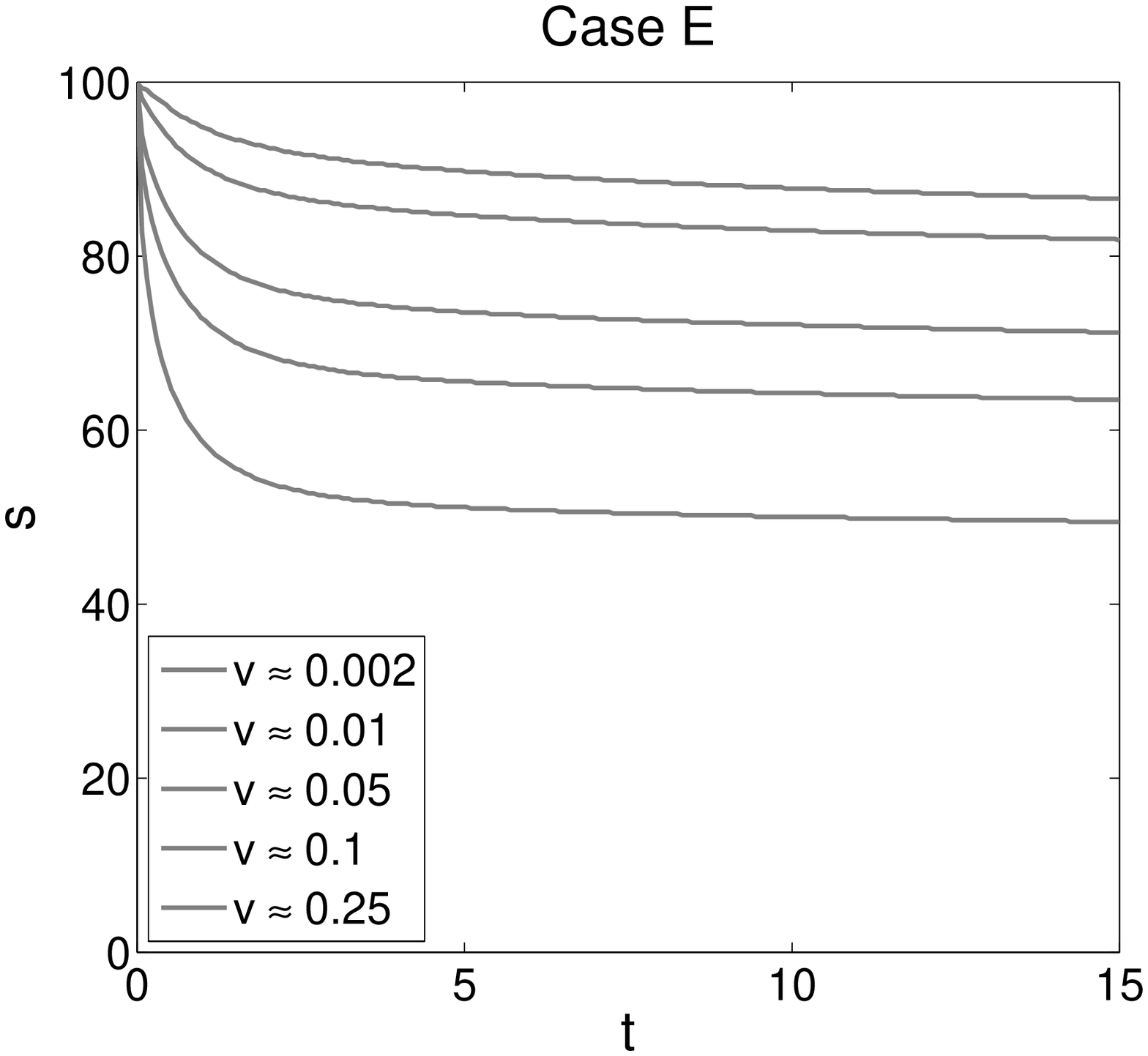}&
         \includegraphics[width=0.5\textwidth]{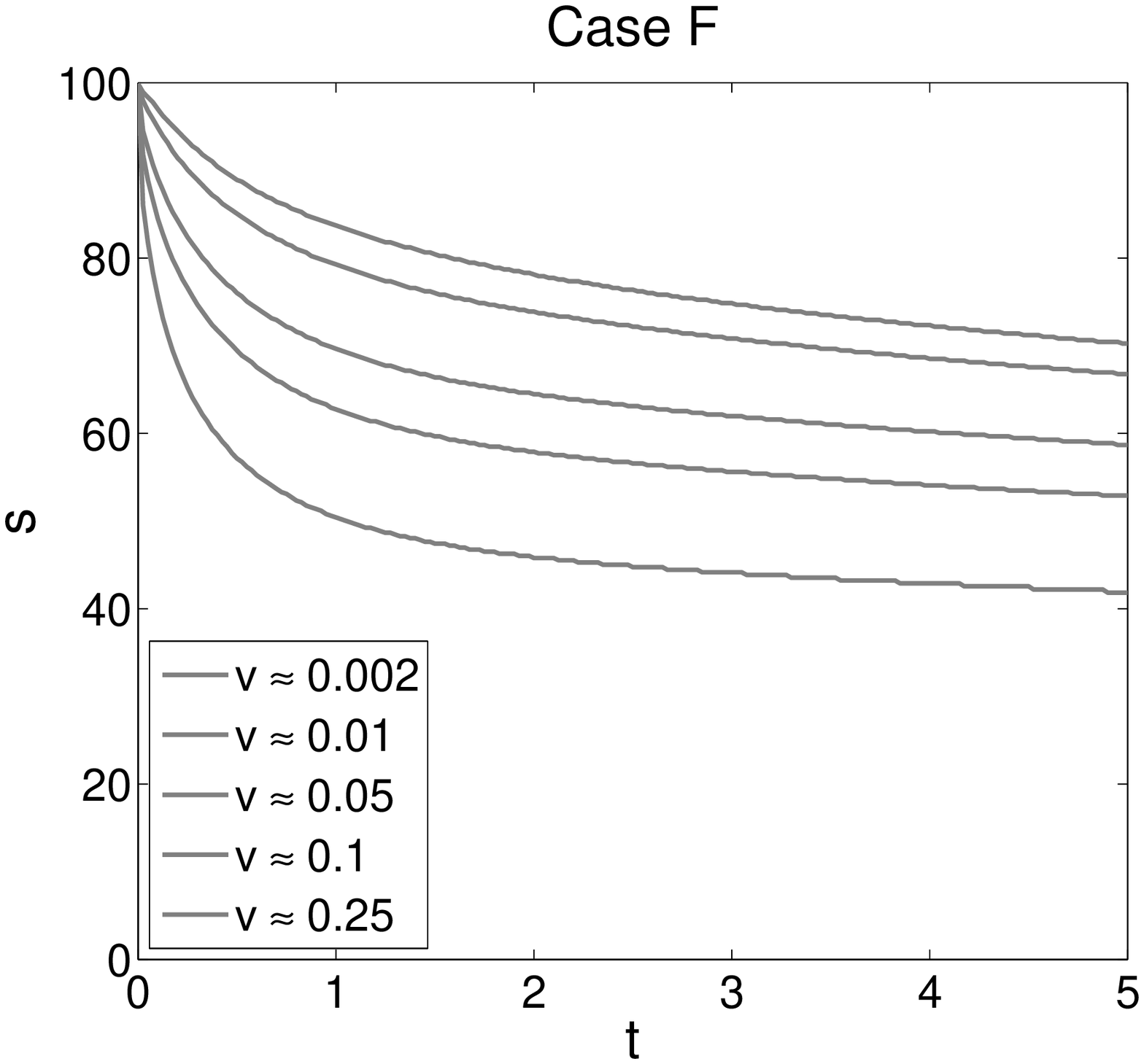}
\end{tabular}
\end{center}
\caption{Approximated free boundaries for vanilla American put options
in all cases of Table~\ref{cases} with $\rho\neq 0$.
From top to bottom: $v\approx 0.0021$, $v\approx 0.0093$, $v\approx 0.0484$,
$v\approx 0.0972$, $v\approx 0.2392$.}
\label{FreeBoundary}
\end{figure}

Next, Figure~\ref{FreeBoundary} shows for a selection
of values $v\approx 0.002, 0.01, 0.05, 0.1, 0.24$ the
corresponding parts of the free boundary in the
$(t,s)$-plane.
With the ADI-IT method proposed in this paper, the
exercise and continuation regions are directly
obtained by determining at each time point $t=t_n$
the two subsets of spatial grid points $(s_i,v_j)$
where the corresponding component of the auxiliary
vector $\hl_{n}$ is strictly positive resp.~equal
to zero.
The part below each curve in Figure~\ref{FreeBoundary}
represents the exercise region and the part above the
continuation region.

Finally, as a more exotic option we consider the capped
American put option.
This is an American-style option with a cap $B<K$ on the
underlying asset price.
If the asset price goes below the cap $B$, then
the option is automatically exercised and an amount of
$K-B$ is paid out to the holder.
The relevant option value function $u$ satisfies the Heston
PDCP (\ref{H_pdcp}) whenever $s>B$, $v>0$, $0<t\leq T$ with
$\phi$ given by (\ref{payoff}).
The boundary condition (\ref{ibcs}a) becomes
\[
    u(B,v,t) = K-B.
\]

\begin{figure}[H]
\begin{center}
\hspace*{-1.2cm}
\begin{tabular}{c c}
         \includegraphics[width=0.6\textwidth]{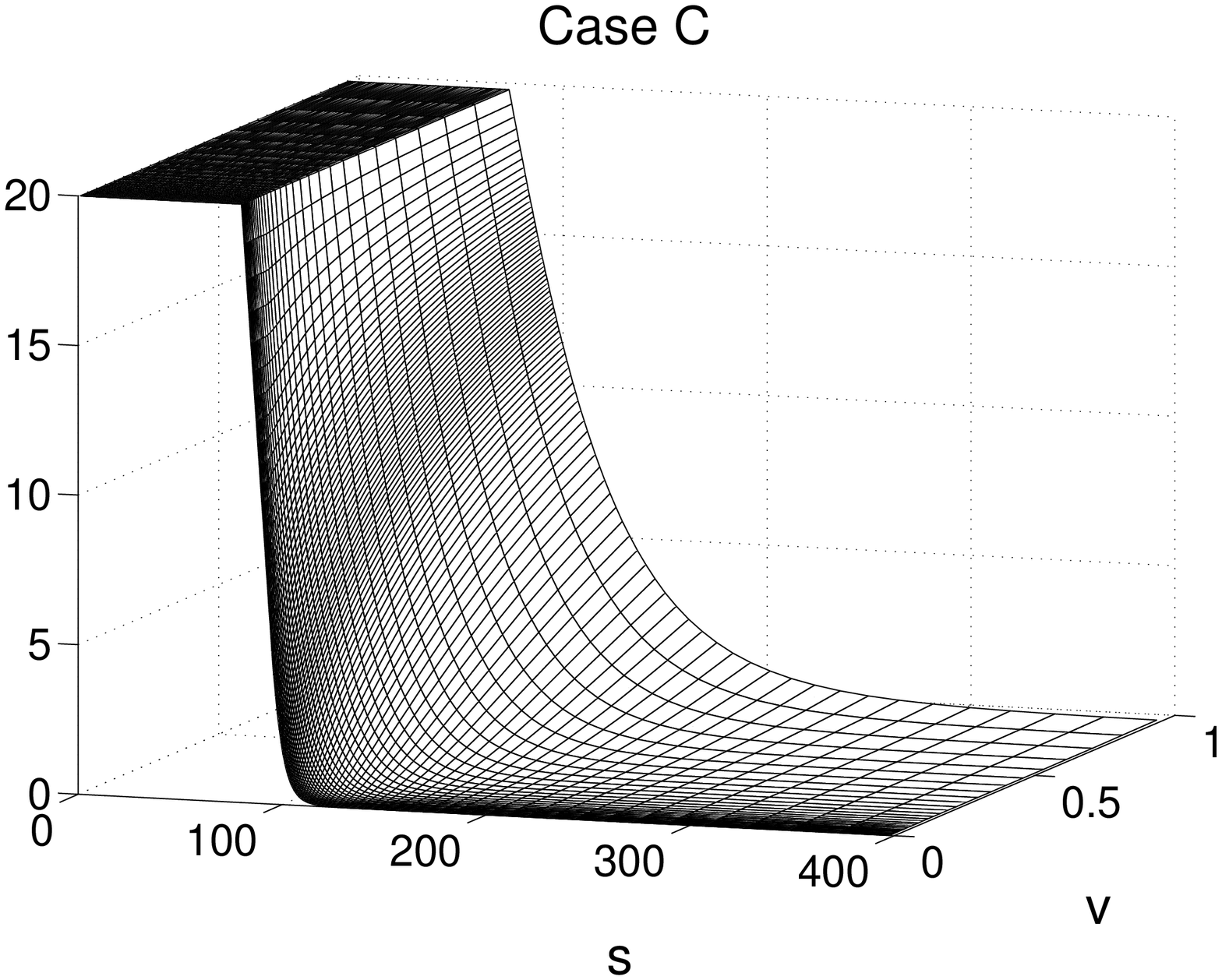}& \hspace*{-0.5cm}
         \includegraphics[width=0.5\textwidth]{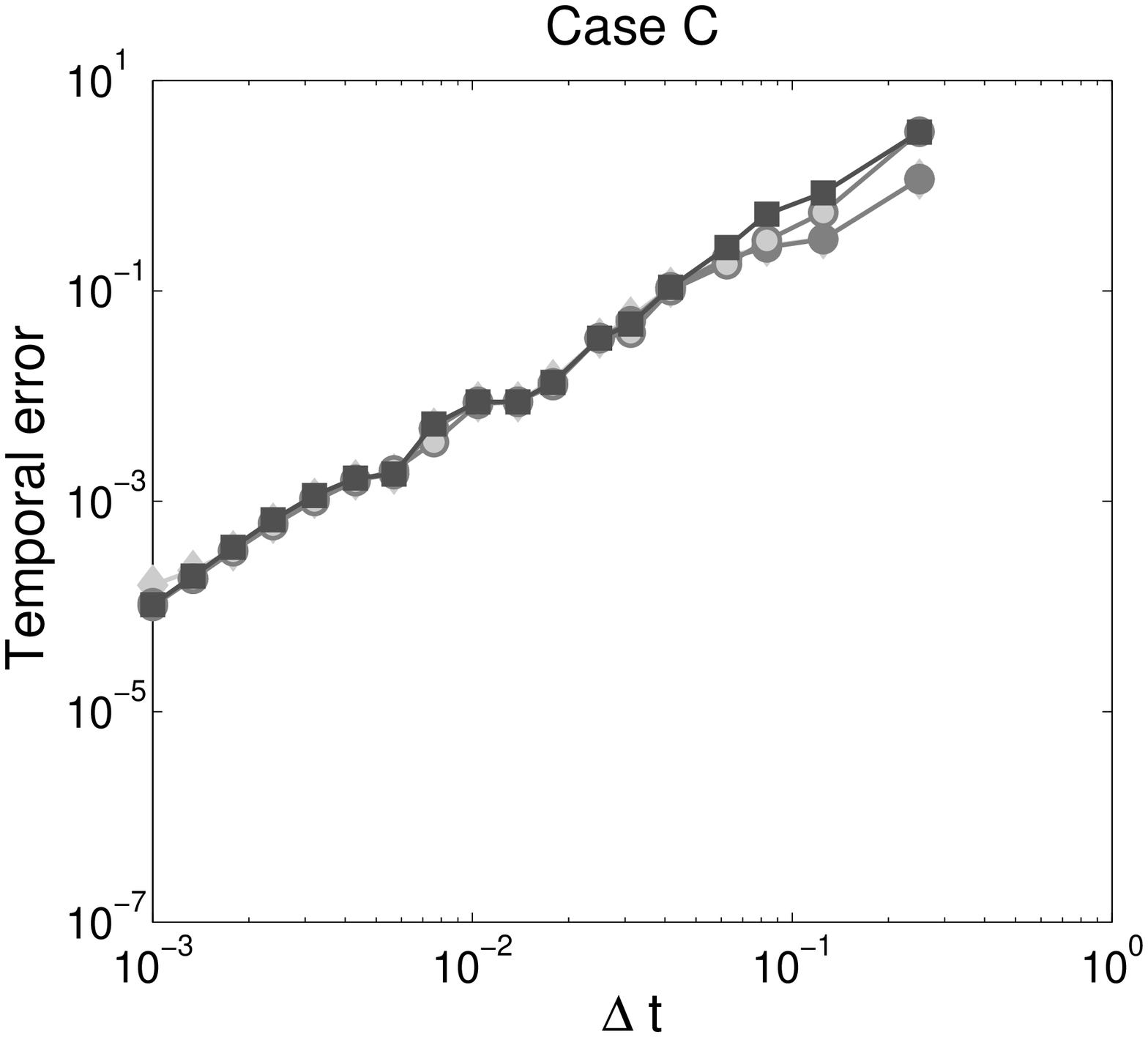}\\
         \includegraphics[width=0.6\textwidth]{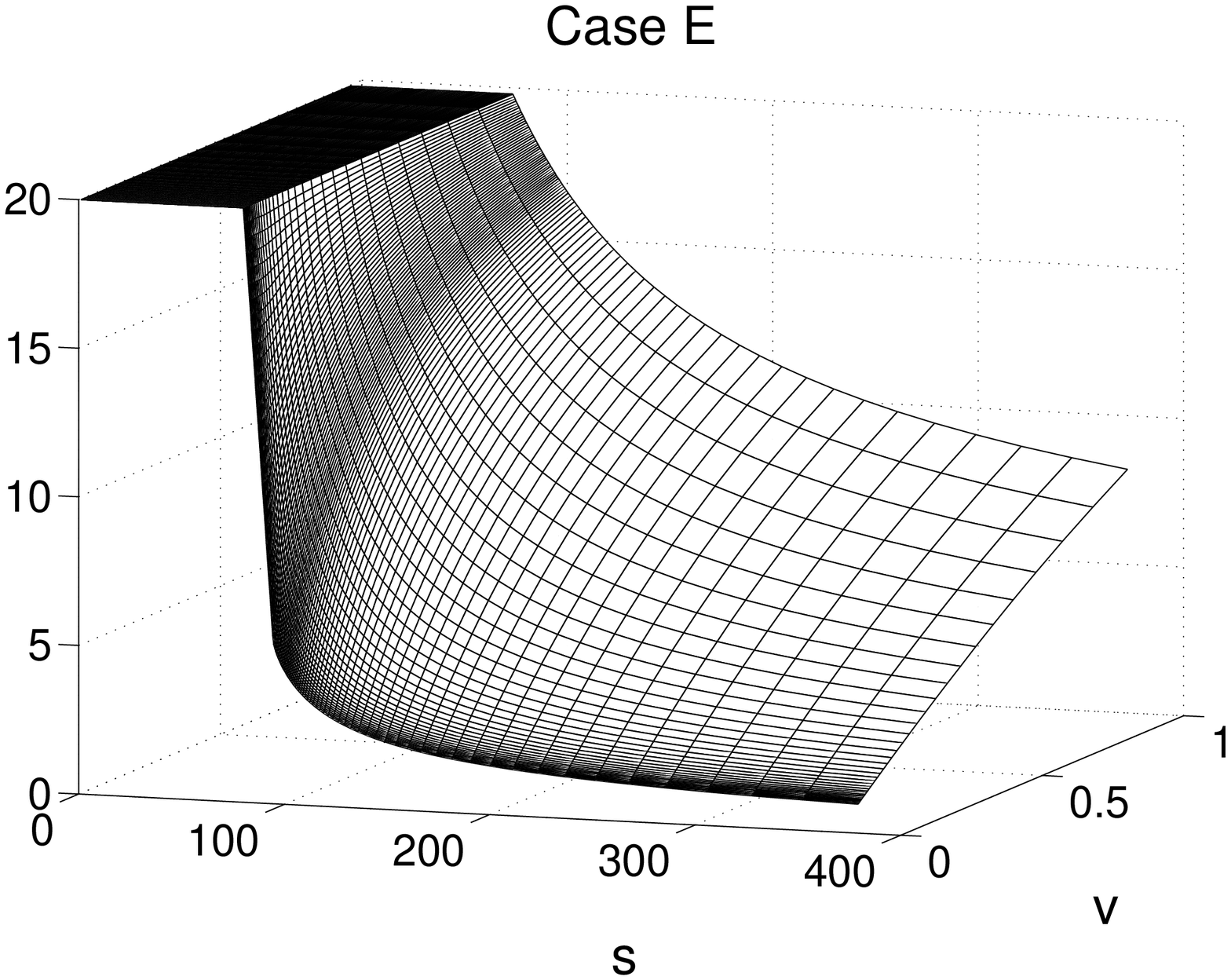}& \hspace*{-0.5cm}
         \includegraphics[width=0.5\textwidth]{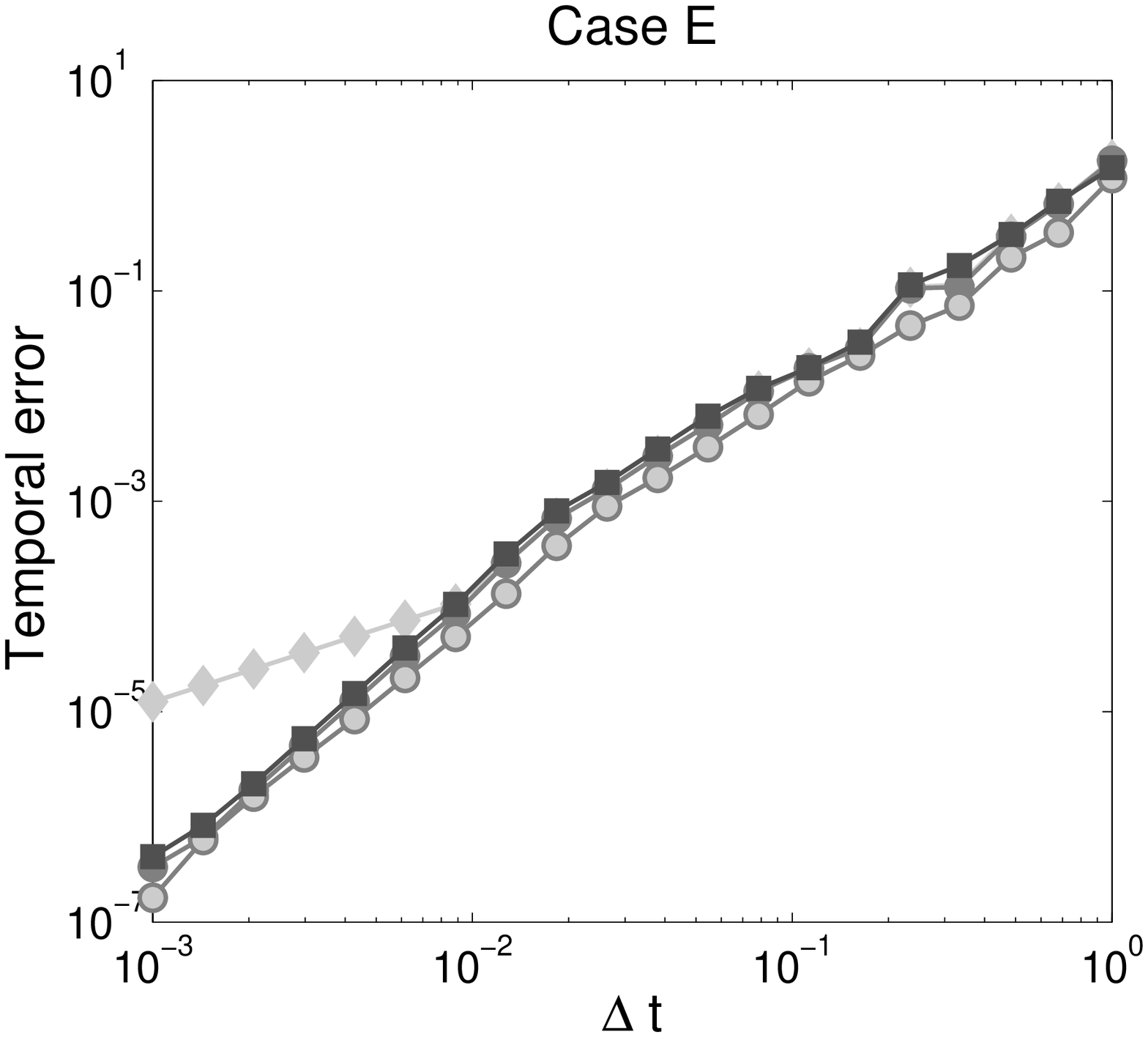}\\
\end{tabular}
\end{center}
\caption{Capped American put options in cases C and E of
Table~\ref{cases} with $\rho\neq 0$ and cap $B=80$.
Left: Option price surfaces.
Right: Temporal errors $\widehat{e}\,(\Delta t;100,50)$ vs.~$\Delta t$.
Four ADI-IT methods: Do-IT with $\theta=\frac{1}{2}$ (light diamond), CS-IT
with $\theta=\frac{1}{2}$ (dark circle), MCS-IT with $\theta=\frac{1}{3}$
(light circle) and HV-IT with $\theta=\frac{1}{2}+\frac{1}{6}\sqrt{3}$ (dark
square). Do-IT and CS-IT with damping - two steps $\Delta t/2$ with BE-IT.}
\label{TemporalErrorCappedPut}
\end{figure}

\noindent
To numerically solve the Heston PDCP for a capped American put
we follow the same approach as above in this paper, with spatial
discretization given in Section~\ref{space} and for the
time discretization the ADI-IT methods defined in
Section~\ref{time_ADI}.
Only two minor modifications are needed: set
\[
\xi_{\min} = \sinh^{-1}\left( \frac{B- S_{\lleft}}{d_1} \right)
\]
in the nonuniform mesh for the $s$-direction and modify
$S_{\lleft}$ to $\max\left(\frac{1}{2}K, e^{-rT}K, B \right)$.
The {\it ROI}\, is naturally truncated at $s=B$.
As an illustration we consider cases C and E with nonzero
correlation and cap $B=80$.
The left-hand side of Figure~\ref{TemporalErrorCappedPut}
displays the capped American put option price surfaces in
these two cases.
On the right-hand side of the figure, the temporal discretization
errors $\widehat{e}\,(\Delta t;100,50)$ for the four ADI-IT methods
are shown.
From these and additional experiments, we obtain similar, positive
results concerning unconditional stability and stiff convergence
as in the case of vanilla American put options.

%%%%%%%%%%%%%%%%%%%%%%%%%%%%%%%%%%%%%%%%%%%%%%%%%%%%%%%%%%%%%%%%%%%%%%%%%%%%%%%%%%%%
%%%%%%%%%%%%%%%%%%   SECTION 6   %%%%%%%%%%%%%%%%%%%%%%%%%%%%%%%%%%%%%%%%%%%%%%%%%%%
%%%%%%%%%%%%%%%%%%%%%%%%%%%%%%%%%%%%%%%%%%%%%%%%%%%%%%%%%%%%%%%%%%%%%%%%%%%%%%%%%%%%

\setcounter{equation}{0}
\section{Conclusions}\label{concl}
We have proposed a simple, effective adaptation
of ADI time discretization schemes to the numerical pricing
via PDCPs of American-style options under the Heston model.
The adaptation has been achieved by invoking a recent splitting
idea due to Ikonen \& Toivanen~\cite{IT09} and we refer to
the acquired methods as ADI-IT methods.
Four ADI-IT methods have been investigated in detail: Do-IT,
CS-IT, MCS-IT and HV-IT.
These are based on the Douglas, Craig--Sneyd, modified
Craig--Sneyd and Hundsdorfer--Verwer schemes.
The favorable result is found that, with properly chosen
values for their parameter $\theta$, all four methods show
an unconditionally stable behavior in the application to a
variety of representative, challenging test cases.
Next, they all exhibit a satisfactory convergence behavior,
provided Do-IT and CS-IT are used with a damping procedure.
In all but one test cases, the MCS-IT and HV-IT methods, as
well as the CS-IT method
applied with damping, show a stiff order of convergence equal
to two.
They always yield about the same size temporal error for
the same time step.
In one test case the observed order was slightly lower,
namely about 1.7, but this is still fine.
The Do-IT method applied with damping was found to perform
equally well as the other three methods for larger time steps,
but for smaller time steps the (expected) lower convergence
order of one sets in for this method whenever the correlation
is nonzero.
In view of the results in this paper, we recommend either
the MCS-IT or HV-IT method in the numerical pricing of
American options under the Heston model.
Also the CS-IT method is a good candidate whenever it is
used with damping.

A theoretical stability and convergence analysis of ADI-IT
methods is still open at this moment.
We proved a relevant, useful theorem for the BE-IT method,
i.e.~the backward Euler scheme combined with IT splitting.
The ideas in this proof may be helpful for an analysis of
ADI-IT methods in the future.

On the practical side, a comparison of the performance
of the ADI-IT approach to other numerical techniques
for the Heston PDCP is of much interest.
This requires an extensive and careful study, however,
and is left for future research.
Nevertheless, from the discussion and results presented
in this paper, we believe it is clear that ADI-IT methods
are expected to be competitive.

Finally, a merit of the ADI-IT approach is the versatility:
it is readily applicable in the case of many other underlying
asset pricing models, other American-style options,
or other semidiscretizations (by finite differences,
volumes or elements) of the pertinent PDCPs.

\section*{Acknowledgements}
This work has been supported financially by the Research
Foundation -- Flanders, FWO contract no.~G.0125.08.

% Figure 5

% Figure 6

% Figure 7

\vfill\eject

\end{document}